\documentclass[11pt]{article}
\pdfoutput=1 
\usepackage{./jheppub}

\usepackage[T1]{fontenc}

\usepackage{graphicx}
\usepackage{amsfonts,amsmath,amssymb}
\usepackage{verbatim}
\usepackage{array}
\usepackage{mathrsfs}
\usepackage{mathrsfs}
\usepackage{yfonts}
\usepackage{dsfont}
\usepackage{bbm}
\usepackage{colonequals}
\usepackage{amscd}

\usepackage{subcaption}

\usepackage{wrapfig}

\usepackage{suffix,mathtools,cancel,bbm}

\usepackage{multirow}

\usepackage{enumerate}

	\usepackage{tikz, tikz-3dplot, pgfplots}
	\usepackage{tkz-graph}
	\usetikzlibrary[positioning,patterns] 
	\usepackage{genyoungtabtikz}

\newcommand{\bn}{\begin{enumerate}}
\newcommand{\en}{\end{enumerate}}

\def\CI{{\cal I}}

\def\CN{{\cal N}}
\def\CO{{\cal O}}

\def\CS{{\cal S}}

\newcommand{\beq}{\begin{equation}}
\newcommand{\eeq}{\end{equation}}

\newcommand\nn{\nonumber}

\newcommand{\cC}{\mathcal{C}}

\newcommand{\cF}{\mathcal{F}}
\newcommand{\cG}{\mathcal{G}}

\newcommand{\cI}{\mathcal{I}}

\newcommand{\cK}{\mathcal{K}}
\newcommand{\cL}{\mathcal{L}}

\newcommand{\cN}{\mathcal{N}}
\newcommand{\cO}{\mathcal{O}}

\newcommand{\cS}{\mathcal{S}}

\newcommand{\cV}{\mathcal{V}}

\newcommand{\cX}{\mathcal{X}}

\numberwithin{equation}{section}

\def\bea{\begin{eqnarray}}
\def\eea{\end{eqnarray}}

\DeclarePairedDelimiterX\MeijerM[3]{\lparen}{\rparen}%
{\begin{smallmatrix}#1 \\ #2\end{smallmatrix}\delimsize\vert\,#3}

\newcommand\MeijerG[8][]{%
  G^{\,#2,#3}_{#4,#5}\MeijerM[#1]{#6}{#7}{#8}}

\WithSuffix\newcommand\MeijerG*[7]{%
  G^{\,#1,#2}_{#3,#4}\MeijerM*{#5}{#6}{#7}}

\def\cN{\mathcal{N}}

\def \beg#1{\begin{#1}} 

\def \bea{\beg{eqnarray}}
\def \eea{\end{eqnarray}}
\def \ee{\end{equation}}

\def \restr#1#2{{\left.\kern-\nulldelimiterspace#1\vphantom{\big|}\right|_{#2}}}

\def \nn{\nonumber}

\usepackage{colortbl}
\definecolor{mygray}{gray}{0.93}

\title{\boldmath 
Anomalies of QFTs from M-theory and Holography}

\author[a]{Ibrahima Bah,}
\author[a]{Federico Bonetti,}
\author[b,c]{Ruben Minasian,}
\author[d]{and Emily Nardoni} 
\affiliation[a]{Department of Physics and Astronomy, Johns Hopkins University, 3400 North Charles Street, Baltimore, MD 21218, USA}
\affiliation[b]{Institut de Physique Th\'{e}orique, Universit\'{e} Paris Saclay, CNRS, CEA, F-91191, Gif-sur-Yvette, France}
\affiliation[c]{School of Physics, Korea Institute for Advanced Study, Seoul 02455, Korea}
\affiliation[d]{Mani L. Bhaumik Institute for Theoretical Physics, Department of Physics and Astronomy, University of California, Los Angeles,  CA 90095, USA}

\emailAdd{iboubah@jhu.edu, fbonett3@jhu.edu, ruben.minasian@ipht.fr, enardoni@ucla.edu}

\abstract{
We describe a systematic way of computing the 't Hooft anomalies for continuous symmetries of Quantum Field Theories in even dimensions that can be geometrically engineered from M5-branes. 
Our approach is based on anomaly inflow, and characterizes the anomaly polynomial of the QFT in terms of the geometric definition of the field theory. 
In particular, when the QFT admits a holographic dual, the topological data of the solution is sufficient to compute the anomalies of the dual field theory, including finite terms in $N$.
We study several classes of examples in four and six dimensions, with or without known M5-brane probe configurations.
}

\usepackage{etoolbox}
\appto\appendix{\addtocontents{toc}{\protect\setcounter{tocdepth}{1}}}

\appto\listoffigures{\addtocontents{lof}{\protect\setcounter{tocdepth}{1}}}
\appto\listoftables{\addtocontents{lot}{\protect\setcounter{tocdepth}{1}}}

\begin{document} 

\setcounter{tocdepth}{2}

\maketitle
\flushbottom



\section{Introduction}
 
Geometric engineering is a powerful way to study Quantum Field Theories  (QFTs) and their various dynamics.
Many interesting QFTs can be explored by studying the low energy limit of branes in string theory backgrounds. 
Some of the most interesting and yet mysterious QFTs have such a definition: for instance, the $A_{N-1}$ $\CN=(2,0)$ superconformal field theories in six dimensions have a description as the worldvolume theories on a stack of $N$ flat M5-branes \cite{Witten:1995zh,Strominger:1995ac}; and a class of 6d theories with reduced $(1,0)$ supersymmetry are obtained from M5-branes probing an orbifold singularity \cite{Blum:1997mm}. Further wrapping the branes on a compact manifold yields a large class of lower dimensional QFTs. For example, by reducing the $(2,0)$ theories on a Riemann surface we obtain a large class of generically strongly coupled 4d QFTs of varying amounts of supersymmetry known as Class $\CS$, first analyzed for $\CN=2$ theories in \cite{Gaiotto:2009we,Gaiotto:2009hg}, and studied for $\CN=1$ cases in \cite{Maruyoshi:2009uk,Benini:2009mz,Bah:2011je,Bah:2011vv,Bah:2012dg}.  
Embedding these systems in string theory provides an organizing principle and geometric toolset for exploring their properties, especially in their strong coupling regimes. 

An important problem in these constructions is to compute the anomalies of the field theories\footnote{Throughout we restrict to the case of anomalies in background rather than dynamical gauge symmetries ('t Hooft anomalies), so that their existence does not render the theory inconsistent.}. 't Hooft anomalies provide a robust set of observables that are useful for probing the dynamics of QFTs.  They are preserved under renormalization group flow, and then can be used to constrain the IR phases of quantum systems via anomaly matching. 
For a superconformal field theory, the 't Hooft anomalies 
involving R-symmetry
are related to central charges by the superconformal algebra \cite{Anselmi:1997am,Kuzenko:1999pi}. 
Anomalies are also naturally geometric quantities. For the case of continuous symmetries in even $d$-dimensional QFTs, they are encoded in the anomaly polynomial, a $(d+2)$-form polynomial in curvature forms associated to gauge and gravity fields \cite{AlvarezGaume:1983ig,AlvarezGaume:1984dr,Bardeen:1984pm}. 
For QFTs obtained by dimensional reduction of a higher dimensional field theory, anomaly matching is implemented by integrating the upstairs anomaly polynomial over the compact manifold in the reduction. However, this prescription only gives the contribution to the lower dimensional anomaly polynomial that derives from symmetries manifest in the higher dimensional theory---it is not sensitive to decoupled sectors, accidental symmetries, and other subtleties. 

The primary objective of this work is to provide a systematic way of computing the anomalies of geometrically engineered QFTs in $d$ dimensions from M5-branes. 
 Our main tool is anomaly inflow in the M-theory background, first studied for M5-branes in \cite{Duff:1995wd,Witten:1996hc,Freed:1998tg,Harvey:1998bx}. The M5-branes act as a singular magnetic source for the M-theory 4-form flux $G_4$. In the supergravity description we excise a small neighborhood of the stack, thus inducing a boundary for 11d spacetime. For the geometries under consideration, this boundary is a fibration of an internal space $M_{10-d}$ over the low energy QFT worldvolume $W_d$.  The global symmetries of the QFT are fixed by isometries of the internal space, as well as gauge symmetries of the three-form potential.   In inflow, the anomalies for degrees of freedom on the branes must cancel the classical anomalous variation of the effective 11d supergravity action localized on the branes. 
The anomalous variation of the action is related by descent relations to a 12-form characteristic class $\CI_{12}$.
 Reducing $\CI_{12}$ along the transverse directions to the QFT worldvolume yields the inflow anomaly polynomial $I_{d+2}^{\text{inflow}}$ associated to the QFT. Then, the anomalies of the QFT are equal to those computed via inflow up to decoupling modes, such that $I_{d+2}^{\text{inflow}} + I_{d+2}^{\text{QFT}} + I_{d+2}^{\text{decoupl}} = 0$.

We provide a general prescription for computing $\CI_{12}$, and describe its properties and uniqueness. The essential point is that $\CI_{12}$ is determined entirely by topological properties of the internal space $M_{10-d}$, 
and equivariant classes constructed from the boundary data of the 4-form flux.  We see in examples of dimensionally reduced theories that the procedure of directly integrating $\CI_{12}$ to compute $I_{d+2}^{\text{inflow}}$ can contain more information than the reduction of the anomaly polynomial of the parent theory.
Further, the embedding in M-theory may allow for a geometric interpretation of the decoupling modes. 

We will then examine the implications of this machinery for holography, {\it i.e.}~when the $d$-dimensional QFT is a conformal field theory with a large-$N$ $AdS_{d+1}$ gravity dual ($N$ being the number of M5-branes).
The dual geometry in M-theory consists of a warped product of $AdS_{d+1}$ times $M_{10-d}^{\text{hol}}$, supported by a $G_4^{\text{hol}}$ flux configuration on the internal space. 
The transverse directions $M_{10-d}$ to the QFT worldvolume are identified with the internal space $M_{10-d}^{\text{hol}}$. 
The main observation is that we can use a known solution of 11d supergravity to infer $\CI_{12}$. This is because (1) the topology of the internal space $M_{10-d}^{\text{hol}}$ is the same as that of $M_{10-d}$, and (2) the holographic $G_4^{\text{hol}}$ flux configuration can be identified with our seed boundary $G_4$-flux utilized in the inflow machinery.

The power of our method in the context of holography is twofold. First, since the seed topological data for inflow can be read off of a known supergravity solution, we can obtain the CFT anomalies even if we don't know the probe M5-brane configuration. We demonstrate this below by applying our method to the $AdS_5\times M_6$ Gauntlett-Martelli-Sparks-Waldram (GMSW) solutions \cite{Gauntlett:2004zh}. Second, this prescription provides a path from the classical solution of 2-derivative supergravity which is valid at large-$N$, to the exact anomaly polynomial via inflow. 
Since our $\CI_{12}$ involves higher derivative terms inherited from the M-theory action, our prescription captures contributions to the anomalies at finite $N$. In  examples, we correctly produce all $N$-dependent anomaly coefficients excluding $\CO(N^0)$ terms which we can identify with decoupling modes.

We demonstrate our method in several examples, some of which have an explicit brane construction, and some cases in which only the holographic solution is known. 
In section \ref{sec_6d}, we focus on QFTs in six dimensions. First we exemplify our method in the case of flat M5-branes. Of course, anomaly inflow for this setup is well-known \cite{Freed:1998tg,Harvey:1998bx}\footnote{The anomaly polynomials for general 6d $(2,0)$ ADE theories can be computed using anomaly matching on the tensor branch \cite{Intriligator:2000eq}, while inflow for the $D_N$ series is studied in \cite{Yi:2001bz}. Also note that the holographic computation of the $c$ central charge is given in \cite{Henningson:1998gx}.}, and we use this analysis mainly to set notation for subsequent examples. We then apply our method to the case of M5-branes probing a $\mathbb{C}^2/\Gamma$ singularity, for $\Gamma$ an ADE subgroup of $SU(2)$. We match the anomaly polynomial for these $\CN=(1,0)$ theories as given in \cite{Ohmori:2014kda}, where an analysis via anomaly matching on the tensor branch as well as an inflow analysis appears. One comment is that our analysis produces an additional term relative to the inflow results from \cite{Ohmori:2014kda} that corresponds precisely to the contribution of the Green-Schwarz term associated to the center of mass mode for the branes. For the case of ${\Gamma} = \mathbb{Z}_k$, we reproduce the full result of \cite{Bah:2017gph}.

In section \ref{sec_4d}, we consider two classes of four-dimensional $\CN=1$ QFTs. First, we reproduce the anomalies of 4d $\CN=1$ SCFTs for which the internal space $M_{6}$ is an $S^4$ fibration over a smooth Riemann surface, analyzed by Bah-Beem-Bobev-Wecht (BBBW) \cite{Bah:2011vv,Bah:2012dg}. We then apply our method to a class of 4d SCFTs where $M_6$ is an $S^2$ fibration over a product of smooth Riemann surfaces, corresponding to the GMSW supergravity solutions \cite{Gauntlett:2004zh}. The M5-brane probe description is generally not known for these solutions, but we are nonetheless able to compute the inflow anomaly polynomial, and show that we match the holographic computation of the central charge \cite{Gauntlett:2006ai}. Our results are the first computation of the subleading-in-$N$ corrections to the central charges of this class of QFTs. 

We conclude with a discussion of our results, as well as several appendices that contain the details of computations that appear in the main text.



\section{General aspects of anomaly inflow in M-theory} \label{sec:strategy}

In this section we discuss general aspects of anomaly inflow
 for M-theory setups with
wrapped M5-branes. We establish a connection
to holography and outline a general recipe for
obtaining
 the  inflow anomaly polynomial. 
The latter is governed by a 4-form $E_4$
that encodes 
topological information about the $G_4$-flux configuration.
In this section we discuss general properties of $E_4$
and all the necessary
ingredients for its construction.

\subsection{Anomaly inflow for wrapped M5-branes}

Let us consider a stack of $N$ M5-branes
with worldvolume $W_6$.
The tangent bundle of the 11d ambient space $M_{11}$,
restricted to the worldvolume $W_6$,
decomposes according to
\beq
TM_{11} \Big|_{W_6} = TW_6 \oplus NW_6 \ ,
\eeq
where $TW_6$ is the tangent bundle to the stack
and $NW_6$ is the normal bundle to the stack.
The latter has structure group $SO(5)$
and encodes the five transverse directions of the M5-branes.

We are interested in setups in which 
\beq
W_6 =W_d \times \cS_{6-d} \ ,
\eeq
where
$W_d$ is external $d$-dimensional spacetime
and
 $\cS_{6-d}$ is a smooth compact 
even-dimensional
internal space.
At low energies the system is described
by a QFT living on $W_d$.
For $d = 6$, the internal space is understood
to be absent. 
For $d<6$,
in order to specify the setup
we have to describe the topology of the normal
bundle $NW_6$ over the internal space 
$\cS_{6-d}$. This amounts to specifying a partial
topological twist of the worldvolume theory
living on the M5-branes upon compactification
on $\cS_{6-d}$.
The topological twist is essential
in preserving  some supersymmetry in the $d$ non-compact
directions.

Given the internal space $\cS_{6-d}$ wrapped by the M5-branes
and the $S^4$ that surrounds the stack in its five transverse directions,
there is a compact $(10-d)$-dimensional space $M_{10-d}$
that encodes the topological twist and governs the anomalies
of the QFT on $W_d$.
The space $M_{10-d}$ is an $S^4$ fibration over 
$\cS_{6-d}$,
\beq \label{M10minusd}
S^4 \hookrightarrow M_{10-d} \rightarrow \cS_{6-d} \ .
\eeq
The structure group of the fibration \eqref{M10minusd}
is a subgroup of the $SO(5)$ structure group
of the normal bundle $NW_6$.

The QFT living on $W_d$ at low energies
can admit  't Hooft anomalies for the   global symmetries
of the theory. In this work we focus on continuous   symmetries.
Their anomalies are diagnosed by coupling the 
 QFT  on $W_d$ to
 background gauge connections and   metric.

Since the full M-theory setup is anomaly-free,
the 't Hooft anomalies in $d$ dimensions must be
counterbalanced by anomaly inflow
from the M-theory bulk,
which is analyzed using the methods of \cite{Freed:1998tg,Harvey:1998bx}.
The M5-brane stack acts as a singular magnetic
source for the M-theory $G_4$ flux.
To describe the setup in supergravity,
the singularity is removed by 
 excising 
a small tubular neighborhood of the 
M5-brane stack. 
As a result, the 11d spacetime $M_{11}$ acquires
a boundary $\partial M_{11} = M_{10}$. If $r$ denotes the radial coordinate
away from the M5-brane stack,
 $M_{10}$ is located at $r = \epsilon$,
 where $\epsilon$ is a small positive constant.
The space $M_{10}$ is a fibration of $M_{10-d}$
over $W_d$, 
\beq \label{M10_fibration}
M_{10-d} \hookrightarrow M_{10} \rightarrow W_d \ .
\eeq
The fibration \eqref{M10_fibration}
is specified by the background gauge connections
for the symmetries of the QFT that originate
from continuous isometries of $M_{10-d}$.
Let us stress that the fibration 
\eqref{M10_fibration}
encodes the gauging of the $d$-dimensional theory
with background gauge fields on $W_d$,
while the fibration \eqref{M10minusd} describes the
topological twist that defines the theory on $W_d$.

The magnetic source for $G_4$ is modeled by
imposing suitable boundary conditions
 at $r = \epsilon$. More precisely, we have
\beq \label{G4vsE4}
\left. \frac{G_4}{2\pi} \right|_{r=\epsilon} = E_4  \ ,
\eeq
where $E_4$ is a closed 4-form on the space $M_{10}$.
The relation \eqref{G4vsE4} is written in conventions
in which $G_4$-flux quantization reads
\beq \label{G4_quantization}
\int_{\cC_4} \bigg(\frac{G_4}{2\pi} - \frac{\lambda}{2} \bigg)\in \mathbb Z \ ,
\eeq
where $\cC_4$ is a 4-cycle  and
$\lambda = p_1(TM_{11})/2$ \cite{Witten:1996md}.
(In all setups we consider $\lambda/2$ is integral.)
The 4-form $E_4$
  has to be globally defined 
and invariant under the structure group
of the fibration \eqref{M10_fibration}, \emph{i.e.}~invariant
under all   symmetries of the $d$-dimensional theory.
Moreover, we have
\beq \label{N_flux}
\int_{S^4} E_4 = N \ ,
\eeq
where $S^4$ is the 4-sphere
surrounding the stack.
Depending on the choice of $\cS_{6-d}$
and the topology of the fibration \eqref{M10minusd},
the 4-form $E_4$ might have additional non-trivial
fluxes through 4-cycles in $M_{10}$,
besides \eqref{N_flux}.

As explained in \cite{Freed:1998tg,Harvey:1998bx},
in the presence of the boundary $M_{10}$
the topological terms in the low-energy effective
action of M-theory $S_{{11d}}$
are no longer invariant under 
gauge transformations in the background connections on $W_d$.
In fact, we have
\beq
\frac{\delta S_{11d}}{2\pi} = \int_{M_{10}} \cI_{10}^{(1)} \ ,
\eeq
where $ \cI_{10}^{(1)}$ is a 10-form on $M_{10}$,
linear in the gauge parameters.
In accordance with the Wess-Zumino
consistency conditions,
$\cI_{10}^{(1)}$ is related by descent to a 
12-form characteristic class~$\cI_{12}$,
\beq
d\cI_{10}^{(1)} = \delta \cI_{11}^{(0)} \ , \qquad
d\cI_{11}^{(0)} = \cI_{12} \ .
\eeq
The class $\cI_{12}$ is given by
\beq \label{I12_def}
\cI_{12} = - \frac 16 \, E_4^3 - E_4 \, X_8 \ , 
\eeq
where we have suppressed wedge products.
The 8-form $X_8$ is given by
\beq \label{X8def}
X_8 = \frac{1}{192} \bigg[ p_1(TM_{11})^2
- 4 \, p_2(TM_{11}) \bigg] \ ,
\eeq
where the quantities $p_{1,2}(TM_{11})$
are the first and second Pontryagin
classes of the 11d tangent bundle,
implicitly pulled back to the boundary at $r = \epsilon$.

If we integrate the class $\cI_{12}$ on the
$M_{10-d}$ fibers of \eqref{M10_fibration},
we obtain the $(d+2)$-form inflow anomaly polynomial
of the $d$-dimensional theory on $W_d$,
\beq \label{Iinflow}
I_{d+2}^{\rm inflow} = \int_{M_{10-d}} \cI_{12} \ .
\eeq
The inflow anomaly polynomial 
\eqref{Iinflow} cancels against the 't Hooft anomalies
of the interacting QFT living on $W_d$ at low energies,
and of the decoupling modes related to the center-of-mass
of the M5-brane stack.
We thus write
\beq
I_{d+2}^{\rm inflow} + I_{d+2}^{\rm QFT} + I_{d+2}^{\rm decoupl} = 0 \ .
\eeq

\subsection{Applications to holography}

One of the main interests of this work is the case in which the interacting QFT 
on $W_d$ is a CFT with a gravity dual.
The dual geometry in M-theory is a warped product
of $AdS_{d+1}$ with an internal $(10-d)$-dimensional
space,
\beq
M_{11} = AdS_{d+1} \times_w M_{10-d}^{\rm hol} \ .
\eeq
This $AdS_{d+1}$ solution is supported 
by a non-trivial $G_4$-flux
configuration $G_4^{\rm hol}$ on the internal space
$M_{10-d}^{\rm hol}$.\footnote{For $d=2$,
$G_4^{\rm hol}$ can have additional terms 
with three external legs, proportional to the
volume form on $AdS_3$. Such terms in $G_4^{\rm hol}$
do not play a role in the following discussion.
}

The main observation is that,
for $AdS_{d+1}$ solutions that correspond to wrapped M5-branes,
 the topology
of the internal space $M_{10-d}^{\rm hol}$
is the same as the topology of the space $M_{10-d}$
defined in \eqref{M10minusd}
\cite{Gauntlett:2006ux}.
By a similar token,
the holographic $G_4$-flux
configuration $G_4^{\rm hol}/(2\pi)$
lies in the same cohomology class
as $E_4$ after all external connections are turned off,
\beq
\frac{G_4^{\rm hol}}{2\pi}  = E_4 \Big|_{\text{external connections = 0}} 
\qquad \text{in cohomology of $M_{10-d}$} \ .
\eeq
Let us emphasize that the topological properties of
$M_{10-d}$ and $E_4$ are the main
ingredients in the implementation of anomaly
inflow for wrapped M5-branes.
The discussion above indicates that these topological
features can be equivalently   extracted from the probe setup
or from the holographic solution.

There exists a larger set of $AdS_{d+1}$ solutions
in M-theory for which the probe 
M5-brane configuration is not known.
In these solutions the internal space
$M_{10-d}^{\rm hol}$
is not necessarily   an $S^4$ fibration over
some $(6-d)$-dimensional space, as in \eqref{M10minusd}.
We expect that our method for the computation
of $I_{d+2}^{\rm inflow}$ applies to such setups.
The anomaly is governed by the topological
properties of 
 $M_{10-d}^{\rm hol}$ and  $G_4^{\rm hol}$,
which  
determine $E_4$.

The general task at hand is the construction of $E_4$
given the 
 topology of the space $M_{10 -d}$.
Recall that $E_4$ is a 4-form on the total space $M_{10}$
of the fibration \eqref{M10_fibration}.
Crucially, we do not assume that $M_{10-d}$ is an $S^4$
fibration over a $(6-d)$-dimensional space.
As a result, the following considerations apply beyond
setups that are realized by wrapping M5-branes 
on a smooth compact internal space.

A local representative for the class $E_4$ is     constrained by the following properties:
\begin{itemize}
\item $E_4$ is globally defined,
\item $E_4$ is closed,
\item $E_4$ is  invariant under all symmetries of $M_{10-d}$.
\end{itemize}
The 4-form $E_4$ is constructed by combining the curvatures
of the background connections on $W_d$ with $p$-forms
in the internal space $M_{10-d}$.
Crucially, since $M_{10-d}$ is fibered over $W_d$,
see \eqref{M10_fibration},
the internal $p$-forms on $M_{10-d}$
must be appropriately ``gauged'', \emph{i.e.}~coupled
to the background connections on $W_d$.

The constraints listed above may not completely
fix the expression of $E_4$.
In section \ref{sec_E4_in_general} we present a general recipe for the 
construction of $E_4$, we characterize its ambiguities,
and we argue that they do not affect the inflow anomaly polynomial.



\section{Construction of $E_4$  } \label{sec_E4_in_general}

In this section we introduce a convenient formalism for the
parametrization of the
4-form $E_4$.  We show 
how to construct a good representative for $E_4$
in terms on $p$-forms in $M_{10-d}$.
The natural language to describe this construction is that of $G$-equivariant
cohomology.
While $E_4$ is generically non-unique,
we argue that
the inflow anomaly polynomial can be extracted unambiguously.

\subsection{Parametrization of $E_4$} \label{sec_general_param}

A local representative for the class $E_4$ is a closed, gauge-invariant,
globally-defined 4-form on the total space $M_{10}$ of the fibration of $M_{10-d}$
over external spacetime $W_d$, see \eqref{M10_fibration}.
The 4-form $E_4$ is constructed using 
 internal $p$-forms on the $M_{10-d}$ fibers,
together with
external curvatures
with legs on $W_d$.

We suppose that $M_{10-d}$ admits a collection of Killing vectors
$k_I^m$, with $m$ a curved  index
on $M_{10-d}$,
and $I$ labeling a basis of Killing vectors.
The latter obey the Lie algebra
\beq
\pounds_I k_J \equiv \pounds_{k_I} k_J = [k_I, k_J] = f_{IJ}{}^K \, k_K \ ,
\eeq
where $\pounds$ denotes Lie derivative.
The non-trivial fibration of $M_{10-d}$ over $W_d$ is encoded
by the gauging of the isometries of $M_{10-d}$.
In what follows, we adopt a notation similar to the one of \cite{Benvenuti:2006xg}.
The gauging
is conveniently described locally by the replacement
\beq \label{eq_replacement}
d\xi^m \;\; \rightarrow \;\;
D\xi^m = d\xi^m + k^m_I \, A^I \ ,
\eeq
where $A^I$ is the external connection associated to the Killing
vector $k^m_I$.
In our conventions, the field strength $F^I$ of the connection
$A^I$ reads
\beq \label{non_abelian_F}
F^I = dA^I - \frac 12 \, f_{JK}{}^I \, A^J \, A^K \ .
\eeq
Let $\omega$ be a $p$-form on $M_{10-d}$, 
\beq
\omega = \frac{1}{p!} \, \omega_{m_1 \dots m_p}  \, d\xi^{m_1} \dots
d\xi^{m_p } \ ,
\eeq
where the components $\omega_{m_1 \dots m_p}$
depend only on the coordinates $\xi$ on $M_{10-d}$.
We use the symbol $\omega^{\rm g}$ for the gauged
version of $\omega$, obtained 
by means of the replacement
\eqref{eq_replacement},
\beq \label{def_gauging}
\omega^{\rm g} = \frac{1}{p!} \, \omega_{m_1 \dots m_p}  \, D\xi^{m_1} \dots
D\xi^{m_p } \ .
\eeq
Further details about this gauging procedure are collected in appendix \ref{app_gauging}.

If we turn off all external connections, $E_4$ reduces to
a 4-form denoted $V_4$. The latter must be closed and invariant
under all the isometries of $M_{10-d}$.
Furthermore, $G_4$-flux quantization requires the class of $V_4$
to be integral.
Let $\cV^\Lambda$, $\Lambda = 1, \dots, b^4(M_{10-d})$ be a basis
of the integral cohomology group $H^4(M_{10-d},\mathbb Z)$.\footnote{For 
$d=4$, the internal space is six dimensional
and 4-cycles are dual to 2-cycles and harmonic 2-forms;
the label $\Lambda$ coincides with $\alpha$ in this case.}
We can expand the cohomology class of $V_4$ as
\beq \label{V4_cohomology}
V_4 =  N_\Lambda \, \cV^\Lambda \ ,  
\eeq
with the integers $N_\Lambda$ specifying the topology of the
$G_4$-flux configuration, which is part of the input data
that defines the setup and 
 is held fixed throughout the construction of $E_4$.

The first step in the parametrization of $E_4$ is 
promoting $V_4$ to a closed and gauge-invariant object.
The completion of $V_4$ is denoted $V_4^{\rm eq}$ and is given by
\beq \label{V4_eq_def}
V_4^{\rm eq} = V_4^{\rm g} + F^I \, \omega_I^{\rm g} + F^I \, F^J \,\sigma_{IJ} \ .
\eeq
In the previous expression, $\omega_I$ are 2-forms on $M_{10-d}$,
while $\sigma_{IJ}$ are 0-forms.
The superscript `g' refers to the gauging prescription defined in \eqref{def_gauging},
while `eq' stands for equivariant, for   reasons outlined below.
The 4-form $V_4^{\rm eq} $ must be invariant under the gauge transformations
associated to the isometries of $M_{10-d}$.
Let  $\delta_\lambda V_4^{\rm eq} $ denote the gauge variation of $V_4^{\rm eq} $,
with $\lambda^I$ denoting the gauge parameters.
In computing $\delta_\lambda V_4^{\rm eq}$,
we encounter variations of external curvatures, and variations of
gauged internal forms.
The former are given by the usual expression,
\beq
\delta_\lambda F^I = - f_{JK}{}^I \, \lambda^J \, F^K \ .
\eeq
The gauge variations of gauged internal forms
are given in terms of the Lie derivative
with respect to the Killing vector fields.
For example,
\beq 
\delta_\lambda (\omega_I^{\rm g}) = \lambda^J \, (\pounds_J \omega_I)^{\rm g} \ .
\eeq
We refer the reader to appendix \ref{app_gauging}
for a more detailed explanation of this point.
The requirement $\delta_\lambda V_4^{\rm eq} = 0$
translates into the conditions
\begin{align}  \label{V4_gauge}
\pounds_I V_4 &= 0 \ , \qquad 
\pounds_I \omega_J = f_{IJ}{}^K \, \omega_K \ , \qquad
\pounds_I \sigma_{J_1 J_2} = f_{IJ_1}{}^K \, \sigma_{KJ_2}
+ f_{IJ_2}{}^K \, \sigma_{J_1 K } \ .
\end{align}

We also have to demand $d(V_4^{\rm eq} )= 0$.
In computing the external derivative of $V_4^{\rm eq}$,
the following identity is useful,
\beq \label{useful_identity}
d (\omega_I^{\rm g} ) + A^J \, (\pounds_J \omega_I)^{\rm g} = (d\omega_I)^{\rm g}
+ F^J \, (\iota_J \omega_I)^{\rm g} \ .
\eeq
The quantity on the LHS is the natural covariant
derivative acting on a gauged internal form,
since the gauge algebra acts via Lie derivative
along Killing vector fields.
Accordingly, the RHS is a manifestly gauge invariant
quantity. Even though we have written \eqref{useful_identity}
for $\omega_I$, a similar identity holds for any
internal gauged form. Further details on this point
can be found in appendix \ref{app_gauging}.
Making use of \eqref{useful_identity} and similar identities,
together with the Bianchi identity
for $F^I$,
we verify that requirement $d(V_4^{\rm eq} )= 0$ implies the following 
conditions,
\begin{align} \label{V4_closure}
dV_4 = 0 \ , \qquad
\iota_I V_4 + d\omega_I = 0 \ , \qquad
\iota_{(I} \omega_{J)} + d\sigma_{IJ} = 0 \ .
\end{align}
The symbol $\iota_I$ is a shorthand notation
for the interior product $\iota_{k_I}$ of a $p$-form
with   the Killing vector $k_I^m$.

So far, we have only discussed the external connections 
associated to isometries of $M_{10-d}$.
Another class of external connections is 
related to harmonic 2-forms on $M_{10-d}$.
If $\omega_\alpha$, $\alpha = 1, \dots, b^2(M_{10-d})$,
is a basis of harmonic 2-forms, expansion of the M-theory 3-form
potential onto $\omega_\alpha$ yields external vectors
$A^\alpha$, according to the schematic relation
$C_3 = A^\alpha \, \omega_\alpha$.
The connections $A^\alpha$ are Abelian and have
field strength $F^\alpha = dA^\alpha$.
Notice that, for $d=4$, one linear combination of the vectors
$A^\alpha$ is massive and does not correspond to a symmetry of the system.
This point is addressed in greater detail in appendix \ref{4d_discussion}.

Each harmonic 2-form $\omega_\alpha$ is closed and invariant under 
the action of all isometries of $M_{10-d}$.\footnote{The
fact that $\pounds_I \omega_\alpha = 0$ can be seen as follows.
From
$d\omega_\alpha=0$ we derive
$\pounds _I \omega_\alpha = d(\iota_I \omega_\alpha)$.
Making use of $\nabla_{(m} k_{I|n)}  = 0$ and  $\nabla^m \omega_{\alpha mn} = 0$,
we verify
$(\pounds_I \omega_\alpha)_{mn} = \nabla^p (k_I \wedge \omega_\alpha)_{pmn}$.
We have thus established that the 2-form $\pounds_I \omega_\alpha$
is both exact and co-exact.
It follows that 
 $\int_{M_{10-d}} (\pounds_I \omega_\alpha) *   (\pounds_I \omega_\alpha) = 0$ (no sum over $\alpha$, $I$), which in turn guarantees
$\pounds_I \omega_\alpha = 0$.
} 
As a result, we can complete $\omega_\alpha$ to a closed and gauge-invariant
object, denoted $\omega_\alpha^{\rm eq}$. It is given by
\beq \label{omega_eq_def}
\omega_\alpha^{\rm eq} = \omega_\alpha^{\rm g} + 2 \, F^I \, \sigma_{I\alpha} \ ,
\eeq
where $\sigma_{I\alpha}$ are suitable 0-forms on $M_{10-d}$,
and the factor of 2 is inserted for later convenience.
We demand $\delta_\lambda \omega_\alpha^{\rm eq} = 0$,
such that 
\beq \label{omega_alpha_gauge}
\pounds_I \omega_\alpha = 0 \ , \qquad
\pounds_I \sigma_{J\alpha} = f_{IJ}{}^K \, \sigma_{K\alpha} \ .
\eeq
Moreover, we need $d(\omega_\alpha^{\rm eq}) = 0$,
which is equivalent to
\beq \label{omega_alpha_closure}
d\omega_\alpha = 0 \ , \qquad
\tfrac 12 \, \iota_I \omega_\alpha + d\sigma_{I\alpha} = 0 \ .
\eeq

We are now in a position to write down the most general parametrization of
$E_4$. It reads
\beq \label{E4_new}
E_4 = V_4^{\rm eq} + \frac{F^\alpha}{2\pi} \, \omega_\alpha^{\rm eq}
+ \gamma_4 \ ,
\eeq
where  $V_4^{\rm eq}$ is as in \eqref{V4_eq_def}, $\omega_\alpha^{\rm eq}$
is as in \eqref{omega_eq_def}, while  
$\gamma_4$ denotes an arbitrary closed and gauge-invariant 4-form
with purely external legs.
In appendix \ref{app_gamma4}
we show that for $d = 2$ and $d = 4$ the inflow
anomaly polynomial is insensitive to the 4-form $\gamma_4$,
which may then be set to zero.
For $d=6$, the inflow anomaly polynomial does depend on $\gamma_4$.
In appendix \ref{app_gamma4} we 
argue that 
the correct value of $\gamma_4$ is obtained
by extremizing the inflow anomaly polynomial.
This prescription is equivalent to demanding that the 8-form
$E_4^2 + 2 \, X_8$ be trivial in the cohomology of $M_4$.
We interpret this requirement on $E_4^2 + 2 \, X_8$
as a tadpole cancellation condition in M-theory,
which must be satisfied in order to have a consistent setup.
In section  \ref{sec_6d} we verify that our treatment of $\gamma_4$
gives the correct inflow anomaly in two examples,
a flat stack of M5-branes, and M5-branes probing a $\mathbb C^2/\Gamma$
singularity.

The quantities $V_4^{\rm eq}$ and $\omega_\alpha^{\rm eq}$
introduced above admit a natural interpretation in terms
of $G$-equivariant cohomology of $M_{10-d}$, where $G$ is
 the isometry group of $M_{10-d}$. 
This justifies the label `eq'. 
 In appendix \ref{app_equiv}
we show how the objects
$V_4^{\rm eq}$, $\omega_\alpha^{\rm eq}$
can be identified with $G$-equivariantly closed
(poly)forms on $M_{10-d}$, which specify non-trivial $G$-equivariant
cohomology classes.

Let us close this section by introducing a more compact notation,
which is sometimes convenient in what follows.
Let $X = (I,\alpha)$ be a collective index 
that unifies external connections originating from isometries of $M_{10-d}$
and external connections associated to harmonic 2-forms on $M_{10-d}$.
We can cast  $E_4$ in \eqref{E4_new}
in the form 
\beq \label{compact_E4}
E_4 = V_4^{\rm g} + F^X \, \omega_X^{\rm g} 
+ F^X \, F^Y \, \sigma_{XY}
+   \gamma_4 \ ,
\eeq
with the identifications
\beq
F^X = (F^I , \tfrac{1}{2\pi} \, F^\alpha )   \ , \qquad
\omega_X = (\omega_I , \omega_\alpha) \ , \qquad
\sigma_{XY} = \begin{pmatrix}
\sigma_{IJ} & \sigma_{I\beta} \\
\sigma_{J\alpha} & 0
\end{pmatrix} \ .
\eeq
By a similar token, we can summarize
\eqref{V4_gauge} and \eqref{omega_alpha_gauge}
by writing
\beq \label{gauge_summary}
\pounds_X V_4 = 0 \ , \qquad
\pounds_X \omega_Y = f_{XY}{}^Z \, \omega_Z \ , \qquad
\pounds_X \sigma_{Y_1 Y_2} = f_{XY_1}{}^Z \, \sigma_{Z Y_2}
+ f_{XY_2}{}^Z \, \sigma_{Y_1 Z} \ ,
\eeq
with the understanding that $\pounds_\alpha \equiv 0$,
and that the only non-zero entry of $f_{XY}{}^Z$ are $f_{IJ}{}^K$.
In a similar fashion, we summarize \eqref{V4_closure},
\eqref{omega_alpha_closure} as
\beq \label{closure_summary}
dV_4 = 0 \ , \qquad
\iota_X V_4 + d\omega_X = 0 \ , \qquad
\iota_{(X} \omega_{Y)} + d\sigma_{XY} = 0 \ ,
\eeq
with the convention $\iota_\alpha \equiv 0$.

It is interesting to note that the compact expression \eqref{compact_E4}
is suggestive of a possible extension
of the $G$-equivariant cohomology interpretation
of appendix \ref{app_equiv}.
Indeed, we can formally augment the isometry group $G$
to a larger group $\widehat G$, by adding an extra $U(1)$
factor for each curvature $F^\alpha$.
The extra $U(1)$ generators act trivially on $M_6$,
because of $\iota_\alpha \equiv 0$.
We can then interpret
the quantity $V_4^{\rm g} + F^X \, \omega_X^{\rm g} 
+ F^X \, F^Y \, \sigma_{XY}$
as a $\widehat G$-equivariant 
(as opposed to $G$-equivariant) completion of~$V_4$.

\subsection{Deformations of $E_4$} \label{sec_remarks_on_E4}

An important question concerning  the construction of $E_4$ is to
determine how uniquely this object is fixed
by the conditions
\eqref{V4_gauge}, \eqref{V4_closure},
\eqref{omega_alpha_gauge}, \eqref{omega_alpha_closure}.
We refer the reader to appendix \ref{app_deformations} for a detailed
analysis of this problem, and here we discuss only 
some salient aspects.

A class of deformations of $E_4$
is of the form
\begin{align} \label{useless_deformation}
V_4 &\rightarrow V_4 + dW_3 \ , &
\omega_I &\rightarrow \omega_I + \iota_I W_3 + d\lambda_I \ , &
\sigma_{IJ} &\rightarrow \sigma_{IJ} + \iota_{(I}\lambda_{J)} \ , \nn \\
&& \omega_\alpha & \rightarrow \omega_\alpha + d\lambda_\alpha \ , &
\sigma_{I\alpha} & \rightarrow \sigma_{I\alpha} + \tfrac 12 \, \iota_I \lambda_\alpha \ .
\end{align}
where $W_3$ is a globally-defined 3-form on $M_{10-d}$,
and $\lambda_I$, $\lambda_\alpha$ are globally-defined
1-forms on $M_{10-d}$. Gauge-invariance requires
\beq
\pounds_I W_3 = 0 \ , \qquad
\pounds_I \lambda_J = f_{IJ}{}^K \, \lambda_K \ , \qquad
\pounds_I \lambda_\alpha = 0 \ ,
\eeq
but $W_3$, $\lambda_I$, $\lambda_\alpha$ are otherwise
arbitrary.
It is easily checked that the new $E_4$ is still closed and gauge-invariant.
Furthermore, we have checked that
the integrals $\int_{M_{10-d}} E_4^3$ 
and $\int_{M_{10-d}} E_4 \, X_8$
are   invariant under
the deformation \eqref{useless_deformation},
which implies that the inflow anomaly polynomial
is unaffected by it.
In the language of $G$-equivariant cohomology,
the deformation \eqref{useless_deformation}
corresponds to adding $G$-equivariantly exact pieces
to $E_4$, which does not change the $G$-equivariant cohomology class of $E_4$.

A more interesting class of ambiguities in the determination of $E_4$ is the following.
For a given $V_4$, we can construct equivariant completions
$V_4^{\rm eq}$ that 
correspond to 
different $G$-equivariant cohomology classes.
More explicitly, we can consider the modification
\beq \label{new_V4eq}
\omega_I \rightarrow \omega_I + c^\alpha_I \, \omega_\alpha \ , \qquad
\sigma_{IJ} \rightarrow \sigma_{IJ} + c^\alpha_{(I} \, \sigma_{J)\alpha}
+ u_{IJ} \ .
\eeq
The quantities $c^\alpha_I$, $u_{IJ}$ are constants.
Compatibility with \eqref{V4_gauge} 
requires
\beq \label{V4_nontrivial}
f_{IJ}{}^K \, c^\alpha_K = 0 \ , \qquad
f_{IJ_1}{}^K \, u_{K J_2}
+ f_{IJ_2}{}^K \, u_{J_1 K} = 0 \ .
\eeq
In other words, $c^\alpha_I$ must be acted upon trivially by the
adjoint representation, which means that $c^\alpha_I$
can be non-zero only if
$I$ labels a generator of an Abelian factor
of the isometry group.
On the other hand,
the constants $u_{IJ}$ must be components of a symmetric invariant tensor
in the adjoint representation of the isometry algebra. 
For instance, 
if the isometry group is simple, $u_{IJ}$
must be a multiple of the Cartan-Killing form.

Contrary to \eqref{useless_deformation},
the modification \eqref{new_V4eq}   changes the
class defined of $V_4^{\rm eq}$, and hence $E_4$.
It should be noted, however,
that the effect of the shift parametrized by the constants $c^\alpha_I$
can always be undone by a linear redefinition 
of the external curvatures $F^I$, $F^\alpha$,
\beq \label{F_rotation}
F^I \rightarrow F^I \ , \qquad
\tfrac{1}{2\pi} \, F^\alpha \rightarrow \tfrac{1}{2\pi} \, F^\alpha  - c^\alpha_I \, F^I \ .
\eeq
In this sense, the only genuine ambiguity in \eqref{new_V4eq}
is the part parametrized by the constants $u_{IJ}$.
We notice that any shift of $\sigma_{IJ}$ with $u_{IJ}$
generates closed and gauge-invariant terms 
with purely external legs, which can always be reabsorbed
in the 4-form $\gamma_4$. The (in)dependence 
of 
the inflow anomaly polynomial on $\gamma_4$ is 
studied in appendix \ref{app_gamma4}.

In a completely similar way, we can choose inequivalent
equivariant completions of $\omega_\alpha$.
Pragmatically, we consider the shifts
\beq \label{new_sigma_Ialpha}
\sigma_{I\alpha} \rightarrow \sigma_{I\alpha} + u_{I \alpha}  \ ,
\eeq
where the constants $u_{I\alpha}$ are constrained by 
\eqref{omega_alpha_gauge},
\beq
\pounds_I  u_{J\alpha} = f_{IJ}{}^K \, u_{K\alpha} \ .
\eeq
Once again, a shift parametrized by $u_{I\alpha}$ 
corresponds to a modification of $\gamma_4$.

We close this section by highlighting an important feature
of the case $d=4$, which is explained in more detail in appendix
\ref{4d_discussion}.
For $d=4$, one of the vectors $A^\alpha$ originating from expansion
of $C_4$ onto a basis $\omega_\alpha$ of harmonic 2-form is massive.
This linear combination does not correspond to a symmetry of the system.
The associated background curvature must then be set to zero.
More precisely, this holds provided we choose the 2-forms
$\omega_I$  in such a way that
\beq \label{omegaI_condition}
\int_{M_6} V_4 \, \omega_I = 0 \ .
\eeq
This condition can always be achieved by shifting $\omega_I$
by an appropriate linear combination of $\omega_\alpha$'s.
If \eqref{omegaI_condition} holds, the linear combination 
of $F^\alpha$ curvatures that must be set to zero is
\beq
N_\alpha \, F^\alpha = 0 \ , \qquad
N_\alpha = \int_{M_6} V_4\, \omega_\alpha \ .
\eeq
The constraint $N_\alpha F^\alpha = 0$ is essential
in order to get the correct anomaly in four dimensions.
It plays a particularly non-trivial role
in the example studied in section \ref{sec_GMSW}.



\section{Examples in six dimensions} \label{sec_6d}

In this section we exemplify the method of section
\ref{sec_general_param} for the construction of $E_4$
in two examples involving a stack of $N$ M5-branes
in six uncompactified dimensions.
In the first example, the branes sit  at a smooth point in the transverse directions,
while in the second example they probe an orbifold singularity.

\subsection{A stack of $N$ M5-branes}
Anomaly inflow for this setup is well-known \cite{Freed:1998tg,Harvey:1998bx}.
The analysis of this section is 
useful for the rest of this paper,
since it allows us to introduce 
the objects \eqref{explicit_S4_objects} 
which are later used
for various applications.

The 4-form $V_4$, which encodes
the background $G_4$-flux configuration,
must be proportional to the volume form on 
the internal space $M_4  = S^4$. We write
\beq
V_4 = N \, {\rm vol}_{S^4} \ .
\eeq
In our normalization, $\int_{S^4} {\rm vol}_{S^4}=1$.
The constant $N$ is an integer by virtue of $G_4$-flux
quantization in M-theory,
and counts the number of M5-branes in the stack.

The internal space $S^4$ admits an $SO(5)$ isometry group.
Let us introduce   constrained coordinates $y^A$, $A = 1, \dots, 5$
with $y^A \, y_A = 1$. 
The gauging of the $SO(5)$ isometry is given by
\beq
(dy^A)^{\rm g} = dy^A - A^{AB} \, y_B \ ,
\eeq
where $A^{[AB]}$ is the external $SO(5)$ connection,
with field strength
\beq
F^{AB} = dA^{AB} - A^{AC} \, A_C{}^B \ . 
\eeq
Notice that $S^4$ does not admit any non-trivial
harmonic 2-form. As a result, we do not have any additional
external connections.

The 4-form $E_4$ is 
determined
by solving \eqref{V4_gauge},
\eqref{V4_closure}.
The result of the analysis takes the form
\beq \label{6d_E4}
E_4 = \overline V_4 + F^{AB} \, \overline \omega_{AB}^{\rm g}
+ F^{AB} \, F^{CD} \,  \overline \sigma_{AB,CD} 
+ \gamma_4 \ ,
\eeq
where $\gamma_4$ is an arbitrary closed, gauge-invariant 4-form
with external legs, and we have defined the following forms on $S^4$,
\begin{align} \label{explicit_S4_objects}
\overline  V_4 & = \frac{3\, N}{8\pi^2} \,  \cdot \, \frac{1}{4!} \,
\epsilon_{A_1 \dots A_5} \,   d  y^{A_1}  \,
d  y^{A_2} \, d  y^{A_3} \, d  y^{A_4} \,   y^{A_5} \ , \nn \\
\overline \omega_{AB} & = \frac{3\, N}{8\pi^2} \,  \cdot \, \frac{-2}{4!} \,
\epsilon_{AB C_1 C_2 C_3} \,   d  y^{C_1}  \,
d  y^{C_2} \,    y^{C_3}   \ , \nn \\
\overline \sigma_{AB,CD} & = \frac{3\, N}{8\pi^2} \,  \cdot \, \frac{1}{4!} \,
\epsilon_{AB CDE} \,      y^{E}  \ .
\end{align}
If we set $\gamma_4 = 0$,  the 4-form $E_4$ in \eqref{6d_E4}
is proportional to the global angular form of $SO(5)$,
which appears in the original analysis
of \cite{Freed:1998tg,Harvey:1998bx}.

The Pontryagin classes of $TM_{11}$ can be computed exploiting
the decomposition
of the 11d tangent bundle restricted on the worldvolume $W_6$
of the brane,
\beq
TM_{11} \rightarrow TW_6 \oplus N_{SO(5)} \ ,
\eeq
where $N_{SO(5)}$ is the bundle encoding the $SO(5)$ gauging.
We have
\begin{align}
p_1(TM_{11}) & = p_1(TW_6) + p_1(SO(5)) \ , \nn \\
p_2(TM_{11}) & = p_2(TW_6) + p_2(SO(5)) + p_1(TW_6) \, p_1(SO(5))  \ ,
\end{align}
and hence
\beq \label{6d_X8}
X_8 = - \frac{1}{48} \Big[ p_2(TW_6) + p_2(SO(5)) \Big]
+ \frac{1}{192} \, \Big[ p_1(TW_6) - p_1(SO(5)) \Big]^2 \ .
\eeq
Notice that $X_8$ has no legs along the internal $S^4$ directions.

We can now compute the inflow anomaly polynomial,
using \eqref{6d_E4} and \eqref{6d_X8}.
The result reads
\begin{align}\label{6dinflow_with_Cu}
I_8^{\rm inflow} & = 
- \frac {1}{24} \,N^3 \, p_2(SO(5))
- \frac 12 \, N \, \gamma_4^2
\nn \\
& + \frac{1}{48}  \, N \, \Big[ p_2(TW_6) + p_2(SO(5)) \Big]
- \frac{1}{192} \, N \, \Big[ p_1(TW_6) - p_1(SO(5)) \Big]^2 \ .
\end{align}
The first line collects the contribution
of the $E_4^3$ term, while the second line contains the contribution
of the $E_4 \, X_8$ term.
To verify 
\eqref{6dinflow_with_Cu}
we can make use of the identities \eqref{vanishing_integrals}, 
 \eqref{nonvanishing_integrals}.
In our conventions,
\beq
p_1(SO(5) ) = - \frac 12 \, \frac{1}{(2\pi)^2} \, {\rm tr} \, F^2 \ , \qquad
p_2(SO(5)) = \frac 18 \, \frac{1}{(2\pi)^2} \, \Big[
({\rm tr} \, F^2)^2 - 2 \, {\rm tr} \, F^4
\Big] \ ,
\eeq
where the trace is in the fundamental representation
of $SO(5)$.

In appendix \ref{app_gamma4} we argue that $\gamma_4$ is fixed
by extremizing $I_8^{\rm inflow}$ with respect to an arbitrary
variation in $\gamma_4$.
In the present situation, we obtain simply
\beq
\gamma_4 = 0 \ .
\eeq
For this value of $\gamma_4$, the result \eqref{6dinflow_with_Cu}
agrees with the original inflow polynomial of \cite{Harvey:1998bx}.

\subsection{M5-branes probing an ADE singularity}

Let us now analyze a setup in which a stack of M5-branes
probes a $\mathbb C^2/\Gamma$ singularity,
where $\Gamma$ is an ADE subgroup of $SU(2)$ \cite{Brunner:1997gk,Blum:1997fw,Blum:1997mm,Intriligator:1997dh,Brunner:1997gf,Hanany:1997gh}.
The probe picture in the UV is as follows.

Let us consider the transverse $\mathbb R^5$ to the M5-branes,
with coordinates $y^1$, \dots, $y^5$. The group $\Gamma$
acts on $\mathbb R^5$ by leaving $y^5$ invariant,
and acting on $\mathbb R^4$ parametrized by $y^1$, \dots, $y^4$.
More precisely, the action of $\Gamma$ is embedded in the factor
$SU(2)_L$ of the isometry group $SO(4) \cong SU(2)_L \times SU(2)_R$
of the $\mathbb R^4$ spanned by $y^{1,2,3,4}$.
All points on the $y^5$ axis are fixed points under the action of $\Gamma$.
In the probe picture, the stack is positioned at the origin
$y^{1,2,3,4,5} =0$. Because of the $\Gamma$ quotient,
supersymmetry is reduced from $(2,0)$ to $(1,0)$.

Before performing the quotient,
the stack is surrounded by the round sphere $S^4 \subset \mathbb R^5$.
After acting with $\Gamma$, $S^4$ is replaced with $S^4/\Gamma$.
The north and south poles of $S^4$, located at $y^5 = \pm 1$, $y^{1,2,3,4} = 0$,
are both fixed points for the $\Gamma$ action.
Locally near each pole we have an orbifold singularity
$\mathbb C^2/\Gamma$. 
The orbifold singularities at each pole can be resolved preserving
6d $(1,0)$ supersymmetry.
If $\mathfrak g_\Gamma$ is the ADE Lie algebra
associated to $\Gamma$, we write
$r_\Gamma = \text{rank}(\mathfrak g_\Gamma)$.
The
 resolution introduces a number $r_\Gamma$ of  $\mathbb C \mathbb P^1$
curves, whose intersection pattern reproduces the  Dynkin diagram
of $\mathfrak g_\Gamma$.
We use the symbol $M_4$ for the smooth space obtained
from $S^4/\Gamma$ by resolving the singularities at the north
and south poles.

The isometry group $SO(5)$ of $S^4$ is reduced
by the action of $\Gamma$. More precisely,
the isometry group of $S^4/\Gamma$ is the subgroup of 
$SU(2)_L \times SU(2)_R$ that commutes with the action of $\Gamma$.
If $\Gamma = \mathbb Z_k$ with $k \ge 3$,
the isometry group is $U(1)_L \times SU(2)_R$,
with $U(1)_L$ the Cartan of $SU(2)_L$.
If $k = 1,2$ the isometry group is the full $SU(2)_L \times SU(2)_R$.
Finally, if $\Gamma$ is of type D, E, the isometry group
is $SU(2)_R$ only.

To treat all cases uniformly, we formally introduce external
connections for the full $SU(2)_L \times SU(2)_R$.
It is understood that the external $SU(2)_L$ connection
is zero for $\Gamma$ of D, E type, it is along the Cartan for 
$\Gamma = \mathbb Z_k$, $k \ge 3$, and it is a full non-Abelian
connection for $k = 1,2$.

The resolved space $M_4$ admits non-trivial 2-cycles,
and hence harmonic 2-forms.
They are dual to the resolution $\mathbb C \mathbb P^1$'s
at the north and south poles.
We write these harmonic 2-forms as
$\omega_{{\rm N} i}$, $\omega_{{\rm S} i}$,
with the index $i = 1, \dots, r_\Gamma$ labelling the Cartan
generators of the ADE Lie algebra $\mathfrak g_\Gamma$.
These harmonic 2-forms correspond to additional
flavor symmetries of the setup.
In the resolved phase, the flavor symmetry
associated to the harmonic 2-forms is
$(U(1)^{r_\Gamma})_{\rm N} \times (U(1)^{r_\Gamma})_{\rm S}$.
If we shrink the resolution cycles to zero size,
we have enhancement to the non-Abelian symmetry
$(G_\Gamma)_{\rm N} \times (G_\Gamma)_{\rm S}$,
where $G_\Gamma$ is the   Lie group
associated to $\Gamma$.

We are now in a position to discuss the 4-form $E_4$
for the setup under examination.
It can be written as
\beq \label{Gamma_E4}
E_4 = \overline V_4^{\rm g} + F^{AB} \, \overline \omega_{AB}^{\rm g}
+ F^{AB} \, F^{CD} \, \overline \sigma_{AB,CD} 
+ \frac{F^{{\rm N} i}}{2\pi} \, \omega_{{\rm N} i}
+ \frac{F^{{\rm S} i}}{2\pi} \, \omega_{{\rm S} i}
+ \gamma_4\ .
\eeq
Several comments are in order.
The above expression is written in terms of the $SO(5)$ curvature
$F^{AB}$. It is understood, however, that $F^{AB}$ is only non-zero
along the generators of the
 subgroup $SU(2)_L \times SU(2)_R \subset SO(5)$.
The quantities $\overline V_4$, $\overline \omega_{AB}$,
$\overline \sigma_{AB,CD}$ are as in \eqref{explicit_S4_objects},
but with the replacement $N \rightarrow N \, |\Gamma|$,
where $|\Gamma|$ is the order of the finite group $\Gamma$.
The extra factor $|\Gamma|$ is needed to compensate for the fact that
the $\Gamma$ action introduces a factor $1/|\Gamma|$ in all
integrals over $S^4/\Gamma$.
The curvatures $F^{{\rm N} i}$, $F^{{\rm S} i}$
are associated to the flavor symmetry
at the poles in the resolved phase.
We stress that the harmonic 2-forms $\omega_{{\rm N} i}$,
$\omega_{{\rm S} i}$ are invariant under the
isometry group of $S^4/\Gamma$, because they are localized
at the poles, which are fixed under $SU(2)_L \times SU(2)_R$.
This is why we do not have to gauge $\omega_{{\rm N} i}$,
$\omega_{{\rm S} i}$
in \eqref{Gamma_E4}.
Finally, $\gamma_4$ is an arbitrary closed, gauge-invariant, external
4-form.

The derivation of the inflow anomaly polynomial
for this setup was discussed in \cite{Ohmori:2014kda},
without the $\gamma_4$ term.
We review the derivation, including  $\gamma_4$,
in appendix \ref{app_Gamma}.
The result reads
\begin{align} \label{Gamma_result}
- I_8^{\rm inflow} & = 
 \frac{N^3 \, |\Gamma|^2}{24} \, \big[ c_2(L) - c_2(R)  \big]^2 
+ \frac 12 \, N \, \gamma_4^2
+ \frac 14 \,  \gamma_4 \, 
\bigg[ 
 \frac{{\rm tr} \, (F^{\rm N})^2}{(2\pi)^2}
- \frac{{\rm tr} \, (F^{\rm S})^2}{(2\pi)^2}
\bigg] 
 \nn  \\
&  + \frac{N \, |\Gamma|}{8} \, 
\big[ c_2(L) - c_2(R)  \big] \, \bigg[ 
  \frac{{\rm tr} \, (F^{\rm N})^2}{(2\pi)^2}
+  \frac{{\rm tr} \, (F^{\rm S})^2}{(2\pi)^2}
\bigg]
\nn \\
& +  \frac{N \, |\Gamma| \, \chi_\Gamma}{48}  \, 
 \big[ c_2(L)
- c_2(R)\big] \, \big[ p_1(TW_6) + 4 \, c_2(R)\big]
 +   
\frac{N}{48} \, c_2(L)\, \big[ p_1(TW_6) + 4 \, c_2(R)\big]
\nn \\
& +  
 \frac{N}{192} \big[  p_1(TW_6)^2 - 4 \, p_2(TW_6)  \big]
+ \frac{N}{48} \, c_2(R) \, p_1(TW_6)
 \ .  
\end{align}
The quantities $c_2(L,R)$ are the second Chern classes of
$SU(2)_{L,R}$,
while $p_{1,2}(TW_6)$ are the Pontryagin classes of the external
6d background metric. We have written the result in terms of the
full non-Abelian flavor symmetry curvatures $F^{\rm N}$, $F^{\rm S}$,
even though only the Cartan curvatures 
$F^{{\rm N} i}$, $F^{{\rm S} i}$
are directly accessible in the supergravity approximation.
The quantity $\chi_\Gamma$ is the Euler characteristic
of the ALE space that resolves the $\mathbb C^2/\Gamma$ orbifold.
It is given by
\beq \label{chi_Gamma}
\chi_\Gamma = r_\Gamma + 1 - \frac{1}{|\Gamma|} \ .
\eeq
The ranks $r_\Gamma$ and orders $|\Gamma|$ for all ADE 
groups are summarized in table \ref{table_orders}.

\begingroup
\renewcommand{\arraystretch}{1.4}

\begin{table}
\centering
 \begin{tabular}{| r || c | c | c | c | c |}
\hline Lie algebra $\mathfrak g_\Gamma$   & $SU(k)$ & $SO(2k)$ & $E_6$ & $E_7$ & $E_8$ \\ \hline \hline
rank $r_\Gamma$ &  $k-1$ & $k$ & 6 & 7 & 8 \\ \hline
order $|\Gamma|$ & $k$ & $4k-8$ & 24 & 48 & 120 \\ \hline
\end{tabular}
\caption{
The rank $r_\Gamma$ and order $|\Gamma|$ for ADE subgroups $\Gamma$
of $SU(2)$.
} 
\label{table_orders}
\end{table}

\endgroup

According to the general discussion of appendix \ref{6d_ambiguity},
the external 4-form $\gamma_4$ is fixed by extremizing $I_8^{\rm inflow}$, which is equivalent to imposing the tadpole cancellation condition in M-theory. 
In the present situation, we obtain
\beq
\gamma_4 = - \frac{1}{4 \, N} \, \bigg[ 
  \frac{ {\rm tr} \, (F^{\rm N})^2}{(2\pi)^2}
-  \frac{{\rm tr} \, (F^{\rm S})^2}{(2\pi)^2}
\bigg]  \ .
\eeq
Plugging this back into \eqref{Gamma_result}, we obtain
\begin{align} \label{Gamma_resultBIS}
- I_8^{\rm inflow} & = 
 \frac{N^3 \, |\Gamma|^2}{24} \, \big[ c_2(L) - c_2(R)  \big]^2 
- \frac{1}{32 \, N} \, 
\bigg[ 
  \frac{ {\rm tr} \, (F^{\rm N})^2}{(2\pi)^2}
-  \frac{{\rm tr} \, (F^{\rm S})^2}{(2\pi)^2}
\bigg] ^2
 \nn  \\
&  + \frac{N \, |\Gamma|}{8} \, 
\big[ c_2(L) - c_2(R)  \big] \, \bigg[ 
  \frac{ {\rm tr} \, (F^{\rm N})^2}{(2\pi)^2}
+  \frac{{\rm tr} \, (F^{\rm S})^2}{(2\pi)^2}
\bigg]
\nn \\
& +  \frac{N \, |\Gamma| \, \chi_\Gamma}{48}  \, 
 \big[ c_2(L)
- c_2(R)\big] \, \big[ p_1(TW_6) + 4 \, c_2(R)\big]
 +   
\frac{N}{48} \, c_2(L)\, \big[ p_1(TW_6) + 4 \, c_2(R)\big]
\nn \\
& +  
 \frac{N}{192} \big[  p_1(TW_6)^2 - 4 \, p_2(TW_6)  \big]
+ \frac{N}{48} \, c_2(R) \, p_1(TW_6)
 \ .  
\end{align}
This result agrees with the analysis of \cite{Ohmori:2014kda}.
It is interesting to point out that, in the computation of \cite{Ohmori:2014kda}, the term 
$- \frac{1}{32 \, N} \, (2\pi)^{-2}
\big[ 
{\rm tr} \, (F^{\rm N})^2
-{\rm tr} \, (F^{\rm S})^2
\big] ^2$
is interpreted as a Green-Schwarz term associated to the
center of mass mode of the M5-brane stack,
and is included by hand. In our derivation,
it is automatically generated by $\gamma_4$-extremization.

Let us consider the case $\Gamma = \mathbb Z_k$.
Using the full anomaly polynomial recorded in \cite{Bah:2017gph}
for the interacting 6d (1,0) SCFT, we can extract
the contribution of decoupling modes related to
the center of mass of the M5-brane stack.
To compare with \cite{Bah:2017gph}, we replace
\beq
c_2(L) \to - c_1(s)^2 \ ,
\eeq
where we are using the notation of \cite{Bah:2017gph}
for the first Chern class $c_1(s)$ of the Cartan $U(1)_L$
of $SU(2)_L$.
Comparing \eqref{Gamma_resultBIS} to the results of \cite{Bah:2017gph},
we infer
\begin{align}
I_8^{\rm decoupl} & = - I_8^{\rm inflow} - I_8^{\rm SCFT}
\nn \\
& = I^{\rm tensor}_8 + \frac 12 \,  I_8^{\rm vec, N}
+ \frac 12 \,  I_8^{\rm vec, S}
- \frac 16 \, k \, c_1(s) \, \bigg[\frac{ {\rm Tr}_{\rm fund}  \, (F^{\rm N})^3}{(2\pi)^3} 
-   \frac{ {\rm Tr}_{\rm fund}  \, (F^{\rm S})^3}{(2\pi)^3}   \bigg] \ .
\end{align}
The quantities $ I^{\rm tensor}_8$, $I_8^{\rm vec, N}$, 
are given by\footnote{Following \cite{Bah:2017gph},
the traces of $F^N$ are defined in such a way that
\beq
\frac{{\rm tr}  \,  (F^{\rm N})^2 }{(2\pi)^2} = - 2 \, \sum_i (n^{\rm N}_i)^2 \ , \qquad
 \frac{{\rm Tr}_{\rm fund} \, (F^{\rm N})^3 }{(2\pi)^3} =  \sum_i (n^{\rm N}_i)^3
\ , \qquad
\frac{{\rm Tr}_{\rm fund}  \, (F^{\rm N})^4}{(2\pi)^4} =  \sum_i (n^{\rm N}_i)^4 \ ,
\eeq
where $n^{\rm N}_i$ are the Chern roots of $SU(k)_{\rm N}$,
 $i = 1, \dots, k$ and $\sum_i n_i^{\rm N} = 0$.
The same conventions hold for~$SU(k)_{\rm S}$.
}
\begin{align}
I_8^{\rm vec,N} & = - \frac{k^2-1}{24} \, c_2(R)^2
- \frac{k^2-1}{48} \, c_2(R) \, p_1(TW_6)
- \frac{k^2-1}{5760} \big[  7 \, p_1(TW_6)^2 - 4 \, p_2(TW_6) \big]
\nn \\
& - \frac k4 \, c_2(R) \, \frac{{\rm tr}  \, (F^N)^2}{(2\pi)^2}
- \frac{k}{48} \, p_1(TW_6) \,   \frac{{\rm tr}  \, (F^N)^2}{(2\pi)^2}
- \frac 16 \, \bigg[    \frac{{\rm tr} \, (F^N)^2}{(2\pi)^2}\bigg]^2
- \frac{k}{12} \,   \frac{{\rm Tr}_{\rm fund} \, (F^N)^4}{(2\pi)^4} \ , \nn \\
I_8^{\rm tensor} & = 
\frac{1}{24} \, c_2(R)^2 + \frac{1}{48} \, c_2(R) \, p_1(TW_6)
+ \frac{23}{5760} \, p_1(TW_6)^2
- \frac{29}{1440} \, p_2(TW_6) \ .
\end{align}
The quantity $I_8^{\rm vec,S}$ is completely analogous to
$I_8^{\rm vec,N}$ given above.



\section{Examples in four dimensions} \label{sec_4d}

In this section we examine two 4d setups
to exemplify our prescription for the computation of $I_6^{\rm inflow}$.
In a first class of examples, the space $M_6$
is an $S^4$ fibration over a smooth Riemann surface.
This case corresponds to the setups analyzed in
BBBW \cite{Bah:2011vv,Bah:2012dg}.
Next, we analyze the geometry $M_6$
that is read off from the GMSW \cite{Gauntlett:2004zh} $AdS_5$ solution
to 11d supergravity.

\subsection{$S^4$ fibrations over a smooth Riemann surface} \label{twisting_on_Sigma}

Let us consider a stack of M5-branes wrapping a genus-$g$ Riemann
surface without punctures $\Sigma_g$. In this setup, the internal space $M_6$ is an $S^4$
fibration over $\Sigma_g$. Upon including external connections,
$M_6$ is fibered over external spacetime $W_4$.
The relevant fibrations are thus
\beq \label{fibration_with_twist}
M_6 \hookrightarrow M_{10} \rightarrow W_4 \ , \qquad
S^4 \hookrightarrow M_6 \rightarrow \Sigma_{g} \ .
\eeq
In order to implement anomaly inflow, we need to study the 
topology and isometries of $M_6$.

\subsubsection{Topology and isometries of $M_6$} 
\label{4d_topology_isometries}

In this work we study a class of fibrations
$S^4 \hookrightarrow M_6 \rightarrow \Sigma_g$
that preserve 4d $\cN= 1$ supersymmetry \cite{Bah:2011vv,Bah:2012dg}.
In terms of the ambient space $\mathbb R^5 \supset S^4$,
we   refer to the decomposition $\mathbb R^5 = \mathbb C_1 
\times \mathbb C_2 \times \mathbb R$.
The topology of $M_6$ is then encoded in 
the two line bundles $\cL_1$, $\cL_2$ that describe the twisting of the two
$\mathbb C_1$, $\mathbb C_2$ factors on the Riemann surface.
Let $q_1$, $q_2$ be the degrees of the line bundles.
In order to preserve supersymmetry,
the total space $\cL_1 \oplus \cL_2 \rightarrow \Sigma_{g}$
has to be a Calabi-Yau threefold,
which amounts to the requirement
\beq \label{susy_condition}
q_1 + q_2 = - \chi(\Sigma_g)  = 2(g-1) \ .
\eeq
Setups in which $g =1$, \emph{i.e.}~the Riemann surface is a torus,
require special care, because of the presence of 
emergent symmetries. For this reason,
we restrict ourselves to the cases of a higher-genus
Riemann surface, $g\ge 2$,
or a sphere, $g=0$.
In our discussion $q_1$ and $q_2$ cannot therefore
be simultaneously zero.
If $q_1=0$ or $q_2=0$, supersymmetry
enhances to $\cN = 2$.

The topology of $M_6$ can be equivalently
described in terms of the background value of 
a non-zero background value for the
$SO(5)$ field strength $F^{AB}$, which is
proportional to the volume form on $\Sigma_g$.
For the setups under examination,
the background $F^{AB}$ takes the form
\beq \label{twist_F} 
F^{AB}_\Sigma = q^{AB} \, V_\Sigma \ , \qquad
q^{AB} =
{ \small \begin{pmatrix}
0 & q_1 &   &   &   \\
-q_1 & 0 &   &    &   \\
  &   & 0 & q_2 &   \\
  &   & -q_2 & 0 &   \\
  &   &   &   & 0 
\end{pmatrix} 
} \ .
\eeq
The subscript $\Sigma$ on $F$ is a reminder that
this is the background part, or twist part,  of the $SO(5)$ field strength,
as opposed to the external 4d gauge part.
The 2-form $V_\Sigma$ is proportional to the volume form on $\Sigma_g$,
and is normalized according to
\beq \label{VSigma_integral}
\int_{\Sigma_g} V_\Sigma = 2\pi \ .
\eeq

In order to apply the recipe of
section \ref{sec_general_param} for the construction of $E_4$,
we need to identify the isometries of $M_6$
that we intend to couple to 4d gauge fields.
We have two distinct classes of isometries:
\begin{enumerate}[(i)]
\item For any genus $g$, a subgroup $SO(2)_1 \times SO(2)_2$
of the $SO(5)$ isometry group of the $S^4$ fiber is preserved by the twist
described by \eqref{twist_F}.
We therefore introduce two Abelian external connections
$A^1$, $A^2$, with field strengths $F^1 = dA^1$, $F^2 = dA^2$,
to gauge this residual isometry.
The embedding of $A^1$, $A^2$ into the full
$SO(5)$ connection, and the analogous relation
for the field strengths,  read
\beq \label{external_stuff}
A_{\rm ext}^{AB} = 
{ \small  \begin{pmatrix}
0 & - A^1 &   &   &   \\
A^1 & 0 &   &   &   \\
  &   &   & -A^2 &   \\
  &   & A^2 & 0 &   \\
  &   &   &   & 0 
\end{pmatrix}
} \ , \qquad
F_{\rm ext}^{AB} = 
{ \small  \begin{pmatrix}
0 & - F^1 &   &   &   \\
F^1 & 0 &   &   &   \\
  &   &   & -F^2 &   \\
  &   & F^2 & 0 &   \\
  &   &   &   & 0 
\end{pmatrix}
} \ ,
\eeq
with the subscript ``ext'' standing for external.
If $q_2 = 0$, the $SO(2)_2$ factor enhances to $SO(3)_2$,
and $F^2$ is replaced by the suitable non-Abelian
$SO(3)_2$ field strength. Similar remarks
apply if $q_1  = 0$.

\item In the special case $g = 0$, $M_6$ possesses additional
isometries that originate from the isometry group $SO(3)_{S^2}$ of the Riemann surface,
which is a 2-sphere endowed with its standard round metric.
As explained in appendix \ref{app_details},
the Killing vectors of the base $\Sigma_{g=0} = S^2$ considered in isolation
extend to \emph{bona fide} Killing vectors
of the entire space $M_6$, for any value of $q_1$, $q_2$.
We use the notation $A^a$, with $a = 1,2,3$,
for the external $SO(3)_{S^2}$ connection that gauges
this additional isometry, and $F^a$ for the corresponding
field strength.

\end{enumerate}

\subsubsection{Aside on terminology: twisting vs gauging}
\label{twisting_vs_gauging}

The nested fibration structure \eqref{fibration_with_twist}
of the setups under examination
allows us to define two distinct 
operations on differential forms, which we refer to as twisting and gauging.

The twisting operation is defined with reference to the fibration
$S^4 \hookrightarrow M_6 \rightarrow \Sigma_g$.
It makes use of the internal field strength 
\eqref{twist_F}, but it does not involve any external 4d gauge connection.
Twisting affects forms on $M_6$ with legs along the $S^4$ fiber
directions, while it has no effect on legs along the Riemann surface.
Operationally, in terms of the constrained coordinates $y^A y_A = 1$
of the $S^4$ fibers, the twisting operation amounts to the replacement
\beq \label{twisting}
dy^A \;\; \rightarrow \;\; (dy^A)^{\rm t} =  dy^A - q^{AB} \, y_B \, A_\Sigma \ , \qquad
dA_\Sigma = V_\Sigma \ .
\eeq
The 1-form $A_\Sigma$ on the Riemann surface is an antiderivative
of the volume form and is only locally defined.
Because of the non-trivial fibration, the untwisted 1-forms
$dy^A$ are not well-defined on $M_6$.
Their twisted counterparts $(dy^A)^{\rm t}$, however,
are good objects in $M_6$.

Let us now turn to the gauging operation.
This is the same operation discussed in section \ref{sec_general_param},
and is based on the isometries of $M_6$. 
Let us first consider a higher-genus Riemann surface.
The only isometries are then of the class (i) above.
Since isometries of class (i) originate from the $S^4$ fiber of $M_6$,
the gauging procedure 
 has no effect on $V_\Sigma$.
We can write
\beq
\text{higher-genus $\Sigma_g$} \, : \qquad
\left\{ 
\begin{array}{rcl}
(dy^A)^{\rm t} & \rightarrow & (dy^A)^{\rm tg} = dy^A - q^{AB} \, y_B \, A_\Sigma
- A_{\rm ext}^{AB} \, y_B  \ , \\[2mm]
V_\Sigma & \rightarrow&  V_\Sigma^{\rm g} = V_\Sigma \ .
\end{array}
\right.
\eeq
Notice that, since the untwisted $dy^A$ 1-forms are not well-defined
on $M_6$, it does not make sense to consider $(dy^A)^{\rm g}$.

If the Riemann surface is a sphere,
we have 
both isometries of class (i) and of class (ii).
To proceed, it is convenient to 
describe the 2-sphere by means of three
constrained coordinates $z^a$, $a=1,2,3$, satisfying $z^a z_a = 1$.
The gauging operation in this case  
  satisfies
\beq
\text{two-sphere} \, : \qquad
\left\{ 
\begin{array}{rcl}
(dy^A)^{\rm t} & \rightarrow & (dy^A)^{\rm tg} = dy^A - q^{AB} \, y_B \, A_\Sigma
- A_{\rm ext}^ {AB} \, y_B
+ \frac 12 \, q^{AB} \, y_B \, z_a \, A^a  \ , \\[2mm]
dz^a & \rightarrow & (dz^a)^{\rm g} = dz^a +  \epsilon^{abc} \, A_b \, z_c
 \ , 
\\[2mm]
V_\Sigma & \rightarrow&  V_\Sigma^{\rm g} = \frac 14 \, \epsilon_{abc} \, 
(dz^a)^{\rm g} \, (dz^b)^{\rm g} \, z^c \ .
\end{array}
\right. \nn
\eeq
Crucially, gauging of the additional $SO(3)_{S^2}$ isometry
of class (ii)  
  involves legs along the Riemann surface.
In appendix \ref{app_details} we collect some
useful formulae that are helpful in checking the above relations.

\subsubsection{Construction of $E_4$}

The first task in the construction of $E_4$ is the identification of the
4-form $V_4$, to which $E_4$ reduces if we turn off all external
4d connections.
We claim that 
\beq \label{4dV4}
V_4 = \overline  V_4^{\rm t} + q^{AB} \, V_\Sigma \, \overline  \omega_{AB}^{\rm t}  \ .
\eeq
The forms $\overline V_4$ and $\overline \omega_{AB}$
have legs on the $S^4$ fibers and were introduced in \eqref{explicit_S4_objects}.
The subscript ``t'' signals the twisting operation 
discussed in the previous subsection.
The untwisted 4-form $\overline  V_4$ is closed,
but it is not well-defined in the total internal space $M_6$.
Its twisted counterpart $\overline  V_4^{\rm t}$ 
is a good object in $M_6$, but it is not closed.
This explains the necessity of the other term in \eqref{4dV4}.
Indeed, to see that   $V_4$ is closed,
we simply observe that it 
is proportional to the global angular form of $SO(5)$,
provided we replace the $SO(5)$ field strength 
$F^{AB}$ with $F^{AB}_\Sigma$ as in \eqref{twist_F}.
(The term  with two $F^{AB}_\Sigma$ factors is then zero because
$V_\Sigma \, V_\Sigma =0$.)

We claim that $V_4$ can be taken to be as in \eqref{4dV4} 
without any loss of generality.
This  can be motivated as follows.
Firstly, we know from section \ref{sec_remarks_on_E4}
that any modification of $V_4$ by an exact 4-form $dW_3$
(compatible with the isometries of $M_6$)
does not have any effect on the inflow
anomaly polynomial. Secondly, 
we observe that we do not have any other closed
but not exact 4-form in $M_6$.

The space $M_6$ admits one non-trivial 4-cycle,
given by the $S^4$ fiber over a generic point on $\Sigma_g$.
We then have one harmonic 2-form $\omega$, Poincar\'e dual to
this  4-cycle. We can write
\beq
\int_{S^4} V_4 = \int_{M_6} V_4 \, \omega = N \ .
\eeq
As discussed in section \ref{4d_discussion}, one linear combination
of the vectors associated to harmonic 2-forms is massive.
Since in this case we only have one harmonic 2-form,
its associated vector is massive, and we can simply ignore it
in the following discussion. 
As a result, all external connections are associated to
the isometries of class (i) and (ii) discussed above.

We are now in a position to apply the recipe
of section \ref{sec_general_param} for the construction of $E_4$.
We refer the reader to appendix \ref{app_details}
for the derivation of $E_4$. The result takes the form
\beq \label{E4_final}
E_4 = \overline V_4^{\rm tg} + \cF^{AB} \, \overline \omega_{AB}^{\rm tg}
+ \cF^{AB} \, \cF^{CD} \, \overline \sigma_{AB,CD}
+ (C_1 \, F^1 + C_2 \, F^2) \,  \left( V_\Sigma^{\rm g}
- \frac 12 \,  F^a \, z_a \right) + \gamma_4 \ .
\eeq
In the previous expression,  we have introduced
the 2-forms
\beq  \label{calF_def}
\cF^{AB} = 
 F^{AB}_{\rm ext}
+ q^{AB} \, \left( V_\Sigma^{\rm g}
- \frac 12 \,  F^a \, z_a \right)  \ .
\eeq
The quantities $C_{1,2}$ are   constants, while $\gamma_4$
is an arbitrary closed, gauge-invariant 4-form
with external legs only.

The values of $C_1$, $C_2$ are actually fixed 
by the following considerations.
In section \ref{4d_discussion} we derived that one linear combination
of the vectors associated to harmonic 2-forms is massive.
This results holds under the assumption that a basis of
connections is chosen, such that \eqref{omegaI_condition} holds.
In order to check whether $E_4$ in \eqref{E4_final}
satisfies \eqref{omegaI_condition}, we need to extract the terms
linear in the isometry curvatures $F^1$, $F^2$, and $F^a$,
\beq
E_4 = V_4^{\rm g} + F^1 \, \omega_1^{\rm g} + F^2 \, \omega_2^{\rm g}
+ F^a \, \omega_a^{\rm g} + \dots 
\eeq
Comparison with \eqref{E4_final}, keeping 
 \eqref{external_stuff} into account,  leads to the identifications
\begin{align}
\omega_1  &= -2 \, ( \overline \omega_{12}^{\rm t} + 2 \, q^{CD} \, V_\Sigma \, \overline \sigma_{12,CD} ) + C_1 \, V_\Sigma \ , \nn \\
\omega_2  &= -2 \, ( \overline \omega_{34}^{\rm t} + 2 \, q^{CD} \, V_\Sigma \, \overline \sigma_{34,CD} )  + C_2 \, V_\Sigma \ , \nn \\
\omega_a &= - \frac 12 \, z_a \, q^{AB} \, ( \overline \omega_{AB}^{\rm t} + 2 \, q^{CD} \, V_\Sigma \, \overline \sigma_{AB,CD} ) \ .
\end{align}
Equivalently, the above equations can be read off
from \eqref{appendix_results_no1}-\eqref{appendix_results_no5}.
Making use of the identities \eqref{vanishing_integrals}, 
we verify that $\int_{M_6} V_4 \, \omega_a = 0$,
while, in order to achieve $\int_{M_6} V_4 \, \omega_{1,2} = 0$,
we must set
\beq
C_1 = 0 \ , \qquad C_2 = 0   \ .
\eeq

As a final remark, we would like to point out that, if the Riemann surface is a 2-sphere, the quantity
\beq
\frac{ V_\Sigma^{\rm g} }{2\pi}
- \frac 12 \,  \frac{F^a}{2\pi} \, z_a 
\eeq
is equal to the global angular form $e_2^{S^2}$ of $SO(3)$. The definition and properties of the latter are reviewed in appendix \ref{app_GMSW}.

\subsubsection{Computation of $X_8$} \label{X8_for_4dtwist}

To compute $X_8$, we adopt the following point of view on the setup under
consideration.
Let $\widetilde W_6$ denote the space obtained
by combining external spacetime $W_4$
with the Riemann surface $\Sigma_g$.
The Pontryagin classes of $\widetilde W_6$
detect the curvature of the background metric on $W_4$.
If the Riemann surface is a sphere,
they also detect 
the gauging of its $SO(3)_{S^2}$ isometries, \emph{i.e.}~the
gauging via the connections $A^a$.
The total space may   be thought of as an $S^4$ fibration over $\widetilde W_6$.
This fibration is encoded in an $SO(5)$ bundle.
Its connection consists of two parts:
one describes the twist of $S^4$ over the Riemann surface,
the other corresponds to the gauging of the isometries
related to the $A^{AB}_{\rm ext}$ vectors.

The considerations of the previous paragraph lead us to write
\begin{align}
p_1(TM_{11}) &=   p_1(T\widetilde W_6) + p_1(SO(5) )  \ , \nn \\
p_2(TM_{11}) &=   p_2(T\widetilde W_6) + p_2(SO(5) ) 
+    p_1(T\widetilde W_6) \, p_1(SO(5) )  \ .
\end{align}
To proceed, we notice that 
\beq
p_1(T\widetilde W_6) = p_1(TW_4) + p_1(SO(3)_{S^2}) \ , \qquad
p_2(T\widetilde W_6) =0 \ ,
\eeq
with   $p_1(SO(3)_{S^2})$    only present 
if the Riemann surface is a sphere.
It is given by
\beq
p_1(SO(3)_{S^2}) =  \frac{1}{(2\pi)^2} \, F_a \, F^a  \ .
\eeq
Notice that any form with more than six
legs on external spacetime can be discarded.

The final task is the computation of the Pontryagin
classes $p_{1,2}(SO(5))$.
They can be written in terms of traces
of powers of the $SO(5)$ field strength,
\beq \label{SO5_Pontryagin}
p_1(SO(5)) = - \frac 12 \, \frac{1}{(2\pi)^2} \, {\rm tr} F_{SO(5)}^2 \  , \qquad
p_2(SO(5)) =  \frac 18 \, \frac{1}{(2\pi)^4} \bigg[
({\rm tr }\, F_{SO(5)}^2)^2 -2 \, {\rm tr} \, F_{SO(5)}^4
\bigg]  \ .
\eeq
In the present situation, $F_{SO(5)}^{AB}$ contains two pieces, 
\beq
F_{SO(5)}^{AB} = F^{AB}_\Sigma + F^{AB}_{\rm ext}  \ ,
\eeq
which are given in \eqref{twist_F}, \eqref{external_stuff} respectively.

We are now in a position to compute 
$X_8$. 
We only need to collect terms linear in $V_\Sigma$.
The result reads
\begin{align}
X_8 & = \frac {1}{48} \left( q_1 \, \frac{F^1}{2\pi} 
+ q_2 \, \frac{F^2}{2\pi} \right) \Big[  p_1(TW_4) + p_1(SO(3)_{S^2}) \Big] \,
\frac{V_\Sigma}{2\pi}
\nn \\
&  - \frac{1}{48} \, \bigg[  \bigg( \frac{F^1}{2\pi} \bigg)^2 
- \bigg( \frac{F^2}{2\pi} \bigg)^2     \bigg] \,
 \left( q_1 \, \frac{F^1}{2\pi} 
- q_2 \, \frac{F^2}{2\pi} \right) \, \frac{V_\Sigma}{2\pi} + \dots
\end{align}

\subsubsection{Inflow anomaly polynomial} 

Our first task is the computation of $\int_{M_6} E_4^3$.
Notice that in \eqref{E4_final} the only objects
with legs along the $S^4$ fibers are
$\overline V_4^{\rm tg}$, $\overline  \omega_{AB}^{\rm tg}$.
The integration of $E_4^3$ along $S^4$ can then
be performed using the identities \eqref{vanishing_integrals}, \eqref{nonvanishing_integrals}.
The integration along the $S^4$ fibers yields
\begin{align}
\int_{S^4} E_4^3 & = \frac 14 \, N^3 \cdot  \frac{1}{8} \, \frac{1}{(2\pi)^4} \,
\bigg[ ({\rm tr} \, \cF^2)^2 - 2 \, {\rm tr} \, \cF^4
\bigg]
+ 3 \, N \, \gamma_4^2 \ .
\end{align}
We now have to integrate over the Riemann surface.
The   term $\gamma_4$, however,
has no legs along $\Sigma_g$, and drops out.
The integral over $\Sigma_g$ is performed recalling the
definition of $\cF^{AB}$ in \eqref{calF_def}.
The result reads
\begin{align}
\int_{M_6} E_4^3 & =  - \frac 18 \, N^3 \,  
 \left( q_1 \, \frac{F^2}{2\pi} 
+ q_2 \, \frac{F^1}{2\pi} \right)  \bigg[
4 \, \frac{F^1}{2\pi} \, \frac{F^2}{2\pi}
+ q_1 \, q_2 \, \frac{F^a \, F_a}{(2\pi)^2}
\bigg] \ .
\end{align}
The terms with $F^a$ are only present is the Riemann surface
is a sphere.

Combining the $E_4^3$ contribution and the $E_4 X_8$
contribution, we get
the total inflow anomaly polynomial.
In order to facilitate comparison with the CFT 
expectation,
we introduce the notation
\beq \label{ni_definition}
\frac{F^1}{2\pi} = - 2 \,  n_1 \ , \qquad
\frac{F^2}{2\pi} = - 2 \,   n_2 \ .
\eeq
We then have 
\begin{align} \label{spherecase}
I_6^{\rm inflow}
& =
- \frac 16 \, N \, \Big(   
q_1 \, n_1^3 + q_2 \, n_2^3
   \Big)
 - \frac 23 \, \bigg( N^3 - \frac 14 \, N \bigg) \, \Big( 
q_1 \, n_1 \, n_2^2 + q_2 \, n_2 \, n_1^2
\Big)  \nn \\
& + \frac{1}{24} \, N \, \Big( q_1 \, n_1 + q_2 \, n_2 \Big) \, p_1(TW_4)
\nn \\
&
- \frac{1}{24} \, \bigg[  (N^3 \, q_2^2 -N) \, q_1 \, n_1 
 + (N^3 \, q_1^2 -N) \, q_2 \, n_2 \bigg] \, p_1(SO(3)_{S^2}) \ .
\end{align}
The first two lines of the previous expression are in accordance with the results 
quoted in \cite{Bah:2012dg,Bah:2018gwc}. The last line is only present when the Riemann surface is a sphere, and at present has not appeared in field-theoretic analyses of this scenario.
The decoupling modes that have to be subtracted to obtain
the anomaly of the interacting SCFT are given by 
dimensional reduction on $\Sigma_g$ of a free 6d (2,0) tensor
multiplet, which corresponds to the center of mass mode of the branes.

\subsection{$S^2$ fibrations over a product of Riemann surfaces} \label{sec_GMSW}

In this subsection we apply the methods of section \ref{sec_general_param}
to construct the inflow anomaly polynomial associated to 
a class of  $AdS_5$ solutions of 11d supergravity first discussed in \cite{Gauntlett:2004zh},
which we refer to as GMSW solutions in this work.
The input data for the construction of the inflow anomaly polynomial
are the geometry of the internal space $M_6$
and the closed, gauge-invariant 4-form $V_4$ which
we use as seed for the construction of $E_4$.
Both $M_6$ and $V_4$ are read off from the supergravity 
solution. The geometry of $M_6$ can be directly
inferred from the 11d line element,
while $V_4$ is identified, up to normalization,
with the $G_4$-flux of the solution.

\subsubsection*{Salient features of the solutions}

Let us now discuss some basic properties of  $M_6$ and $V_4$
in the GMSW solutions.
We refer the reader to appendix \ref{app_GMSW}
for  a more detailed review.

The line element of $M_6$ is of the form
\begin{align} \label{metric_for_M6}
ds^2(M_6) = h^2_{S^2} \, ds^2(S^2) + h^2_\Sigma \, ds^2(\Sigma_g)
+ h^2_y \, dy^2 + h_\psi^2 \, D\psi^2 \ .
\end{align}
Some comments on our notation are in order.
The coordinate $y$ parametrizes an interval,
$y \in [y_{\rm min}, y_{\rm max}]$,
and the metric functions $h_{S^2}$, $h_\Sigma$, $h_y$, 
$h_\psi$
are functions of $y$ only.
Their explicit expressions can be extracted from \eqref{GMSW_review_metric}.
The symbol $ds^2(S^2)$ denotes the line element on a round $S^2$
with unit radius,
while $ds^2(\Sigma_g)$ is the line element on a Riemann surface
of genus $g$ equipped with a constant curvature metric.
We only consider the cases $g = 0$ or $g \ge 2$,
and we normalize the metric in such a way that the Ricci
scalar is $R = \pm 2$.
The angle $\psi$ has periodicity $2\pi$.
The circle $S^1_\psi$ is twisted over $S^2$ and $\Sigma_g$,
with\footnote{Compared to \cite{Gauntlett:2004zh}, we have flipped the sign of $\psi$.}
\beq
dD\psi = -  2\,  V_{ S^2} - \chi \, V_\Sigma \  .
\eeq
The quantity $\chi$ is the Euler characteristic of $\Sigma_g$.
The 2-form $V_{ S^2}$ is proportional to the volume form of $S^2$,
while $V_\Sigma$ is proportional to the volume form on $\Sigma_g$.
We use the normalization conventions
\beq
\int_{S^2} V_{S^2} = 2\pi \ , \qquad
\int_{\Sigma_g} V_\Sigma = 2\pi \ .
\eeq

The metric functions $h_{S^2}$, $h_\Sigma$ are smooth and strictly
positive on the entire $y$ interval.
The metric function $h^2_y$ 
is everywhere positive on the interior of the $y$ interval,
with simple poles at the endpoints.
The function $h^2_\psi$, on the other hand, is 
everywhere positive on the interior of the $y$ interval,
with simple zeros at the endpoints.
The 2d space obtained combining the $y$ interval with the $\psi$
circle is topologically a 2-sphere, which we denote $S^2_{y\psi}$.
The behavior of  $h_y$, $h_\psi$
at the endpoints of the $y$ interval is such that $S^2_{y\psi}$ is free of conical
singularities.

The angle $\psi$ is an isometry direction for $M_6$. 
The dual 1-form reads
\beq
k_\psi = h_\psi^2 \, D\psi \ .
\eeq
The space $M_6$ admits additional $SO(3)$ isometries
originating from $S^2$. 
These isometries are preserved by the  
$S^2_{y\psi}$
fibration on top of $S^2$. The corresponding Killing 1-forms are
\beq \label{S2_base_isom}
k_a =  h_{S^2}^2  \, \epsilon_{abc} \, z^b \, dz^c 
+  \, z_a \, h_\psi^2 \, D\psi \ .
\eeq
The scalars $z^a$, $a = 1,2,3$ are constrained coordinates
on $S^2$, satisfying $z^a \, z_a = 1$.
If the Riemann surface $\Sigma_g$ is also a 2-sphere,
it give rise to a completely analogous set of Killing vectors,
generating an extra $SO(3)$ factor in the isometry group.
For simplicity, in the rest of this section
we focus on the isometries associated to the angle $\psi$
and the $S^2$, and we do not consider the additional isometries
that emerge if $\Sigma_g$ is also a 2-sphere.

The form $V_4$, which is going to be used as seed in the
construction of $E_4$ below,
is extracted from the expression of the $G_4$-flux in the GMSW solution.
The form $V_4$ can be written as
\beq \label{GMSW_V4}
V_4 = 
\bigg[  d\gamma_\Sigma \, \frac{V_\Sigma}{2\pi}
+ d\gamma_{S^2} \, \frac{V_{S^2}}{2\pi} \bigg] \, \frac{D\psi}{2\pi}
- \Big[ 2\, \gamma_\Sigma + \chi  \, \gamma_{S^2}   \Big] \,
\frac{V_\Sigma}{2\pi}  \, \frac{V_{ S^2}}{2\pi}  \ .
\eeq
In the previous equation, $\gamma_\Sigma$ and $\gamma_{S^2}$ are functions of $y$
only.
The   expressions for 
$\gamma_\Sigma$, $\gamma_{S^2}$ can be extracted from  in  \eqref{fi_identifications}, \eqref{fi_expr}.
 
Let us stress that the presentation \eqref{GMSW_V4} of $V_4$
in terms of $\gamma_{S^2}$, $\gamma_\Sigma$
is subject to a redundancy.
More precisely, there is a 1-parameter family
of redefinitions of the functions $\gamma_{\Sigma}$, $\gamma_{S^2}$
that leave $V_4$ invariant,
\beq
\gamma_{S^2} \rightarrow \gamma_{S^2} + 2 \, K \ , \qquad
\gamma_\Sigma \rightarrow \gamma_\Sigma - \chi \, K \ ,
\eeq
where $K$ is an arbitrary constant.
This redundancy will be fixed below when we construct $E_4$
and impose the condition \eqref{omegaI_condition}.

\subsubsection*{Flux quantization}

We can extract the flux quantum numbers of the 
setup by integrating $V_4$ on suitable 4-cycles 
in $M_6$ \cite{Gauntlett:2006ai}. 
If we integrate $V_4$ along the Riemann surface $\Sigma_g$ and 
$S^2_{y \psi}$, we obtain
\begin{align}
N_\Sigma := \int_{\Sigma_g \times S^2_{y \psi}} V_4 = \Big[\gamma_\Sigma
\Big ]_{
y = y_{\rm min}}^{y = y_{\rm max}}
=  \gamma_\Sigma^{\rm N} - \gamma_\Sigma^{\rm S} \ .
\end{align}
The superscripts `N', `S' denote evaluation at $y = y_\text{max, min}$,
respectively.
We can also integrate $V_4$ along $S^2_{y\psi}$ and $S^2$,
\begin{align} \label{north_and_south_fluxes}
N_{S^2} := \int_{ S^2 \times S^2_{y \psi}} V_4 =  
\Big[ \gamma_{S^2} \Big ]_{
y = y_{\rm min}}^{y = y_{\rm max}}
=  \gamma_{S^2}^{\rm N} - \gamma_{S^2}^{\rm S}  \ .
\end{align}
Finally, we can integrate $V_4$ over $\Sigma_g \times S^2$
at $y = y_{\rm max}$ or $y = y_{\rm min}$,
\begin{align}
N_{\rm N} & := \int_{y = y_{\rm max}} V_4  = - \Big[
2 \, \gamma_\Sigma^{\rm N} + \chi \, \gamma_{S^2}^N 
\Big] \ , \nn \\
N_{\rm S}& := \int_{y = y_{\rm min}} V_4  =- \, \Big[
2\, \gamma_\Sigma^{\rm S} + \chi  \, \gamma_{S^2}^S 
 \Big] \ .
\end{align}
The four quantities $N_\Sigma$, $N_{S^2}$, $N_{\rm N}$, $N_{\rm S}$
are all integers, but they are not independent,
since 
\beq
N_{\rm N} - N_{\rm S} + 2 \, N_{\Sigma} + \chi  \, N_{S^2} = 0 \ .
\eeq
Since $\chi$ is   an even integer,
the difference $N_{\rm N} - N_{\rm S}$ is  
an even integer.
It follows that the sum $N_{\rm N} + N_{\rm S}$
is also an even integer, and we can thus define the integer $M$ via
\beq \label{M_def}
M = \frac 12  \, (N_{\rm N} + N_{\rm S} ) \ .
\eeq 
The integers $(N_{S^2}, N_{\Sigma}, M)$ can be taken to be
the independent quanta specifying the $G_4$-flux configuration.

\subsubsection*{Harmonic 2-forms on $M_6$}

The space $M_6$ admits three independent harmonic 2-forms.
This is in accordance with the fact that we have three independent flux quanta,
associated to the three independent 4-cycles of $M_6$.
The harmonic 2-forms are denoted $\omega_\alpha$ and can be parametrized as
\beq \label{omega_alpha_param}
\omega_\alpha = dH_\alpha \, \frac{D\psi}{2\pi}
+ ( t_{\alpha S^2}  - 2 \, H_\alpha) \,\frac{V_{S^2}}{2\pi}
+ ( t_{\alpha \Sigma}  - \chi \, H_\alpha) \,\frac{V_\Sigma}{2\pi} \ ,
\eeq
where $H_\alpha$ is a function of $y$ and $t_{\alpha S^2}$, $t_{\alpha \Sigma}$
are suitable constants.
This parametrization is subject to a 1-parameter family of redefinitions,
corresponding to shifts of $H_\alpha$ by an arbitrary constant.
For definiteness, we fix this ambiguity by demanding that
\beq
H_\alpha^{\rm N} + H_\alpha^{\rm S} = 0 \ .
\eeq
The quantities $H_\alpha$, $t_{\alpha S^2}$, $t_{\alpha \Sigma}$ may  be fixed
 in terms of the metric functions in \eqref{metric_for_M6}
by requiring that $\omega_\alpha$ be co-closed. This would require solving
and ODE for $H_\alpha$. To proceed, however, we do not need to
find the explicit  form of the function $H_\alpha$.
It is sufficient to demand that the three $\omega_\alpha$'s be Poincar\'e
dual to the three 4-cycles associated to the flux quanta
$(N_\Sigma, N_{S^2}, M)$.
More precisely, we require
\beq \label{our_basis}
\int_{M_6} V_4 \, \omega_1 = N_{S^2} \ , \qquad
\int_{M_6} V_4 \, \omega_2 = N_\Sigma \ , \qquad
\int_{M_6} V_4 \, \omega_3 = M \ .
\eeq
We compute
\begin{align} \label{intermediate}
\int_{M_6} V_4 \, \omega_\alpha &= \Big[ 
t_{\alpha S^2} \, \gamma_\Sigma + t_{\alpha \Sigma} \, \gamma_{S^2}
- 2 \, \gamma_\Sigma \, H_\alpha - \chi \, \gamma_{S^2} \, H_\alpha
\Big]_{\rm S}^{\rm N} \nn \\
& = N_\Sigma \,  t_{\alpha S^2} 
+ N_{S^2} \,   t_{\alpha \Sigma}  
+ 2 \, M \, H_\alpha^{\rm N} \ ,
\end{align}
where we have expressed $\gamma_{S^2,\Sigma}^{\rm N,S}$ in terms of the flux
quanta.
From \eqref{intermediate}  we see that \eqref{our_basis}
implies
\beq \label{H_table}
\begin{array}{l  || c | c | ccc}
  & \phantom{l} H_\alpha^{\rm N} \phantom{l} & \phantom{l} t_{\alpha S^2} \phantom{l} &\phantom{l}  t_{\alpha \Sigma} \phantom{l} \\[1mm] \hline
\alpha = 1\phantom{l}  & 0 & 0 & 1 \\[1mm]
\alpha = 2\phantom{l}  & 0 & 1 & 0 \\[1mm]
\alpha = 3\phantom{l}  & \frac 12 & 0 & 0
\end{array}
\eeq
The above table contains all information we need  about $\omega_\alpha$
to compute the inflow anomaly polynomial.

\subsubsection*{Construction of $E_4$}

In the construction of $E_4$ we introduce background connections
for the $U(1)_\psi$ isometry as well as the $SO(3)$ isometry of $S^2$.
We also have three background connections $A^\alpha$ associated
to the three harmonic 2-forms $\omega_\alpha$, even though one combination
of these vectors is massive, as discussed in more detail later.
The construction of $E_4$ proceeds according to the general
recipe of section \ref{sec_general_param}. The details of the derivation can be 
found in appendix \ref{app_GMSW}.

The 4-form $E_4$ can be written as
\begin{align}
E_4 & = V_4^{\rm eq} + \frac{F^\alpha}{2\pi} \, \omega_\alpha^{\rm eq} \ , \nn  \\
V_4^{\rm eq} & = \bigg(  d\gamma_\Sigma \, \frac{V_\Sigma}{2\pi}
+ d\gamma_{S^2} \, e_2^{S^2}  \bigg) \, \frac{( D \psi)^{\rm g}}{2\pi}
+ \bigg( \gamma_\Sigma \, \frac{V_\Sigma}{2\pi}
+ \gamma_{S^2} \, e_2^{S^2}  \bigg)
\bigg(   - 2 \, e_2^{S^2} - \chi \, \frac{V_\Sigma}{2\pi} 
+ 2 \, \frac{F^\psi}{2\pi}  \bigg) \ , \nn \\
\omega_\alpha^{\rm eq} &= 
 dH_\alpha \, \frac{(D\psi)^{\rm g}}{2\pi}
+ ( t_{\alpha S^2}  - 2 \, H_\alpha) \, e_2^{S^2}
+ ( t_{\alpha \Sigma}  - \chi \, H_\alpha) \,\frac{V_\Sigma}{2\pi}  
+ 2 \, H_\alpha \, \frac{F^\psi}{2\pi}  \ .
\end{align}
The 1-form $(D\psi)^{\rm g}$ is the gauged version
of $D\psi$. It is computed in appendix \ref{app_GMSW}, and satisfies the property
\beq \label{gauged_psi_identity2}
\frac{d   ( D \psi)^{\rm g}}{2\pi} = - 2 \, e_2^{S^2} - \chi \, \frac{V_\Sigma}{2\pi} 
+ 2 \, \frac{F^\psi}{2\pi}  \ .
\eeq
The quantity $F^\psi = dA^\psi$ is the external connection associated to the
$U(1)_\psi$ isometry.
The 2-form $e_2^{S^2}$ is the closed and $SO(3)$-invariant completion
of $V_{S^2}/(2\pi)$,
\beq
d e_2^{S^2} = 0 \ , \qquad
\int_{S^2} e_2^{S^2} = 1 \ .
\eeq
The explicit expression of $e_2^{S^2}$ can be found in appendix 
\ref{app_GMSW}.

Recall from section \ref{sec_remarks_on_E4} that 
we must impose the relation \eqref{omegaI_condition}
in order to be able to set to zero
the combination $N_\alpha F^\alpha$ of background field strengths
associated to harmonic 2-forms.
As detailed in appendix \ref{app_GMSW},
imposing \eqref{omegaI_condition} allows us to write down the
values of the functions $\gamma_{S^2}$, $\gamma_\Sigma$
at the endpoints of the $y$ interval
in terms of the three flux quanta $N_{S^2}$, $N_\Sigma$, $M$,
\beq \label{gamma_sol}
\gamma_{S^2}^{\rm N,S} = \frac{M \, N_{S^2}}{2 \, N_{\Sigma}  - \chi \, N_{S^2}} \pm \frac 12 \, N_{S^2} \ , \qquad
\gamma_{\Sigma}^{\rm N,S} = - \frac{M \, N_{\Sigma}}{2 \, N_{\Sigma}  - \chi \, N_{S^2}} \pm \frac 12 \, N_{\Sigma}  \ .
\eeq

\subsubsection*{Computation of $X_8$}
The first Pontryagin class $p_1(TM_{11})$ takes the form
\beq
p_1(TM_{11}) = p_1(TW_4) + p_1(SO(3)) + \bigg[
- 2 \, e_2^{S^2} - \chi \, \frac{V_\Sigma}{2\pi} 
+ 2 \, \frac{F^\psi}{2\pi} \bigg]^2 \ .
\eeq
The above relation is justified as follows. The internal space $M_6$ is an $S^1_\psi$
fibration over a 5d space. Moreover, $M_6$ is in turn fibered over external spacetime
$W_4$. The terms $p_1(TW_4) + p_1(SO(3))$ capture the
first Pontryagin class of the 5d space fibered over $W_4$.
The class $p_1(SO(3))$ is associated to the $SO(3)$ isometry of $S^2$.
The final contribution is equal to $[d(D\psi)^{\rm g}/(2\pi)]^2$.
It accounts for the Chern root associated to the $S^1_\psi$ fibration,
whose connection has both internal legs (on $S^2$ and $\Sigma_g$)
as well as external legs on $W_4$.
By a similar token, the second Pontryagin class of the total geometry
reads
\beq
p_2(TM_{11}) = \Big[ p_1(TW_4)+  p_1(SO(3)) \Big] \, \bigg[
- 2 \, e_2^{S^2} - \chi \, \frac{V_\Sigma}{2\pi} 
+ 2 \, \frac{F^\psi}{2\pi} \bigg]^2 \ .
\eeq
Notice that we can drop any term in $p_2(TM_{11})$ with
more than six external legs.
In summary, the class $X_8$ for the setup under examination
takes the form
\beq
X_8 = \frac{1}{192} \, \bigg\{  
p_1(TW_4) + p_1(SO(3)) - \bigg[
- 2 \, e_2^{S^2} - \chi \, \frac{V_\Sigma}{2\pi} 
+ 2 \, \frac{F^\psi}{2\pi} \bigg]^2
\bigg\}^2 \ .
\eeq

\subsubsection*{Inflow anomaly polynomial}
We can now compute $\int_{M_6} E_4^3$ and $\int_{M_6} E_4 \, X_8$
and extract the inflow anomaly polynomial.
Integrals over $S^2$ are conveniently performed with the help
of the Bott-Cattaneo formula, reviewed in appendix \ref{app_GMSW}.
We also need \eqref{H_table} and \eqref{gamma_sol}.

The curvatures associated to the three harmonic 2-forms
$\omega_\alpha$ are subject to the constraint
\beq
N_\alpha \, F^\alpha = N_{S^2} \, F^1 + N_\Sigma \, F^2 + M \, F^3 = 0 \ .
\eeq
We choose to give the result in terms of $F^2$ and $F^3$,
solving the above constraint for $F^1$.

The  inflow anomaly polynomial reads
\begin{align} \label{total_answer}
(2\pi)^3 \, I_6^{\rm inflow} & = 
- \frac{1}{24} \, (\chi N_{S^2} + 2 N_\Sigma  ) \, p_1(TW_4) \, F^\psi
+ \frac{1}{24} \, \chi \, p_1(TW_4) \, F^3
\nn \\
& + \bigg[
\frac{N_{S^2}^2 \left(12 M^2+4 \chi  N_{S^2} N_{\Sigma }+\chi ^2 N_{S^2}^2-12 N_{\Sigma
   }^2\right)}{24 \left(\chi  N_{S^2}-2 N_{\Sigma }\right)}
 + \frac{1}{12} \, \chi \, N_{S^2}  
   \bigg] \, p_1(SO(3)) \, F^\psi
   \nn \\
& - \frac 14 \, M \, N_{S^2} \, p_1(SO(3)) \, F^2    
-\frac{1}{8} N_{S^2} \left(\chi  N_{S^2}+2 N_{\Sigma }\right)
\, p_1(SO(3)) \, F^3    \nn \\
& + \frac{1}{6} \left(\chi  N_{S^2}+2 N_{\Sigma }\right) \, (F^\psi)^3
\nn \\
& + \bigg[
-\frac{N_{S^2} N_{\Sigma } \left(-2 M-\chi  N_{S^2}+2 N_{\Sigma }\right) \left(2 M-\chi  N_{S^2}+2
   N_{\Sigma }\right)}{\left(2 N_{\Sigma }-\chi  N_{S^2}\right){}^2}
- \frac 12 \, \chi
   \bigg] \, (F^\psi)^2 \, F^3
   \nn \\
& + \frac{4 M N_{\Sigma }}{2 N_{\Sigma }-\chi  N_{S^2}}    \, F^\psi \, F^2 \, F^3
+ \frac{4 M^2-\chi ^2 N_{S^2}^2+4 N_{\Sigma }^2}{2 \left(2 N_{\Sigma }-\chi  N_{S^2}\right)} \, F^\psi \, (F^3)^2
\nn \\
& + \frac{N_{\Sigma }}{N_{S^2}} \, (F^2)^2 \, F^3
 + \frac{M}{N_{S^2}} \, F^2  \, (F^3)^2  
 - \frac 16 \, \chi \, (F^3)^3 \ .
\end{align}

An alternative presentation is based on 
a different choice of basis of harmonic 2-forms,
which we denote $\omega_{\rm C}$, $\omega_{\rm N}$, $\omega_{\rm S}$.
These combinations of the three $\omega_\alpha$'s are defined by
\beq
\int V_4 \, \omega_{\rm C} = N_{S^2} \ , \qquad
\int V_4 \, \omega_{\rm N,S} = N_{\rm N,S} \ ,
\eeq
where $N_{\rm N,S}$ were defined in \eqref{north_and_south_fluxes}.
More explicitly,
\beq
\omega_{\rm C} = \omega_1 \ , \qquad
\omega_{\rm N,S} = \mp \frac \chi 2 \, \omega_1 \mp \omega_2 + \omega_3 \ .
\eeq
Correspondingly, we have the identifications
\beq
F^2 =  - F^{\rm N} + F^{\rm S} \ , \qquad
F^3 = F^{\rm N} + F^{\rm S} \ .
\eeq
If desired, it is straightforward to rewrite the anomaly
polynomial \eqref{total_answer} in terms of $F^{\rm N,S}$.

\subsubsection*{Exact superconformal R-symmetry and central charge at large $N$}

To identify the superconformal R-symmetry we use $a$-maximization \cite{Intriligator:2003jj}.
The non-Abelian flavor symmetry $SO(3)$ cannot participate to $a$-maximization.
As a result, we simply turn off the associated background curvature.
At the level of the anomaly polynomial, we perform the replacements
\beq
F^\psi \rightarrow F^{\rm R} \ , \qquad
F^2 \rightarrow s^2 \, F^{\rm R} \ , \qquad
F^3 \rightarrow s^3 \, F^{\rm R}  \ ,
\eeq
with unspecified coefficients $s^{2,3}$.
For simplicity, we work in the large $N$ limit, with the scalings
\beq
N_{S^2} \sim N_\Sigma \sim M \sim F^2 \sim F^3 \sim \cO(N) \ .
\eeq
In the large $N$ approximation, our task is to 
maximize the coefficient of
$(F^{\rm R})^3$. There are four branches of solutions for $s^{2,3}$.
In two branches, the coefficient of $(F^{\rm R})^3$ attains the value 0;
these branches are not acceptable. On the other two branches, we find
\beq
(2\pi)^3 \,    I_6^{\rm CFT}  =
(2\pi)^3 \, (-  I_6^{\rm inflow} )= \frac 16 \, {\rm tr} \, {\rm R}^3 \, (F^{\rm R})^3 \ ,
\eeq
with
\begin{align}
{\rm tr} \, {\rm R}^3   &=  \pm 
\frac{
8 \, N_{S^2}^2 \, N_\Sigma^2 \, ( 4 \, N_\Sigma^2 + 2 \, \chi \, N_{S^2} \, N_\Sigma
+ \chi^2 \, N_{S^2}^2 - 3 \, M^2 )^{3/2}
}{ ( 2 \, \chi \, N_{S^2} \, N_\Sigma + 3 \, M^2) ^2}
\nn \\
& - \frac{
4 \, N_{S^2}^2 \, N_\Sigma^2 \, (2 \, N_\Sigma + \chi \, N_{S^2}) \, (
8 \, N_\Sigma^2 + 2 \, \chi \, N_{S^2} \, N_\Sigma
+ 2 \, \chi^2 \, N_{S^2}^2 - 9 \, M^2
)
}{( 2 \, \chi \, N_{S^2} \, N_\Sigma + 3 \, M^2) ^2} \ .
\end{align}
At large $N$,
\beq
a = c = \frac{9}{32} \, {\rm tr} \, {\rm R}^3 \ .
\eeq
If we select the branch with the plus sign, we find
\begin{align} \label{central_charge}
c  &=    
\frac{
9 \, N_{S^2}^2 \, N_\Sigma^2 \, ( 4 \, N_\Sigma^2 + 2 \, \chi \, N_{S^2} \, N_\Sigma
+ \chi^2 \, N_{S^2}^2 - 3 \, M^2 )^{3/2}
}{ 4\, ( 2 \, \chi \, N_{S^2} \, N_\Sigma + 3 \, M^2) ^2}
\nn \\
& - \frac{
9 \, N_{S^2}^2 \, N_\Sigma^2 \, (2 \, N_\Sigma + \chi \, N_{S^2}) \, (
8 \, N_\Sigma^2 + 2 \, \chi \, N_{S^2} \, N_\Sigma
+ 2 \, \chi^2 \, N_{S^2}^2 - 9 \, M^2
)
}{8\, ( 2 \, \chi \, N_{S^2} \, N_\Sigma + 3 \, M^2) ^2} \ .
\end{align}
We verify in appendix \ref{app_GMSW}
that this result agrees with the holographic central charge
computed in \cite{Gauntlett:2006ai}.
More precisely, the explicit formula given in 
\cite{Gauntlett:2006ai} applies to 
solutions with  $M = 0$.
The formula \eqref{central_charge} 
can be regarded as 
the generalization to the case  $M \neq 0$,
which is harder to tackle directly in holography.

Here we focused on a large-$N$ test of our result. Nonetheless, we expect the inflow anomaly polynomial \eqref{total_answer} to be exact in $N$, but to also contain contributions from decoupled modes. A field-theoretic understanding of the latter would allow us to repeat the $a$-maximization analysis to obtain corrections to the central charge of the CFT \eqref{central_charge}.



\section{Discussion}

We have presented a systematic method of computing anomalies of QFTs that are geometrically engineered in M-theory, using anomaly inflow in the M-theory background.  As we have described, there are two main pieces of data which determine the inflow analysis ({\it i.e.}, ingredients of $\CI_{12}$): the value of $G_4$ on the boundary of 11d spacetime, which we denote by $E_4$,  and the topology of the space $M_{10-d}$ corresponding to the transverse directions to the $d$-dimensional QFT worldvolume. We presented a general recipe for constructing $E_4$ in terms of forms in $M_{10-d}$, and characterized its ambiguities. This is naturally done using the language of $G$-equivariant cohomology, where $G$ is the isometry group of $M_{10-d}$. We have argued that the inflow anomaly polynomial can be extracted unambiguously in $d=2,4,6$. For the remainder of this discussion, we elaborate on some of the results we have obtained by applying this formalism. 

All the ambiguity in our construction of $E_4$ can be encapsulated by a single external 4-form $\gamma_4$. We argue in appendix  \ref{app_gamma4} that inflow anomalies in two and four dimensions are
independent of $\gamma_4$. For $d = 6$, the inflow anomaly
polynomial $I_8^{\rm inflow}$ depends on $\gamma_4$, which is fixed by extremizing
$I_8^{\rm inflow}$. This prescription is equivalent to imposing that the
8-form $E_4^2 + 2 \, X_8$ be trivial in the cohomology of the internal
space $M_4$. We interpret this requirement on $E_4^2 + 2 \, X_8$
as a consequence of tadpole cancellation, which is necessary
to have a well-defined M-theory setup.
Let us stress that our prescription for fixing $\gamma_4$
in $d = 6$ is such that 
we obtain the correct answer for the anomaly polynomial 
for M5-branes probing $\mathbb C^2/\Gamma$ (given in \eqref{Gamma_resultBIS}). In particular, the inclusion of $\gamma_4$ generates  an additional term relative to the inflow analysis of \cite{Ohmori:2014kda}, which provides precisely the contribution of the Green-Schwarz term of the center of mass mode for the M5-branes. Previously this term had only been fixed via anomaly matching on the tensor branch.

Turning to the 4d SCFTs corresponding to BBBW solutions, we have noted that our inflow analysis yields a new set of terms  in the anomaly polynomials for the case of M5-branes compactified on a sphere (the last line of \eqref{spherecase}). These terms are due to the additional $\mathfrak{su}(2)$ isometry algebra of the sphere.
This has not previously been discussed in the literature---such a    symmetry does not appear from reducing the anomaly polynomial from 6d \cite{Bah:2011vv,Bah:2012dg}, and is missing in the analysis of the flow to these theories from closing punctures on the surface  \cite{Agarwal:2015vla,Fazzi:2016eec,Nardoni:2016ffl}. 
From the latter point of view, this $\mathfrak{su}(2)$ is an accidental symmetry in the IR.
It would be interesting to understand if some of the subtleties regarding decoupled operators in these theories are clarified with the knowledge of this IR symmetry enhancement.  

We have also outlined a connection between the data of the holographic supergravity solutions and the input to the inflow anomaly polynomial, and  we demonstrated the utility of this observation by computing the anomaly polynomial of the 4d field theories dual to the GMSW solutions (with result given in \eqref{total_answer}).
Let us contrast our method to the standard application of the AdS/CFT dictionary.
In the latter, anomalies are extracted by computing Chern-Simons
coefficients in the bulk; subleading terms in $N$ require computing
higher derivative corrections to the supergravity action and loops in AdS.  
In our approach, the inflow anomaly polynomial
is expected to be exact in $N$, but
it contains   contributions both from the  interacting CFT of interest
and from  decoupled sectors.
Our analysis of GMSW solutions gives us
strong hints for a UV realization of the dual SCFTs
in terms of M5-branes probing a $\mathbb C^2/\mathbb Z_2$ singularity,
compactified on a Riemann surface with a suitable flavor twist \cite{toappear}.
The field theory picture will shed light 
on decoupling modes for these setups.

There are many interesting future directions 
to explore. We expect our methods to be applicable to a wider
class of 6d theories constructed in M-theory,
including (2,0) theories of type $D_N$
and (1,0) E-string theories.
We also believe that our approach can be extended to setups
with M5-branes wrapped on a Riemann surface with defects.
This analysis has been performed in \cite{Bah:2018jrv,Bah:2019jts}
for regular punctures in 4d $\cN = 2$ theories,
and it would be interesting to study $\cN = 2$ irregular punctures
and $\cN =1$ punctures for general compactifications
of 6d theories on a Riemann surface.
From a broader perspective, 
it would be useful to develop systematic geometric
tools for the computation of 't Hooft anomalies of theories
engineered in  type IIA, type IIB string theories  and F-theory.
Moreover, the methods of this work can be straightforwardly
generalized to include anomalies for continuous higher-form symmetries
\cite{Gaiotto:2014kfa,Cordova:2018cvg}.


\section*{Acknowledgments}

We would like to thank 
Fabio Apruzzi,
Nikolay Bobev,
Simone Giacomelli,
Ken Intriligator,
Craig Lawrie,
Raffaele Savelli,
Sakura Sch\"afer-Nameki,
Alessandro Tomasiello
for interesting conversations and correspondence. The work of IB and FB is supported in part by NSF grant PHY-1820784. RM is supported in part by ERC Grant 787320 - QBH Structure. We gratefully acknowledge the Aspen Center for Physics, supported by NSF grant PHY-1607611, for hospitality during part of this work.  


\appendix



\section{Some useful identities} \label{app_gauging}

\subsection{Identities for gauging of isometries}

Let $k_I^m$ denote the Killing vectors of the internal
space $M_{10-d}$, satisfying
$\pounds_I k_J = f_{IJ}{}^K \, k_K$.
Given a $p$-form $\omega$ on $M_{10-d}$,
its gauged counterpart is denoted $\omega^{\rm g}$
and is defined by
\beq
\omega  = \frac{1}{p!} \, \omega_{m_1 \dots m_p} \, d\xi^{m_1} \,
\dots \, d\xi^{m_p}  \qquad \Rightarrow \qquad
\omega^{\rm g} = \frac{1}{p!} \, \omega_{m_1 \dots m_p} \, D\xi^{m_1} \,
\dots \, D\xi^{m_p}    \ ,
\eeq
where $D\xi^m = d\xi^m + k^m_I \, A^I$.
An alternative equivalent presentation of $\omega^{\rm g}$
is
\beq \label{alternative_omegag}
\omega^{\rm g} = \sum_{M=0}^p \frac{1}{M!} \, A^{I_1} \dots A^{I_{M}} \,
\iota_{I_M} \dots \iota_{I_1} \omega \ ,
\eeq
where $\iota_I$ denotes interior product with the Killing vector
$k_I^m$,
\beq
\iota_I \omega = \frac{1}{(p-1)!} \, k_I^n \, \omega_{n m_1 \dots m_{p-1} } \,
d\xi^{m_1} \dots d\xi^{m_{p-1}} \ .
\eeq

A natural notion of gauge transformation on $\omega^{\rm g}$
can be defined as follows.
Let $\lambda^I$ be a set of scalar functions depending on the
external coordinates only, and consider the vector field $\Xi(\lambda)$
in the total space $M_{10}$ specified by
$\Xi (\lambda)= \lambda^I \, k_I^m \, \partial_{\xi^m}$.
We may then define\footnote{With reference to \eqref{alternative_omegag},
we have   explicitly
\beq
\delta_\lambda A^I \, \frac{\delta}{\delta A^I}
(\omega^{\rm g}) 
= 
 \sum_{M=1}^p \frac{1}{(M-1)!} \delta _\lambda A^{I_1} \, A^{I_2} \dots
A^{I_M} \,
\iota_{I_M} \dots \iota_{I_1} \omega \ . \nn
\eeq}
\beq
\delta_\lambda (\omega^{\rm g}) = \pounds_{\Xi(\lambda)} (\omega^{\rm g})
+ \delta_\lambda A^I \, \frac{\delta}{\delta A^I}
(\omega^{\rm g}) 
\ ,  
\eeq
where $\delta_\lambda A^I$ denotes the standard 
gauge transformation of a connection,
\beq \label{gauge_transf_A}
\delta_\lambda A^I = - D\lambda^I \ , \qquad
D\lambda^I = d\lambda^I - f_{JK}{}^I \, A^J \, \lambda^K \ .
\eeq
Making use of the identity
\beq \label{insertion_and_Lie}
\pounds_I \iota_J - \iota_J \pounds_I = f_{IJ}{}^K \, \iota_K \ ,
\eeq
we verify the relation
\beq \label{gauge_transf_omegag}
\delta_\lambda (\omega^{\rm g}) = \lambda^I \, (\pounds _I \omega)^{\rm g} \ .
\eeq

Let us now suppose that the form $\omega$ is invariant
under Lie derivative with respect to all isometry directions,
\beq
\pounds_I \omega  = 0 \ .
\eeq
Under this assumption, the following identity holds,
\beq \label{identity_singlet}
d(\omega^{\rm g}) = (d\omega)^{\rm g} + F^I \, (\iota_I \omega)^{\rm g} \ ,
\eeq
where
\beq 
F^I = dA^I - \frac 12 \, f_{JK}{}^I \, A^J \, A^K \ .
\eeq
We may now consider a collection of $p$-forms $\omega_I$
that satisfies
\beq
\pounds_I \omega_J = f_{IJ}{}^K \, \omega_K \ .
\eeq
In other words, $\omega_I$ transform in the adjoint representation
of the isometry algebra. For such a collection of $p$-forms,
 one has
\beq  \label{identity_one_index}
d(\omega^{\rm g}_I) = (d\omega_I)^{\rm g} + F^J \, (\iota_J \omega_I)^{\rm g} 
+ f_{IJ}{}^K \, A^J \, \omega_K^{\rm g}  \ .
\eeq
A similar formula holds for a two-indexed collection of $p$-forms:
under the assumption that
\beq
\pounds_I \omega_{J_1 J_2} = f_{IJ_1}{}^K \, \omega_{KJ_2}
+  f_{IJ_2}{}^K \, \omega_{K_1 K}    \ ,
\eeq
one has the identity
\beq  \label{identity_two_indices}
d(\omega^{\rm g}_{I_1 I_2}) = (d\omega_{I_1 I_2})^{\rm g} 
+ F^J \, (\iota_J \omega_{I_1 I_2})^{\rm g} 
+ f_{I_1J}{}^K \, A^J \, \omega_{KI_2}^{\rm g}
+ f_{I_2J}{}^K \, A^J \, \omega_{I_1 K }^{\rm g}  \ .
\eeq
The relations \eqref{identity_singlet}, \eqref{identity_one_index}, \eqref{identity_two_indices}
are all examples of the general identity
\beq
d(\Lambda_{\rm g}) + A^I \, (\pounds_I \Lambda^{\rm g}) = (d\Lambda)^{\rm g}
+ F^I \, (\iota_I \Lambda)^{\rm g} \ ,
\eeq
where $\Lambda$ is a $p$-form on $M_{10-d}$ in an arbitrary
representation of the isometry algebra.



\subsection{Identities for $SO(5)$ isometry of $S^4$} \label{app_6d}
The forms $\overline V_4$, $\overline \omega_{AB}$, $\overline \sigma_{AB,CD}$
are defined in \eqref{explicit_S4_objects}, repeated here for convenience,
\begin{align} 
\overline  V_4 & = \frac{3\, N}{8\pi^2} \,  \cdot \, \frac{1}{4!} \,
\epsilon_{A_1 \dots A_5} \,   d  y^{A_1}  \,
d  y^{A_2} \, d  y^{A_3} \, d  y^{A_4} \,   y^{A_5} \ , \nn \\
\overline \omega_{AB} & = \frac{3\, N}{8\pi^2} \,  \cdot \, \frac{-2}{4!} \,
\epsilon_{AB C_1 C_2 C_3} \,   d  y^{C_1}  \,
d  y^{C_2} \,    y^{C_3}   \ , \nn \\
\overline \sigma_{AB,CD} & = \frac{3\, N}{8\pi^2} \,  \cdot \, \frac{1}{4!} \,
\epsilon_{AB CDE} \,      y^{E}  \ .
\end{align}
Some useful integral identities involving $\overline V_4$, $\overline \omega_{AB}$, $\overline \sigma_{AB,CD}$
are
\beq \label{vanishing_integrals}
\int_{S_4} \overline \omega_{AB} \, \overline \omega_{CD}
 = 0 \ , \qquad
\int_{S^4} \, \overline V_4 \, \overline \sigma_{AB,CD} = 0 \ , 
\eeq
as well as
\begin{align}\label{nonvanishing_integrals}
& \alpha^{A_1 A_2 A_3 A_4} \, \beta^{B_1 B_2 B_3 B_4}  \, \int_{S^4} \overline V_4 \, \overline \sigma_{A_1 A_2,A_3 A_4} \, \overline \sigma_{B_1 B_2 B_3 B_4}   = \nn \\
& \qquad   \qquad \qquad  
=  N^3 \, \bigg[ \frac{3 }{8\pi^2}\bigg]^2 \, \bigg\{
\frac{1}{360} \, \alpha^{ABCD} \, \beta_{ABCD}
- \frac{1}{180} \, \alpha^{ABCD} \, \beta_{ACBD}
\bigg\} \ , \nn \\
& \alpha^{A_1 A_2} \, \beta^{B_1 B_2} \,
\gamma^{C_1 C_2 C_3 C_4} \, 
 \int_{S^4} \overline \omega_{A_1 A_2} \, \overline \omega_{B_1 B_2} \, 
\overline \sigma_{C_1 C_2,C_3 C_4}
\nn \\
& \qquad   \qquad \qquad  
= N^3 \, \bigg[ \frac{3 }{8\pi^2}\bigg]^2 \, \bigg\{
\frac{1}{540} \, \alpha^{AB} \, \beta^{CD} \, \gamma_{ABCD}
- \frac{1}{270} \, \alpha^{AB} \, \beta^{CD} \, \gamma_{ACBD}
\bigg\}  \ .
\end{align}
In the last expressions, the quantities $\alpha$, $\beta$, $\gamma$
are arbitrary tensors used as placeholders for $SO(5)$ indices.

In the main text we described $S^4$ in terms of embedding
coordinates $y^A$, $A = 1,\dots,5$. We can also describe $S^4$ in terms of four 
local coordinates $\xi^m$, $m = 1,\dots,4$.
We can write
\beq
dy^A = \partial_m y^A \, d\xi^m \ , \qquad
(dy^A)^{\rm g} = \partial_m y^A \, D\xi^m 
\ , \qquad
D\xi^m = d\xi^m + k^m_{AB} \, A^{AB} \ ,
\eeq
where $k^m_{AB}$ are the Killing vectors
of the $SO(5)$ isometries. They are given by
\beq \label{mykAB}
k^{mAB} = g^{mn} \, y^{[A} \partial_{n} y^{B]} \ , 
\eeq
where  $g_{mn}$ is the round metric on $S^4$
induced from the flat metric on $\mathbb R^5$,
\beq
g_{mn} = \partial_m y^A \, \partial_n y_A \ .
\eeq
Let us record the useful identities
\beq \label{some_S4_identities}
g^{mn} \, \partial_m y^A \, \partial_n y^B = \delta^{AB} - y^A \, y^B \ , \qquad
\iota_{AB} \, dy^C = y_{[A} \, \delta^C_{B]} \ .
\eeq



\section{The $G$-equivariant cohomology class defined by $E_4$
} \label{app_equiv}

\subsection{Relation between $V_4^{\rm eq}$, $\omega_\alpha^{\rm eq}$
and $G$-equivariant polyforms}

The discussion of section \ref{sec_general_param} fits naturally into the
language of $G$-equivariant cohomology, see \emph{e.g.}~\cite{libine2007lectureBIS}
for a review.
The group $G$ in our discussion is the isometry group
of $M_{10-d}$, acting on $M_{10-d}$ infinitesimally via 
Lie derivative.
The objects of interest are maps from the Lie algebra 
$\mathfrak g$ of $G$ into polyforms on $M_{10-d}$, \emph{i.e.}~formal
linear combinations of differential forms of various degrees,
\beq \label{alpha_map}
\begin{array}{ccccl}
f & : & \mathfrak g & \rightarrow & \Omega^*(M_{10-d}) \\
  &   & \cX & \mapsto & \alpha(\cX) \ .
\end{array}
\eeq
We  fix a basis $\{t_I\}$ of $\mathfrak g$, so that we can write
\beq
\mathfrak g \ni \cX = \cX^I \, t_I \ .
\eeq
The map $f$ must be $G$-equivariant,
which, at the infinitesimal level,
amounts to the property
\beq \label{alpha_equivariance}
\pounds_I f(\cX) = f_{IJ}{}^K \, \cX^J \, \frac{\partial}{\partial \cX^K}\,
 f(\cX) \ .
\eeq
The equivariant differential acting on $\alpha$ is defined by
\beq \label{twisted_deRham}
(d_{\rm eq} f) (\cX) = d \big( f(\cX) \big) + \iota_{\cX} f(\cX) \ ,
\eeq
where the operation $\iota_\cX$ amounts to $\cX^I \iota_I$.
Crucially, $(d_{\rm eq})^2 = 0$.
The $G$-equivariant cohomology 
of $M_{10-d}$ is then realized by considering the set of
$d_{\rm eq}$-closed polyforms, 
modulo $d_{\rm eq}$-exact polyforms.

Let us now revisit the expression \eqref{V4_eq_def}
for the object $V_4^{\rm eq}$.
If we identify the external connections $F^I$
with the abstract variables $\cX^I$ parametrizing
$\mathfrak g$,
we can reinterpret $V_4^{\rm eq}$ as a map 
of the form \eqref{alpha_map},
\beq
f_{V_4} \; : \;\; \cX^I \; \; \mapsto \;\; f_{V_4}(\cX) = V_4 + \cX^I \, \omega_I + \cX^I \, \cX^J \, \sigma_{IJ} \ .
\eeq
We then verify that the conditions \eqref{V4_gauge}
are precisely equivalent  to the equivariance
of $f_{V_4}$ as in \eqref{alpha_equivariance}.
Furthermore, the conditions \eqref{V4_closure}
are equivalent to $d_{\rm eq} f_{V_4} = 0$.
The object $V_4^{\rm eq}$ thus amounts to an
equivariantly closed form, hence the label `eq'.
In a completely analogous fashion, 
the object $\omega_\alpha^{\rm eq}$ in \eqref{omega_eq_def}
corresponds to the map
\beq
f_{\omega_\alpha} \; : \;\;  \cX^I \;\; \mapsto \;\; f_{\omega_\alpha}(\cX) = \omega_\alpha + \cX^I  \, 2\sigma_{I\alpha} \ .
\eeq 
We verify that \eqref{omega_alpha_gauge} is equivalent to
the equivariance of $f_{\omega_\alpha}$,
and that \eqref{omega_alpha_closure}
is equivalent to $d_{\rm eq} f_{\omega_\alpha} = 0$.

Incidentally, we notice that both $f_{V_4}$ and $f_{\omega_\alpha}$
are polynomials in $\cX^I$. 
 The natural notion of degree for each monomial
 in  $f_{V_4}$ or $f_{\omega_\alpha}$ is
\beq
\text{(differential form degree)} + 2 \, \text{(polynomial degree)} \ .
\eeq
It follows that $f_{V_4}$
is homogeneous of degree 4,
and $f_{\omega_\alpha}$ is homogeneous of degree 2.



\subsection{Deformations of $E_4$ and $G$-equivariant cohomology} \label{app_deformations}

In section \ref{sec_general_param} we have demonstrated that
constructing a good representative for $E_4$ amounts to
solving the conditions \eqref{gauge_summary} and \eqref{closure_summary},
repeated here for convenience,
\begin{align} \label{condition_summary}
dV_4 & = 0 \ ,  & \pounds_X V_4 & = 0 \ , \nn \\
\iota_X V_4 + d\omega_X &= 0 \ , & \pounds_X \omega_Y &= f_{XY}{}^Z \, \omega_Z \ , \nn \\
\iota_{(X} \omega_{Y)} + d\sigma_{XY} &= 0 \ , &
\pounds_X \sigma_{Y_1 Y_2} &= 
f_{X Y_1}{}^Z \, \sigma_{Z Y_2}
+ f_{XY_2}{}^Z \, \sigma_{Y_1 Z} \ .
\end{align}
We are using a collective index $X= (I,\alpha)$ that enumerates
all external connections. By definition, $\iota_\alpha = \pounds_\alpha = 0$,
and the only non-zero components of $f_{XY}{}^Z$ are $f_{IJ}{}^K$.
In this appendix, we suppose  to fix 
a reference solution $(V_4, \omega_X, \sigma_{XY})$
to \eqref{condition_summary},
and we investigate the most general deformation 
to a different solution.

\subsection*{The most general deformation}
The outcome of the analysis is as follows. The new forms are given by
\begin{align}
V_4 & \rightarrow   V_4 + dW_3 \ , \nn \\
\omega_X & \rightarrow 
   \omega_X + \iota_X W_3 - Z_{2X} + d\lambda_X + H_{2X}
 \ , \nn \\
\sigma_{XY} &\rightarrow      \sigma_{XY}
+ \iota_{(X} \lambda_{Y)}
 - Z'_{0(XY)}
 - Z_{0(XY)}
 + u_{XY} \ .
\end{align}
The 3-form $W_3$ must be chosen in such a way that
there exists a 2-form $Z_{2X}$ such that
\beq \label{pounds_W3}
\pounds_X W_3 = dZ_{2X} \ .
\eeq 
The 2-form $Z_{2X}$ in turn determines the 1-forms
$Z_{1[XY]}$ and the harmonic 2-forms $H'_{2[XY]}$ via
\beq
f_{XY}{}^Z \, Z_{2Z} - \pounds_X Z_{2Y} + \pounds_Y Z_{2X} = 
  dZ_{1XY} + H'_{2XY} \ .
\eeq
The harmonic forms $H_{2X}$ must be chosen  
compatibly with the constraint
\beq
 f_{XY}{}^Z \, H_{2Z} + H'_{2XY}  = 0\ .
\eeq
Once the harmonic 2-forms $H_{2X}$ are chosen,
they determine the 0-forms $Z'_{0XY}$ and the harmonic 
1-forms $H'_{1XY}$ via
\beq
\iota_X H_{2Y} = dZ'_{0XY} + H'_{1XY} \ .
\eeq
The 1-forms $\lambda_X$ must be chosen in such a way that
there exist 
a   0-form $Z_{0XY}$  
such that
\beq \label{pounds_lambda}
 \pounds_X \lambda_Y = 
   f_{XY}{}^Z \, \lambda_Z   
 + \iota_Y Z_{2X}
 - Z_{1XY}
+ dZ_{0XY}  - H'_{1XY}  \ .
\eeq
Finally, the constants $u_{XY}$ must be chosen
in such a way that
\begin{align}
0 & =  
  \pounds_{(Y_1} Z_{0 X|Y_2)} 
  - \pounds_X Z_{0(Y_1 Y_2)} 
  + f_{X Y_1}{}^Z \, Z_{0(ZY_2)}
      + f_{X Y_2}{}^Z \, Z_{0(ZY_1)}
    \nn \\
&   - \pounds_X Z'_{0(Y_1 Y_2)}
  + f_{X Y_1}{}^Z \, Z'_{0(ZY_2)}
    + f_{X Y_2}{}^Z \, Z'_{0(ZY_1)}
\nn \\
    &  
  - \iota_{(Y_1}  H'_{1X |Y_2)}
     -  \iota_{(Y_1}Z_{1X|Y_2)}
   - f_{X Y_1}{}^Z \, u_{ ZY_2 }  
   - f_{X Y_2}{}^Z \, u_{ ZY_1 }  \ .
\end{align}

\subsection*{Manifestly symmetric deformations and $G$-equivariant cohomology}
 
It is natural to consider deformations that are parametrized by quantities
that are manifestly symmetric under the action of the isometry group $G$.
More explicitly, we impose
\begin{gather} \label{manifest_symmetry}
\pounds_X W_3 = 0 \ , \qquad
\pounds_X \lambda_Y = f_{XY}{}^Z \, \lambda_Z \ , 
\qquad 
\pounds_X H_{2Y} = f_{XY}{}^Z \, H_{2Z} \ ,
\nn \\
 \pounds_X u_{Y_1 Y_2} = f_{XY_1}{}^Z \, u_{ZY_2}
+ f_{XY_2}{}^Z \, u_{Y_1 Z} \ .
\end{gather}
We notice that, in the above equations, we actually have
$\pounds_X H_{2Y}  = 0$ (because $H_{2Y}$ is harmonic)
and $\pounds_X u_{Y_1 Y_2} = 0$ (because $u_{Y_1 Y_2}$ is constant).
Under the additional assumptions \eqref{manifest_symmetry}, the most general
deformation described above takes a simpler form.
Comparison of \eqref{manifest_symmetry} with \eqref{pounds_W3}
shows that we can take $Z_{2X} = 0$, and therefore also
$Z_{1XY} = 0$, $H'_{2XY} = 0$.
Contrasting  \eqref{manifest_symmetry} and \eqref{pounds_lambda}
we infer $0 = dZ_{0XY} - H'_{1XY}$, which implies 
$Z_{0XY} = \text{const}$ and $H'_{1XY} = 0$.
The constant $Z_{0XY}$ can be reabsorbed in $u_{XY}$.

After these simplifications, the deformations parametrized by $W_3$ 
and $\lambda_I$ take the form
\begin{align}
V_4 & \rightarrow   V_4 + dW_3 \ , \nn \\
\omega_I & \rightarrow 
   \omega_I + \iota_I W_3    + d\lambda_I
 \ , 
 & \omega_\alpha & \rightarrow \omega_\alpha \nn \\
\sigma_{IJ} & \rightarrow   \iota_{(I} \lambda_{J)} \ ,
&  
\sigma_{I\alpha} & \rightarrow   0 \ ,
& 
\sigma_{\alpha \beta} & \rightarrow   0 \ .
\end{align}
This is equivalent to adding to the polyform $f_{V_4}(\cX)$
a $G$-equivariantly exact polyform,
\beq
f_{V_4}  \rightarrow f_{V_4} + d_{\rm eq} f_{W_3} \ , \qquad
f_{W_3}(\cX) = W_3  + \cX^I \, \lambda_I \ ,
\eeq
while leaving the polyforms $f_{\omega_\alpha}$ unaffected.

If we focus instead on the deformation parametrized by $\lambda_\alpha$,
we have
\begin{align}
V_4 & \rightarrow   V_4 \ , \nn \\
\omega_I & \rightarrow 
   \omega_I  
 \ , 
 & \omega_\alpha & \rightarrow \omega_\alpha + d\lambda_\alpha \nn \\
\sigma_{IJ} & \rightarrow   0 \ ,
&  
\sigma_{I\alpha} & \rightarrow   \tfrac 12 \, \iota_I \lambda_\alpha \ ,
& 
\sigma_{\alpha \beta} & \rightarrow   0 \ .
\end{align}
In this case, the polyform $f_{V_4}$ is unaffected,
while the polyform $f_{\omega_\alpha}$ are shifted
by $G$-equivariantly exact terms,
\beq
f_{\omega_\alpha} \rightarrow f_{\omega_\alpha} + d_{\rm eq} f_{\lambda_\alpha} \ , \qquad
f_{\lambda_\alpha} (\cX) = \lambda_\alpha \ .
\eeq

Let us now discuss the deformation parametrized 
by $H_{2I}$. Since this object is a harmonic 2-form, we must have
\beq
H_{2I} = c_I{}^\alpha \, \omega_\alpha \  ,
\eeq
with invariance under $G$ imposing $f_{IJ}{}^K \, c_K^\alpha = 0$.
This is a shift of $\omega_I$ by a combination of $\omega_\alpha$'s,
which is discussed in the main text around \eqref{new_V4eq}.

A deformation parametrized by $H_{2\alpha}$ is a change of basis
for the harmonic 2-forms, hence contains no interesting information.

Finally, constant shift by $u_{IJ}$, $u_{I\alpha}$, 
are discussed around \eqref{new_V4eq} and \eqref{new_sigma_Ialpha},
while a shift by $u_{\alpha \beta}$ only affects the purely external,
closed 4-form $\gamma_4$.



\section{(In)dependence of $I_{d+2}^{\rm inflow}$ on $\gamma_4$} \label{app_gamma4}

Throughout this appendix we make use of the compact notation
with collective index $X = (I,\alpha)$ introduced at the end of section
\ref{sec_general_param}. 
The most general $E_4$ in this language is given in 
 \eqref{compact_E4}, repeated here for convenience,
\beq
E_4 = V_4^{\rm g} + F^X \, \omega_X^{\rm g} 
+ F^X \, F^Y \, \sigma_{XY}
+   \gamma_4 \ .
\eeq 
Notice that in the main text we have set by definition $\sigma_{\alpha \beta } = 0$.
In this appendix, it is convenient to relax this assumption,
and let $\sigma_{\alpha \beta}$ be an unspecified constant.
Turning on $\sigma_{\alpha \beta}$ amounts to 
shifting $\gamma_4$ as
\beq
\gamma_4 \rightarrow \gamma_4 + \tfrac{1}{(2\pi)^2} \, F^\alpha \, F^\beta \, 
\sigma_{\alpha \beta} \ .
\eeq 
Since $\gamma_4$ is an arbitrary closed and gauge-invariant external
4-form, this shift is immaterial.

The goal of this appendix is to analyze the dependence of 
$I_{d+2}^{\rm inflow} = \int_{M_{10-d}} \cI_{12}$  on $\gamma_4$
and on constant shifts in $\sigma_{XY}$.

\subsection{The case $d=6$} \label{6d_ambiguity}

Let us first focus on the contribution of the $E_4^3$ term 
in $\cI_{12}$ to the inflow
anomaly polynomial.
Making use of the parametrization \eqref{compact_E4},
we   compute
\begin{align} \label{6d_E4_cube}
- \frac 16 \int_{M_4} E_4^3 & =  - \frac 12 \, F^{X_1} \, F^{X_2} \, F^{X_3} \, F^{X_4} \, \int_{M_4} \Big[   \omega_{X_1} \, \omega_{X_2} \, \sigma_{X_3 \, X_4}
+ V_4 \, \sigma_{X_1 X_2} \, \sigma_{X_3 X_4} \Big]
  \\
& - \frac 12 \, F^X \, F^Y \,   \gamma_4 \, \int_{M_4} \Big[\omega_X \, \omega_Y
+ 2 \, V_4 \, \sigma_{XY} \Big]
- \frac 12 \,   \gamma_4^2 \, \int_{M_4} V_4 \ .
\end{align}
Let us now turn to the term $E_4 \, X_8$ in $\cI_{12}$.
We are only interested here in keeping track
of terms with $\sigma_{XY}$, $\gamma_4$. 
The quantity $E_4 \, X_8$  is necessarily linear
in these parameters.
In order to have a $\sigma_{XY}$ or $\gamma_4$ factor,
we must select the part of $E_4$ with four
external legs, which means that $X_8$ must saturate the integration
over $M_4$. 
The relevant terms in $X_8$ can be written as
\beq
X_8 = Z \, V_4^{\rm g} \, p_1(TW_6) + Z_{XY} \, V_4^{\rm g} \, F^X \, F^Y  + \dots \ ,
\eeq
where $Z$ and $Z_{XY}$ are constants.
Notice that $X_8$ is not expected to receive any contribution
proportional to the curvatures $F^\alpha$
associated to the harmonic 2-forms $\omega_\alpha$.
As a result, the only non-zero components of
$Z_{XY}$ are $Z_{IJ}$.
Keeping nonetheless the collective indices $X$, $Y$, we have
\begin{align}
- \int_{M_4} E_4 \, X_8 &= - \Big[ Z \, p_1(TW_6) + Z_{XY} \, F^X \, F^Y \Big] \, \int_{M_4} \Big[  \gamma_4 + F^X \, F^Y \, \sigma_{XY} \Big] V_4 \nn \\
& + \text{terms without $\sigma_{XY}$, $  \gamma_4$} \ .
\end{align}

We propose the following prescription to fix $\gamma_4$:
extremize $I_8^{\rm inflow}$ with respect to arbitrary variations of $\gamma_4$.
We compute
\begin{align}
\delta I_8^{\rm inflow}   = \delta \gamma_4 \,  \bigg\{
&- \frac 12 \, F^X \, F^Y \, \int_{M_4} \Big[ \omega_X \, \omega_Y + 2 \, V_4 \, \sigma_{XY} \Big] \nn \\
&
- \Big[ \gamma_4  + Z \, p_1(TW_6) + Z_{XY} \, F^X \, F^Y \Big] \, \int_{M_4} V_4
\bigg\} \ .
\end{align}
The quantity $\gamma_4$ is then be fixed to be
\beq \label{our_rule}
\gamma_4 = - \frac 12 \, F^X \, F^Y \, \frac{
\int_{M_4} ( \omega_X \, \omega_Y + 2 \, V_4 \, \sigma_{XY} )
}{\int_{M_4}V_4}
 - Z \, p_1(TW_6) - Z_{XY} \, F^X \, F^Y \ .
\eeq

To further elucidate the prescription \eqref{our_rule},
let us consider the 8-form $E_4^2 + 2 \, X_8$.
The relevance of this combination stems from the fact that 
it corresponds to the combination 
$G_4^2/(2\pi)^2 + 2 \,X_8$ that governs the M2-brane
tadpole cancellation in M-theory compactifications.
Let us focus on the part of $E_4^2 + 2 \, X_8$
with four legs on $M_4$,
\begin{align} \label{E4_squared}
\Big[ E_4^2  + 2 \, X_8 \Big]_\text{4 legs on $M_4$}
& =   F^X \, F^Y \,  \Big[ 
\omega_X \, \omega_Y + 2 \, V_4 \, \sigma_{XY}
\Big]  
+ 2 \, V_4 \, \Big[ \gamma_4 + Z \, p_1(TW_6) + Z_{XY} \, F^X \, F^Y \Big]
\ .
\end{align}
The RHS is a sum of terms, each given by an external 4-form
wedge a 4-form on $M_4$. Let us demand
\beq
\Big[ E_4^2  + 2 \, X_8 \Big]_\text{4 legs on $M_4$}
 = 0 \qquad \text{in cohomology of $M_4$} \ . 
\eeq
Since $h^4(M_4) = 1$, the above requirement is equivalent to
\beq
\int_{M_4 }( E_4^2  + 2 \, X_8 ) = 0 \ .
\eeq
Making use of \eqref{E4_squared}, we see that this 
selects exactly the same  $\gamma_4$
as in \eqref{our_rule}.
This observation allows us to interpret
the prescription 
\eqref{our_rule} as 
an M2-brane tadpole cancellation condition.

\subsection{The case $d=4$}
\label{4d_discussion}

In the case $d=4$, not all harmonic 2-forms $\omega_\alpha$
are associated to  global symmetries of the setup.
To clarify this point, let us consider the low-energy effective action
for the compactification of M-theory on the internal space
$M_6$. Assuming supersymmetry is not completely broken,
the low-energy theory is a supergravity theory in five
dimensions. 
One linear combination of the vectors
$A^\alpha$ associated to $\omega_\alpha$ gets massive
because of its coupling to a 5d axion.  

The 11d background metric for the compactification is of the form
\beq \label{background_lambda}
ds^2(M_{11}) = e^{2\lambda} \, ds^2(W_5) + ds^2(M_6) \ ,
\eeq
where $\lambda$ is a warp factor and $W_5$ denotes the 5d spacetime
where the low-energy supergravity is defined.
We refrain from a full analysis of the low-energy dynamics,
and only focus on the relevant couplings.
The $G_4$-flux consists of a background part,
together with fluctuations.
Let us write
\beq
\frac{G_4}{2\pi} = V_4^{\rm g} + F^X \, \omega_X^{\rm g}
+ F^X \, F^Y \, \sigma_{XY}
+ g_4  + \dots \ .
\eeq 
In the previous expression, $V_4$ is the $G_4$-flux
configuration in the background. The gauging procedure
couples it to the 5d vectors associated to isometries of $M_6$.
The term $F^X \omega_X^{\rm g}$ contains both
the vectors associated to isometries, and the vectors associated to
harmonic 2-forms $\omega_\alpha$.
The term $F^X \, F^Y \, \sigma_{XY}$ is higher-order in external
fluctuations, but we have included because it 
is needed for closure of $G_4$. Finally, $g_4$ 
is a 5d field, independent of the internal coordinates.
It is  
the  zeromode 
in the Kaluza-Klein expansion of $G_4$ onto scalar harmonics on $M_6$.
This 5d field satisfies
\beq
dg_4 = 0 \ , \qquad
g_4 = dc_3 \ ,
\eeq
with $c_3$ a 3-form potential in five dimensions.
Notice that a 3-form potential in five dimensions is dual to a 0-form
potential, \emph{i.e.}~an axion.

The topological couplings in the 11d action induce
Chern-Simons couplings in the low-energy 5d supergravity theory.
A convenient way to perform the dimensional reduction
is to write the Chern-Simons interactions in one dimension higher.
We thus introduce $M_{12}$ with $\partial M_{12} = M_{11}$,
as well as $W_6$ with $\partial W_6 = W_5$.
The $C_3G_4G_4$ term in $M_{11}$
is reformulated as $G_4G_4G_4$ in $M_{12}$,
and upon reduction on $M_6$ yields the couplings
\begin{align}
\int_{M_6} - \frac 16 \, \left[ \frac{G_4}{2\pi} \right]^3
& = - \frac 12 \, g_4 \, F^X  \int_{M_6} V_4 \, \omega_X
- \frac 16 \, F^X \, F^Y \, F^Z \int_{M_6} ( \omega_X \, \omega_Y \, \omega_Z
+ 3 \, V_4 \, \omega_X \, \sigma_{YZ} )\ .
\end{align}
The second coupling is a Chern-Simons coupling in five dimensions,
and is not instrumental for our analysis.
The first coupling is the essential ingredient in what follows.

Recall from the discussion around \eqref{F_rotation}
that we are free to shift $\omega_I$ with $\omega_\alpha$'s, 
if we perform a compensating redefinition of the curvatures
$F^\alpha$. In particular, we can always shift the forms $\omega_I$
with linear combinations of $\omega_\alpha$
in such a way as to obtain
\beq \label{good_4d_basis}
\int_{M_6} V_4 \, \omega_I = 0 \ .
\eeq
With this choice of basis,
 $g_4$ is only coupling to the vectors $F^\alpha$
associated to the harmonic 2-forms $\omega_\alpha$.

The considerations of the previous paragraph
show that the terms in the 5d low-energy effective action 
involving $g_4$
are
\beq \label{some_5d_couplings}
S_{\rm 5d} = \int_{W_5} \bigg[
- \frac 12 \, \cG \, g_4 * g_4 - \cK_\alpha \, g_4 \, A^\alpha
+ \dots
\bigg] \ , \qquad
\cK_\alpha = \frac {1}{4\pi}   \int_{M_6} V_4 \, \omega_\alpha   \ .
\eeq
The Hodge star is computed with the external metric on $W_5$.
The quantity $\cG$ in the kinetic term for $g_4$
can be fixed by reducing $G_4 *_{11} G_4$ on the background
\eqref{background_lambda}. We do not need the precise expression of $\cG$
in what follows.
In the action \eqref{some_5d_couplings}, $g_4$ is the field strength
of the 3-form potential
$c_3$, which is regarded as dynamical field.
We can alternatively dualize, by adding a 0-form Lagrange multiplier $\Phi$
to impose the Bianchi identity for $g_4$,
\beq
\Delta S_{\rm 5d} = -\int_{W_5} g_4 \, d\Phi \ .
\eeq
If we eliminate $g_4$ using its equation of motion,
we obtain
\beq
S_{\rm 5d} + \Delta S_{\rm 5d} = \int_{W_5}
- \frac{1}{2 \cG} \, D\Phi * D\Phi + \dots \ , \qquad
D\Phi = d\Phi + \cK_\alpha \, A^\alpha \ .
\eeq
The scalar $\Phi$ has a shift symmetry coupled
to the combination $\cK_\alpha A^\alpha$,
which is thus rendered massive, as anticipated.

In the computation of the inflow anomaly
polynomial, the connections $A^I$, $A^\alpha$
are background fields coupled to the global symmetries
of the system.
We have argued that the combination $\cK_\alpha A^\alpha$
does not correspond to a symmetry.
As a result, we must set it to zero 
in the computation of the anomaly,
\beq \label{linear_combo_gone}
F^\alpha \, \int_{M_6} V_4 \, \omega_\alpha = 0    \ .
\eeq
Since we work in a basis such that \eqref{good_4d_basis} holds,
the condition \eqref{linear_combo_gone} is equivalent to 
\beq \label{linear_combo_goneBIS}
F^X \, \int_{M_6} V_4 \, \omega_X = 0   \ .
\eeq

We are now in a position to analyze how ambiguities in $E_4$
affect the inflow anomaly polynomial.
First of all, let us study the $E_4^3$ term in $\cI_{12}$.
We compute
\begin{align} \label{E4cube_4d_general}
- \frac 16 \, \int_{M_6} E_4^3 & = - \frac 16 \, F^X \, F^Y \, F^Z \, \int_{M_6}
\Big[ \omega_X \, \omega_Y \, \omega_Z
+ 6 \, V_4 \, \omega_{X} \, \sigma_{YZ} \Big]
 - F^X \, \gamma_4 \, \int_{M_6} V_4 \, \omega_X \ .
\end{align}
The dependence on $\gamma_4$ immediately 
drops away thanks to \eqref{linear_combo_goneBIS}.
The same holds true for any dependence on shifts of $\sigma_{XY}$.
Indeed, we can write
\beq
\sigma_{XY} = \overline \sigma_{XY} + u_{XY} \ ,
\eeq
where $\overline \sigma_{XY}$ is a reference choice
for $\sigma_{XY}$, and $u_{XY}$ are arbitrary constant.
The dependence on $u_{XY}$ in \eqref{E4cube_4d_general}
disappears, thanks to
\beq
- \frac 16 \, F^X \, F^Y \, F^Z \, \int_{M_6}
  6 \, V_4 \, \omega_{X} \, u_{YZ} 
  = - u_{YZ} \, F^Y \, F^Z \, \bigg( F^X \, \int_{M_6} V_4 \, \omega_X \bigg) =0 \ .
\eeq
In conclusion, the value of $\int_{M_6} E_4^3$ is 
insensitive to ambiguities in $E_4$.

The term $E_4 X_8$ can be sensitive to ambiguities
in $E_4$ if $X_8$ can saturate the integral over $M_6$.
For this to be possible, $X_8$ must contain a term of the form
\beq \label{bad_X8_terms}
X_8 = Z_{6I}^{\rm g} \, F^I + \dots \ ,
\eeq
where $Z_{6I}$ is a 6-form on $M_6$
and the label $I$ refers to a $U(1)$ factor in the isometry group.
In all examples we consider in this work,
however, $X_8$ does not contain any terms of the form
\eqref{bad_X8_terms}.
While we do not have a general proof,
we suspect that this feature should hold in general.
As a result, the term $E_4 X_8$ is insensitive to
ambiguities in $E_4$, and the full inflow
anomaly polynomial is determined unambiguously.

\subsection{The case $d=2$}

The contribution to the inflow anomaly polynomial
coming from the $E_4^3$ term in $\cI_{12}$ reads
\begin{align}
- \frac 16 \, \int_{M_8} E_4^3 & = 
- \frac 12 \, F^X \, F^Y \, \int_{M_8} \Big[
V_4 \, \omega_X \, \omega_Y + V_4^2 \, \sigma_{XY}
\Big]
- \frac 12 \, \gamma_4 \, \int_{M_8} V_4^2 \ .
\end{align}
As far as the $E_4 X_8$ term is concerned,
it can depend on ambiguities in $E_4$ only
if $X_8$ can saturate the integration
in the internal space $M_8$.
The part of $X_8$ with 8 internal legs
is the 8-form
\beq
Z_8 = X_8^{\rm background} = \frac{1}{192} \bigg[ p_1(TM_{8})^2
- 4 \, p_2(TM_{8}) \bigg] \ ,
\eeq
where the label ``background'' refers to the fact that
$Z_8$ is the value of $X_8$ when all external curvatures
are turned off.
The terms in $E_4 X_8$ that are potentially
ambiguous are then  
\beq
- \int_{M_8} E_4 \, X_8   = - F^X \, F^Y \, \int_{M_8} Z_8 \, \sigma_{XY}
- \gamma_4 \, \int_{M_8} Z_8
+ \dots \ .
\eeq
In summary, the terms in the inflow anomaly
polynomial containing $\gamma_4$ or $\sigma_{XY}$ are
\begin{align} \label{2d_dangerous_terms}
I_4^{\rm inflow} & = - \frac 12 \, F^X \, F^Y \, \int_{M_8} (
V_4^2 + 2 \, Z_8
) \, \sigma_{XY}
- \frac 12 \, \gamma_4 \, \int_{M_8} (V_4^2 + 2 \,Z_8)
+ \dots
\end{align}
We argue, however, that a good M-theory background
necessarily requires
\beq \label{tadpole}
\int_{M_8} (V_4^2 + 2 \,Z_8) = 0 \ .
\eeq
This condition is the tadpole 
cancellation condition that must hold 
for any compactification of M-theory on an 8-manifold
in absence of localized M2-brane sources.
As we can see, thanks to \eqref{tadpole}
the inflow anomaly polynomial
\eqref{2d_dangerous_terms} is independent on $\gamma_4$
and on constant shifts of $\sigma_{XY}$.
As a result, it is completely determined.



\section{Details on branes at an orbifold singularity} \label{app_Gamma}

\subsection*{Premilinaries}
When the M5-brane stack probes an orbifold singularity,
the isometry of $S^4$ is reduced to a subgroup of $SU(2)_L \times SU(2)_R$
of $SO(5)$.
Under the reduction $SO(5) \rightarrow SU(2)_L \times SU(2)_R$, we find
\begin{align} \label{split_relations}
p_1(SO(5)) & =  - 2 \, \big[ c_2(L) - c_2(R) \big] \ , \nn \\
p_2(SO(5)) & = \big[ c_2(L) - c_2(R) \big] ^2 \ , \nn \\
F^{AB} \, F^{CD} \, \overline \sigma_{AB,CD} & = \frac 12 \, y^5 \, N \, |\Gamma| \, 
 \big[ c_2(L) - c_2(R) \big]  \ .
\end{align}
In the above expressions $c_2(L,R) \equiv c_2(SU(2)_{L,R})$.
We have used the expression for $\overline \sigma_{AB,CD}$ in 
\eqref{explicit_S4_objects}, keeping in mind that $N$ is now replaced by $N\, |\Gamma|$.

The quantity $c_2(L)$ contains both internal and external contributions.
The external contribution is related to isometries of $S^4/\Gamma$,
and is only present if $\Gamma$ is of A type.
The internal contributions are present for any $\Gamma$ and are localized at the north and south poles.
They measure the curvature of the ALE spaces that resolve the orbifold
singularities $\mathbb C^2/\Gamma$ at each pole.
Since the two singularities are identical,
\beq \label{Euler_integrals}
\int_{\rm N} c_2(L) = \int_{\rm S} c_2(L) = \chi_\Gamma \ .
\eeq
In the above relations the symbols $\int_{\rm N,S}$ denote schematically
integration on the ALE space at the north, south pole respectively.
The Euler characteristic $\chi_\Gamma$ is given in \eqref{chi_Gamma}.

\subsection*{The form $E_4$}
The 4-form $E_4$ is given in \eqref{Gamma_E4},
repeated here for convenience, 
\beq 
E_4 = \overline V_4^{\rm g} + F^{AB} \, \overline \omega_{AB}^{\rm g}
+ F^{AB} \, F^{CD} \, \overline \sigma_{AB,CD} 
+ \frac{F^{{\rm N} i}}{2\pi} \, \omega_{{\rm N} i}
+ \frac{F^{{\rm S} i}}{2\pi} \, \omega_{{\rm S} i}
+ \gamma_4\ .
\eeq
It is important to stress that all curvatures in $E_4$, including
the curvature of $SU(2)_L$, are understood to have purely
external legs. In other words, the term 
$F^{AB} \, F^{CD} \, \overline \sigma_{AB,CD}$ does \emph{not}
contain the internal part of $c_2(L)$ that integrates to $\chi_\Gamma$
at the ALE spaces near the poles.
This observation is crucial to obtain the correct result.
The fact that $E_4$ only contains the external connections
is due to the fact that it is built gauging the isometries
of $S^4/\Gamma$.

\subsection*{The form $X_8$}
The Pontryagin classes of the total space can be written as
\begin{align}
p_1(TM_{11}) & = p_1(TW_6) + p_1(SO(5)) \ ,  \nn \\
p_2(TM_{11}) & = p_2(TW_6) + p_2(SO(5)) + p_1(TW_6) \, p_1(SO(5)) \ ,
\end{align}
where $p_{1,2}(SO(5))$ are given as in \eqref{split_relations}.
Expressing $X_8$ in terms of $p_{1,2}(TW_6)$, $c_2(L,R)$, we arrive at
\begin{align}
X_8 = 
\frac{1}{48} \, c_2(L) \, \big[ p_1(TW_6) + 4 \, c_2(R)\big]
+ \frac{1}{192} \big[  p_1(TW_6)^2 - 4 \, p_2(TW_6)  \big]
+ \frac{1}{48} \, c_2(R) \, p_1(TW_6)
 \ . \nn
\end{align}
Let us stress that here $c_2(L)$ contains both internal and external parts,
while all other curvatures have legs along $W_6$ only.

\subsection*{Integral of $E_4 X_8$}
The 4-form $E_4$ is given in \eqref{Gamma_E4}.
The integral $\int_{M_4} E_4 \, X_8$ receives two types of contributions.
Firstly, we can consider the purely external part of $X_8$,
and saturate the integral over $S^4$ using the part of $E_4$
with four internal legs, which is $\overline V_4$.
This contribution is
\beq
\int_{M_4} E_4 \, X_8 \supset \frac{N \, |\Gamma|}{|\Gamma|}
\bigg\{ 
\frac{1}{48} \, c_2(L)^{\rm ext} \, \big[ p_1(TW_6) + 4 \, c_2(R)\big]
+ \frac{1}{192} \big[  p_1(TW_6)^2 - 4 \, p_2(TW_6)  \big]
+ \frac{1}{48} \, c_2(R) \, p_1(TW_6) 
\bigg\} \ , \nn
\eeq
where the factor $N|\Gamma|$ originates from the new normalization
of $\overline V_4$, the factor $1/|\Gamma|$ originates
from the integral over $S^4/\Gamma$ (as opposed to $S^4$).
The superscript ``ext'' on $c_2(L)$
denotes its external part.

The other contribution to $\int_{M_4} E_4 \, X_8$ 
is obtained by saturating the integration over $M_4$ with
the internal part of $c_2(L)$ inside $X_8$.
We then consider the purely external part of $E_4$.
We find
\begin{align}
\int_{M_4} E_4 \, X_8 & \supset
\frac{1}{48} \,  \big[ p_1(TW_6) + 4 \, c_2(R)\big] \,
\int_{M_4} c_2(L) \, \big[
F^{AB} \, F^{CD} \, \overline \sigma_{AB,CD} + \gamma_4
 \big] \ .
\end{align}
The integral over $M_4$ localizes at the two poles.
More precisely, we get integrals over the ALE spaces that resolve
the orbifold singularities at each pole.
Taking into account the opposite orientation of the two poles, we can write
\begin{align}
\int_{M_4} c_2(L) \, \big[
F^{AB} \, F^{CD} \, \overline \sigma_{AB,CD} + \gamma_4
 \big] 
 &= \int_{\rm N} c_2(L) \, \big[
F^{AB} \, F^{CD} \, \overline \sigma_{AB,CD} + \gamma_4
 \big]^{\rm N} 
 \nn \\
& - \int_{\rm S} c_2(L) \, \big[
F^{AB} \, F^{CD} \, \overline \sigma_{AB,CD} + \gamma_4
 \big]^{\rm S}
 \ .
\end{align}
Since $\gamma_4$ is independent on the coordinates on $S^4$,
it drops away.
In contrast, $F^{AB} \, F^{CD} \, \overline \sigma_{AB,CD} ^{\rm N}
= - F^{AB} \, F^{CD} \, \overline \sigma_{AB,CD}^{\rm S}$,
and the two terms add to
\begin{align}
\int_{M_4} c_2(L) \, \big[
F^{AB} \, F^{CD} \, \overline \sigma_{AB,CD} + \gamma_4
 \big] 
 &=N \, |\Gamma| \, \chi_\Gamma \, \big[ 
 c_2(L)^{\rm ext} - c_2(R)
 \big]
 \ ,
\end{align}
where we have used \eqref{split_relations} and \eqref{Euler_integrals}.

In conclusion, the integral of $E_4 \, X_8$ is given by
\begin{align} \label{E4X8_result}
\int_{M_4} E_4 \, X_8 & = \frac{N \, |\Gamma| \, \chi_\Gamma}{48}  \, 
 \big[ c_2(L)^{\rm ext}
- c_2(R)\big] \, \big[ p_1(TW_6) + 4 \, c_2(R)\big]
 +  
\frac{N}{48} \, c_2(L)^{\rm ext} \, \big[ p_1(TW_6) + 4 \, c_2(R)\big]
\nn \\
&
+ \frac{N}{192} \big[  p_1(TW_6)^2 - 4 \, p_2(TW_6)  \big]
+ \frac{N}{48} \, c_2(R) \, p_1(TW_6)
 \ .
\end{align}

\subsection*{Integral of $E_4^3$}

A first set of contributions to $\int_{M_4} E_4^3$ originates
from the region away from the north and south poles.
These contributions are given by
 \begin{align} \label{part_of_E4cube}
\frac 16 \int_{M_4}  E_4^3 & \supset \int_{S^4/\Gamma} \bigg[
\frac 12 \, (FF\overline \sigma) \, (F\overline \omega)^2
+ \frac 12 \, (FF\overline \sigma)^{ 2} \overline V_4
+ \frac 12 \, (F\overline \omega)^2 \, \gamma_4
+ (FF\overline \sigma)  \, \overline V_4 \, \gamma_4
+ \frac 12 \, \overline V_4 \, \gamma_4^2
\bigg] \ ,
\end{align}
where we are suppressing $SO(5)$ indices for brevity.
The integral over $S^4/\Gamma$ can be computed with the
identities \eqref{vanishing_integrals},
\eqref{nonvanishing_integrals}. We must keep in mind,
however, that the quotient by $\Gamma$ generates
an additional factor $1/|\Gamma|$.
We then verify that the RHS of \eqref{part_of_E4cube}
is equal to
\beq
\frac{N^3 \, |\Gamma|^2}{24} \, p_2(SO(5))
+ \frac 12 \, N \, \gamma_4^2  \ .
\eeq

Let us now discuss the contributions to $\int_{M_4} E_4^3$ coming
from the harmonic 2-forms localized at the north and south poles.
These terms are  
\begin{align} \label{other_part_of_E4cube}
\frac 16 \int_{M_4} E_4^3 & \supset \frac 12 \, \int_{M_4} \big[
(FF \overline \sigma)+ \gamma_4 \big] \bigg[
\frac{ (F^{{\rm N}i} \, \omega_{{\rm N}i})^2 }{(2\pi)^2}
+ \frac{ (F^{\mathrm S i} \, \omega_{\mathrm S i})^2 }{(2\pi)^2}
\bigg] \ .
\end{align}
To proceed, we make use of
\begin{align}
\int_{M_4} (FF \overline \sigma + \gamma_4) \, \frac{(F^{\rm N} \omega_{\rm N})^2}{(2\pi)^2}
& = + (FF \overline \sigma^{\rm N} + \gamma_4) 
\, \frac{F^{{\rm N} i}}{2\pi}
\, \frac{F^{{\rm N} j}}{2\pi}
 \int_{\rm N} \omega_{{\rm N} i} \, \omega_{{\rm N}j} \ , \nn \\
\int_{M_4} (FF \overline \sigma + \gamma_4) \, \frac{(F^{\rm S} \omega_{\rm S})^2}{(2\pi)^2}
& = -  (FF \overline \sigma^{\rm S} + \gamma_4) 
\, \frac{F^{{\rm S} i}}{2\pi}
\, \frac{F^{{\rm S} j}}{2\pi}
 \int_{\rm N} \omega_{{\rm S} i} \, \omega_{{\rm S}j}
 \ .
\end{align}
The relative sign is due to the different orientation
of the two ALE spaces near the north and south poles.
The integral $ \int_{\rm N} \omega_{{\rm N} i} \, \omega_{{\rm N}j} $
is proportional to the entries of the Cartan matrix of the ADE Lie
algebra $\mathfrak g_\Gamma$ associated to $\Gamma$,
and similarly for $ \int_{\rm N} \omega_{{\rm S} i} \, \omega_{{\rm S}j}$.
We know that, at the conformal point where all resolution
$\mathbb C \mathbb P^1$'s are shrunk to zero size,
we have a non-Abelian enhancement of the flavor symmetry
at each pole.
In light of this observation,
we make the replacements
\beq
\, \frac{F^{{\rm N} i}}{2\pi}
\, \frac{F^{{\rm N} j}}{2\pi}
 \int_{\rm N} \omega_{{\rm N} i} \, \omega_{{\rm N}j}
 \rightarrow  \frac 14 \, \frac{{\rm tr} (F^\mathrm N)^2}{(2\pi)^2} \ ,
 \qquad
\, \frac{F^{{\rm S} i}}{2\pi}
\, \frac{F^{{\rm S} j}}{2\pi}
 \int_{\rm S} \omega_{{\rm S} i} \, \omega_{{\rm S}j}
 \rightarrow  \frac 14 \, \frac{{\rm tr} (F^\mathrm S)^2}{(2\pi)^2}    \ .
\eeq
Recalling \eqref{split_relations}, it follows that the RHS of \eqref{other_part_of_E4cube} is equal to
\begin{align}
 \frac{N \, |\Gamma|}{8} \, 
\big[ c_2(L)^{\rm ext} - c_2(R)  \big] \, \bigg[ 
\frac{{\rm tr} (F^\mathrm N)^2}{(2\pi)^2}
+ \frac{{\rm tr} (F^\mathrm S)^2}{(2\pi)^2}
\bigg]
& +\frac 14 \,  \gamma_4 \, 
\bigg[ 
\frac{{\rm tr} (F^\mathrm N)^2}{(2\pi)^2}
- \frac{{\rm tr} (F^\mathrm S)^2}{(2\pi)^2}
\bigg] \ .
\end{align}

In conclusion, the integral of $E_4^3$ yields
\begin{align}
\frac 16 \int_{M_4} E_4^3 & = 
 \frac{N^3 \, |\Gamma|^2}{24} \, \big[ c_2(L)^{\rm ext} - c_2(R)  \big]^2 
+ \frac 12 \, N \, \gamma_4^2
\nn \\
&  + \frac{N \, |\Gamma|}{8} \, 
\big[ c_2(L)^{\rm ext} - c_2(R)  \big] \, \bigg[ 
 \frac{ {\rm tr} \, (F^{\rm N})^2}{(2\pi)^2}
+ \frac{{\rm tr}\, (F^{\rm S})^2}{(2\pi)^2}
\bigg]
\nn \\
& + \frac 14 \,  \gamma_4 \, 
\bigg[ 
 \frac{ {\rm tr} \, (F^{\rm N})^2}{(2\pi)^2}
- \frac{{\rm tr}\, (F^{\rm S})^2}{(2\pi)^2}
\bigg]  \ .
\end{align}
The sum of the above quantity with $\int_{M_4} E_4 \, X_8$
given in \eqref{E4X8_result}
gives $- I_8^{\rm inflow}$ as quoted in the main text in \eqref{Gamma_result}.



\section{Details on the BBBW setup}\label{app_details}

This appendix is devoted to some derivations regarding the setups 
studied in section \ref{twisting_on_Sigma}.
The relevant space $M_6$ is an $S^4$ fibration over $\Sigma_g$,
specified by the background flux \eqref{twist_F},
repeated here for the reader's convenience,
\beq
F_\Sigma^{AB} = q^{AB} \, V_\Sigma  \ , \qquad
\int_{\Sigma_g} V_\Sigma = 2\pi   \ .
\eeq
The matrix $q^{AB}$ is given in \eqref{twist_F}.
All the following results, however, hold
for any constant antisymmetric $q^{AB}$.

\subsubsection*{Additional isometries in the case $g=0$}

In the case $g=0$, the line element of $M_6$ reads
\beq
ds^2(M_6) = ds^2(S^2) + ds^2(S^4)^{\rm t} = 
ds^2(S^2) + (dy^A - q^{AB} \, y_B \, \cV) \,
(dy_A - q_{AC} \, y^C \, \cV) \ .
\eeq
Recall that $y^A$, $A = 1,\dots,5$ are constrained coordinates for the $S^4$ fiber,
$y^A y_A = 1$.
The 1-form $\cV$ is defined only locally, and is an antiderivative
of $V_\Sigma$,
\beq
d\cV = V_\Sigma \ .
\eeq
We find it convenient to 
 parametrize the base $S^2$ in terms of three constrained 
coordinates $z^a \, z_a = 1$, $a = 1,2,3$.
By means of a direct computation,
one verifies that the following triplet of 1-forms on $M_6$
are such that the dual contravariant vectors are Killing,
\beq \label{extra_isom} 
k_a =   \epsilon_{abc} \, z^b \, dz^c
- \frac 12 \, z_a  \, q^{AB} \, y_A \, (dy^B  - q^{BC} \, y_C \, \cV) \ .
\eeq
The term $\frac 12 \, \epsilon_{abc} \, z^b \, dz^c$
is the expression of the Killing 1-forms of a round $S^2$
considered in isolation.
The other terms in \eqref{extra_isom}
demonstrate how these 1-forms are extended to the total
space $M_6$, depending on the twist data $q^{AB}$. 

The explicit expression \eqref{extra_isom} of the Killing 1-forms
$k_a$ is useful in  checking the following identities,
\begin{align} \label{insertion_identities}
\iota_{a} (dy^A)^{\rm t} &= \frac 12  \, z_a \, q^{AB} \, y_B \ , &
\iota_{a} dz^b & = -  \epsilon_{a}{}^{bc} \, z_{c}  \ , \nn \\
V_\Sigma & = 
\frac 12 \cdot \frac{1}{2} \, \epsilon_{abc} \, dz^a \, dz^b \, z^c \ , &
\iota_{a} V_\Sigma
&  
= \frac 12 \,   dz_a  \ .
\end{align}
We can now compute 
the interior product of $k_{a}$ with $V_4$
given in  \eqref{4dV4}.
Two useful partial results are
\begin{align} \label{partial_identities}
\iota_{a} (\overline V_4)^{\rm t} & = 
\frac{3N}{8\pi^2} \cdot \frac{1}{12} \, 
z_a \, 
\epsilon_{ABCDE} \, 
(q^{AA'} y_{A'}) \, (dy^B)^{\rm t}\, (dy^C)^{\rm t} \, (dy^D)^{\rm t} \, y^E \nn \\
& = \frac{3N}{8\pi^2} \cdot \frac{-1}{24} \, 
z_a \, 
\epsilon_{ABCDE} \, 
q^{AB} \, (dy^C)^{\rm t}\, (dy^D)^{\rm t}\, (dy^E)^{\rm t}
\ ,  \nn \\
q^{AB} \, \iota_{a} \,  \overline \omega_{AB}^{\rm t}& = 
\frac{3N}{8\pi^2} \cdot \frac{-1}{12} \, 
z_a \, 
\epsilon_{AB CDE} \, q^{AB} \, (q^{CC'} y_{C'}) \, (dy^D)^{\rm t} \, y^E \nn \\
& = \frac{3N}{8\pi^2} \cdot \frac{1}{48} \, 
z_a \, 
\epsilon_{AB CDE} \, q^{AB} \, q^{CD} \, (dy^E)^{\rm t}
 \ ,
\end{align}
where we used two Schouten identities
deriving from $\delta_{B[A_1} \, \epsilon_{A_2 A_3 A_4 A_5 A_6]} = 0$.
Combining all elements, we verify
the identity
\begin{align} \label{iotaV_as_tot_der}
\iota_{a} V_4 & =  \frac{3N}{8\pi^2} \cdot \frac{1}{24} \, d\bigg\{ z_a  \, 
\Big[
V_\Sigma \, \epsilon_{ABCDE} \, q^{AB} \, q^{CD} \, y^E
- \epsilon_{ABCDE} \, q^{AB} \, (dy^C)^{\rm t} \, (dy^D)^{\rm t} \, y^E
\Big] \bigg\} 
\nn \\
& =  d\bigg\{ z_a \,
\Big[
V_\Sigma \, q^{AB} \, q^{CD} \, \overline \sigma_{AB,CD}
+ \frac 12 \,  q^{AB} \, \overline \omega_{AB}^{\rm t}
\Big] \bigg\}  \ .
\end{align}

\subsubsection*{Derivation of $E_4$}

The 4-form $E_4$ is constructed
as
\beq  \label{E4_blueprint}
E_4 = V_4^{\rm g} + F^I \, \omega_I^{\rm g} 
+ F^I \, F^J \, \sigma_{IJ}
+ C \, p_1(TW_4) \ .
\eeq
Here we have used \eqref{compact_E4},
combined with the observation that $M_6$
has no harmonic 2-forms, so that the collective index $X$
reduces to the isometry label $I$.
The latter refers both to isometries of class (i) and 
isometries of class (ii), in the terminology of section \ref{twisting_on_Sigma}.
Accordingly, we split the $I$ index
as 
\beq
I = (\hat I , a) \ , \qquad \hat I = 1,2 \ , \qquad a = 1,2,3 \ .
\eeq 
We have already introduced the $SO(3)_{S^2}$
index $a$ above.
The new index $\hat I$ refers to isometries of class (i).
More precisely, 
we describe the 
external connections $A^{AB}_{\rm ext}$ of \eqref{external_stuff}
by writing
\beq
A^{AB}_{\rm ext} = A^{\hat I} \, M_{\hat I}{}^{AB} \ ,
\eeq
where the index $\hat I$ labels the two generators
of the class (i) isometry $SO(2)_1 \times SO(2)_2$,
$A^{\hat I} = (A_1, A_2)$.
The matrices $M_{\hat I}{}^{AB}$ are constant and  readily
read off from \eqref{external_stuff},
\beq \label{M_matrices}
M_1{}^{AB} = 
{ \small
\begin{pmatrix}
0 & -1 & & & \\ 
1 & 0 & & & \\ 
  &   & 0 & & \\ 
  &   & & 0& \\ 
  &   & & &0 
\end{pmatrix}
} \ , \qquad
M_2{}^{AB} = 
{ \small
\begin{pmatrix}
0 &  & & & \\ 
  & 0 & & & \\ 
  &   & 0 & -1& \\ 
  &   &1 & 0& \\ 
  &   & & &0 
\end{pmatrix}
}  \ .
\eeq
The Killing vectors associated to $A^{\hat I}$
are linear combinations of the Killing vectors $k_{AB}$
of the round $S^4$, 
\beq
k_{\hat I}  = M_{\hat I}{}^{AB} \, k_{AB} \ ,
\eeq 
with $k_{AB}$ as in \eqref{mykAB}.

Determining $E_4$ amounts to solving the following equations
for $\omega_I$, $\sigma_{IJ}$,
\beq
\iota_I V_4 + d\omega_I =  0 \ , \qquad
\iota_{(I} \omega_{J)} + d \sigma_{IJ} = 0 \ ,
\eeq
where $I = (\hat I, a)$ and $V_4$ is given in \eqref{4dV4}.

\paragraph{Solution for $\omega_I$.}
Let us first discuss the forms $\omega_{\hat I}$
associated to isometries from the $S^4$ fiber.
A natural ansatz for $\omega_{\hat I}$ is
\beq \label{omega_ansatz}
\omega_{\hat I} = \widetilde \omega_{\hat I}^{\rm t} + V_\Sigma \, g_{\hat I} \ ,
\eeq
where $ \widetilde \omega_{\hat I} $ is a 2-form with two legs
along the $S^4$ fibers, while $g_{\hat I}$ are 0-forms.
The equation that determines $\omega_{\hat I}$ is
$d\omega_{\hat I} + \iota_{\hat I} V_4 = 0$.
Upon using \eqref{omega_ansatz}
and separating terms with and without
$V_\Sigma$, we find the relations
\beq \label{after_omega_ansatz}
d\widetilde \omega_{\hat I} + \iota_{\hat I} \overline V_4 = 0 \ , \qquad
dg_{\hat I} + p^{AB} \, \iota_{AB} \widetilde \omega_{\hat I} + q^{AB} \,
\iota_{\hat I} \, \overline  \omega_{AB} =0    \ .
\eeq
The first equation is readily solved by setting
\beq
\widetilde  \omega_{\hat I} =M_{\hat I}{}^{AB} \,  \overline \omega_{AB} \ .
\eeq
Since by assumption $\widetilde  \omega_I$
is a 2-form in the $S^4$ fiber,
there is no non-trivial closed but not exact form that we can add to it.
Adding an exact piece would have no effect on the
inflow anomaly polynomial.
The second equation in \eqref{after_omega_ansatz}
  becomes
\beq
0 = dg_{\hat I} + q^{AB} \, M_{\hat I}{}^{CD} \, \Big( \iota_{AB} \overline \omega_{CD}
+ \iota_{CD} \overline \omega_{AB}   \Big) = 
dg_{\hat I} - 2 \, q^{AB} \, M_{\hat I}{}^{CD}  \, d\overline  \sigma_{AB,CD} \ ,
\eeq
where we have used    \eqref{some_S4_identities}.
As we can see, $g_{\hat I}$ is fixed up to a constant.
More precisely, the second equation
in \eqref{after_omega_ansatz}
only has to hold when wedged with $V_\Sigma$.
In summary, $\omega_{\hat I}$ is given by
\beq
\omega_{\hat I} = M_{\hat I}{}^{AB} \, \overline \omega_{AB}^{\rm t}
+ 2 \, V_\Sigma \, q^{AB} \, M_{\hat I}{}^{CD} \, \overline \sigma_{AB,CD}
+ C_{\hat I} \, V_\Sigma \ ,
\eeq
where $C_{\hat I}$ are arbitrary functions depending on $\Sigma$
only. The term $C_{\hat I} \, V_\Sigma$ is thus closed but not 
necessarily exact. We can be more precise:
since $C_{\hat I} \, V_\Sigma$ is a closed form on $S^2$,
it can be decomposed as a sum of an exact form and a harmonic
form. The exact piece can be disregarded. The harmonic piece
must be a \emph{constant} multiple of $V_\Sigma$.
It follows that, without any loss of generality,
we can take $C_{\hat I}$ to be constant.

In the case $g=0$ we also have to construct
$\omega_{a}$, which must satisfy
$d\omega_{a} + \iota_{a} V_4 = 0$.
Thanks to \eqref{iotaV_as_tot_der},
we know how to write $ \iota_{a} V_4$
as a total derivative.
As a result, 
$\omega_{a}$ is given by
\begin{align}
\omega_{a} &= 
-  \, z_a
\bigg[
V_\Sigma \, q^{AB} \, q^{CD} \, \overline \sigma_{AB,CD}
+ \frac 12 \,  q^{AB} \, \overline \omega_{AB}^{\rm t}
\bigg]
 + C_{a} \, V_\Sigma  \ .
\end{align}
Once again, we have not included an exact piece,
because it would have no effect on the inflow anomaly
polynomial.
\emph{A priori}, the 0-form
$C_{a}$ is allowed to have an arbitrary dependence on $S^2$.
Using arguments similar to those of the previous
paragraphs, however, we argue that we can take $C_a$
constant without loss of generality.
A constant $C_a$, however, is incompatible
with the fact that $\omega_a$ must be covariant
with respect to its $SO(3)_{S^2}$ index.
In other words, there is no invariant tensor of $SO(3)_{S^2}$
with one index. We conclude $C_a = 0$.

\paragraph{Solution for $\sigma_{IJ}$.}
Let us now turn to the determination of $\sigma_{IJ}$.
We first focus on the components $\sigma_{\hat I \hat J}$.
The equation to solve is
\begin{align}
0 & = d\sigma_{\hat I \hat J} + \iota_{(\hat I} \omega_{\hat J)}
 = d\sigma_{\hat I \hat J}
 + M_{(I}{}^{AB} \,  \iota_{J)} \overline \omega_{AB} ^{\rm t}
 = d\sigma_{\hat I \hat J} 
 + M_{(\hat I}{}^{AB} \, M_{\hat J)}{}^{CD} \, \iota_{CD}  \,
 \overline \omega_{AB}^{\rm t }  \ .
\end{align}
Making use of  \eqref{some_S4_identities},
we see that 
\beq
\sigma_{\hat I \hat J} = M_{(\hat I}{}^{AB} \, M_{\hat J)}{}^{CD} \, \overline \sigma_{AB,CD}
+ u_{\hat I \hat J} \ ,
\eeq
where $u_{\hat I \hat J} = u_{\hat J \hat I}$ are constants.

For $g=0$, we also have to determine $\sigma_{  \hat I a}$ 
and $\sigma_{ab}$. The former is determined by the requirement
\beq
0 = d\sigma_{\hat I a} + \frac 12 \, \iota_a \, \omega_{\hat I}
+ \frac 12 \, M_{\hat I}{}^{AB} \, \iota_{AB} \, \omega_{a} \ .
\eeq
Using the formulae for $\omega_{\hat I}$, $\omega_a$
given above,
as well as the expression of $\mathring \omega_{AB}$
and $\mathring \sigma_{AB,CD}$ given in \eqref{explicit_S4_objects},
one verifies the identity
\begin{align}
\frac 12 \, \iota_a \, \omega_{\hat I}
+ \frac 12 \, M_{\hat I}{}^{AB} \, \iota_{AB} \, \omega_{a}
= \frac 14 \, C_{\hat I} \, dz_a
+ d\bigg[
\frac{3N}{8\pi^2} \cdot \frac{1}{48} \, z_a \, 
\epsilon_{ABCDE} \, M_{\hat I}{}^{AB} \, q^{CD} \, y^E
\bigg] \ .
\end{align}
It follows that
we can write
\beq
\sigma_{\hat I a} = - \frac{3N}{8\pi^2} \cdot \frac{1}{48} \, z_a \, 
\epsilon_{ABCDE} \, M_{\hat I}{}^{AB} \, q^{CD} \, y^E
- \frac 14 \, C_{\hat I} \, z_a + u_{\hat I a} \ ,
\eeq
where $u_{\hat I a}$ is an arbitrary constant.
Once again, however, 
we must conclude $u_{\hat I a} = 0$,
because there is no $SO(3)_{S^2}$ invariant object
with one index $a$.
Our final task is the determination of $\sigma_{ab}$ in
\beq
0 = d\sigma_{ab} + \iota_{(a} \omega_{b)} \ .
\eeq
We have the identity
\begin{align}
\iota_{(a} \omega_{b)} 
& = 
- d\bigg[ 
 \frac {1}{4} \, z_a \, z_b \, q^{AB} \, q^{CD} \, \overline \sigma_{AB,CD}
\bigg] \ ,
\end{align}
which implies
\beq
\sigma_{ab} =  \frac {1}{4} \, z_a \, z_b \, q^{AB} \, q^{CD} \, \overline \sigma_{AB,CD}
 + u_{ab} \ ,
\eeq
for some constant $u_{ab}$.
This time there is a natural candidate for a constant $u_{ab}$
compatible with $SO(3)_{S^2}$ symmetry,
\beq
u_{ab} = u \, \delta_{ab} \ .
\eeq

\paragraph{Summary.}
The solution for all components of $\omega_I$,
$\sigma_{IJ}$
is summarized as follows,
\begin{align} 
\omega_{\hat I} &= M_{\hat I}{}^{AB} \, \overline \omega_{AB}^{\rm t}
+ 2 \, V_\Sigma \, q^{AB} \, M_{\hat I}{}^{CD} \, \overline \sigma_{AB,CD}
+ C_{\hat I} \, V_\Sigma \ , \label{appendix_results_no1} \\
\omega_{a} &= 
-    z_a
\bigg[
V_\Sigma \, q^{AB} \, q^{CD} \, \overline \sigma_{AB,CD}
+ \frac 12 \,  q^{AB} \, \overline \omega_{AB}^{\rm t}
\bigg]
  \ , \label{appendix_results_no2} \\
 \sigma_{\hat I \hat J} & = M_{(\hat I}{}^{AB} \, M_{\hat J)}{}^{CD} \, \overline \sigma_{AB,CD}
+ u_{\hat I \hat J} \ ,
\label{appendix_results_no3} \\
\sigma_{\hat I a} &= -\frac 12 \, z_a \, 
M_{\hat I}{}^{AB} \, q^{CD} \,  \overline \sigma_{AB,CD}
- \frac 14 \, C_{\hat I} \, z_a   \ ,
\label{appendix_results_no4} \\
\sigma_{ab} &=  \frac {1}{4} \, z_a \, z_b \, q^{AB} \, q^{CD} \, \overline \sigma_{AB,CD}
 + u \, \delta_{ab} \label{appendix_results_no5}  \ .
\end{align}
If we plug the above relations into
\eqref{E4_blueprint}, after some manipulations
we recover the expression \eqref{E4_final} for $E_4$
given in the main text.
The constants $C_1$, $C_2$ in \eqref{E4_final}
are identified with the components of $C_{\hat I}$.
All terms with $u_{\hat I \hat J}$, $u$
are absorbed into 
  $\gamma_4$ in  \eqref{E4_final}.



\section{Details on the GMSW setup} \label{app_GMSW}

\subsection{Brief review of the solutions}  

In this appendix we review a class of  M-theory 
solutions with 4d $\cN = 1$ superconformal symmetry,  first described in GMSW \cite{Gauntlett:2004zh}.
The 11d metric reads
\begin{align} \label{GMSW_review_metric}
ds^2_{11} & = L^2 \, e^{2\lambda} \, \Big[ ds^2(AdS_5) 
+  ds^2(M_6) \Big] \ ,  \nn  \\
ds^2(M_6) & = e^{-6\lambda} \Big[ F_1 \, d  s^2(C_1)
+ F_2 \, d  s^2(C_2) \Big]
+ \frac{e^{-6\lambda}}{\cos^2\zeta} dy^2
+ \frac{\cos^2 \zeta}{9} D\psi^2 \ , \nn \\
dD\psi &= - \chi_1 \, V_{C_1} - \chi_2 \, V_{C_2} \ .  
\end{align}
The constant $L$ is the overall length scale of the solution.
The metric on $AdS_5$ is normalized in such a way that 
the Ricci scalar is $R_{\rm AdS_5} = -20$.
The spaces $C_1$, $C_2$ are two Riemann surfaces, of arbitrary genus.
If $C_i$, $i = 1,2$, is not a torus, the metric $ds^2(C_i)$
is normalized so that the Ricci scalar is $R_{C_i}= 2 \, k_i$,
with $k_i = \pm 1$. The symbol $\chi_i$ denotes the Euler
characteristic of $C_i$,
while $V_{C_i}$ is proportional to the volume form on $C_i$.
If $C_i$ is not a torus, $V_{C_i}$ is normalized according to
$\int_{C_i} V_{C_i}  = 2\pi$,
with no sum over $i$.
Notice that, compared to \cite{Gauntlett:2004zh},
we have reversed the sign of $\psi$.

The quantities $\lambda$, $F_i$ 
depend on $y$ only and are given by
\begin{align}
e^{6\lambda} &= 
\frac{2   \left(a_1-k_1 \,  y^2\right) \left(a_2-k_2 \ 
   y^2\right)}{a_2  \, k_1+a_1 \,  k_2+2  \, y \, k_1 \, k_2 \left(  y - 3 \, \gamma_0
    \right)}
    \ , \qquad 
F_i   = \frac 13 (a_i - k_i \,  y^2)    \ ,
\end{align}
where  $a_i$, $\gamma_0$ are constants.
The quantity $\zeta \in [0, \pi/2]$   is determined by
\beq
e^{3\lambda} \, \sin \zeta = 2  \, y \  \ .
\eeq
The $G_4$ flux takes the form
\beq \label{GMSW_review_flux}
G_4 = L^3 \, \Big[  
(d\gamma_1 \, V_{C_1} + d\gamma_2 \, V_{C_2} )\, D\psi
- (\chi_1 \, \gamma_2 + \chi_2 \, \gamma_1 +
\chi_1 \, \chi_2 \, \gamma_0) \, V_{C_1} \, V_{C_2}
\Big] \ ,
\eeq
with the functions $\gamma_i$  given as
\begin{align} \label{fi_expr}
\gamma_1 & = \frac{
2  \, a_2 \,  k_1 \,  k_2 \,  y - 6  \, a_2 \,  k_1 \, k_2 \, \gamma_0 +a_1 \,  y+k_1  \, y^3
}{
9  \, k_1 \, k_2 \,   \left(a_2 - k_2 \,  y^2 \right)
}  \, \chi_1
\ , \nn \\
\gamma_2 & = \frac{
2 \,  a_1 \,  k_1 \,  k_2 \,  y -  6  \, a_1 \, k_1 \, k_2 \, \gamma_0 +a_2 \,  y+k_2 \,  y^3
}{
9 \, k_1 \, k_2 \,  \left(a_1 - k_1 \, y^2\right)
}  \, \chi_2 
\ .
\end{align}
We are adopting conventions in which the quantization of $G_4$ flux 
reads
\beq \label{G4_quantization_convention}
\int_{\cC_4} \frac{G_4}{(2\pi \ell_p)^3} \in \mathbb Z \ ,
\eeq
where $\cC_4$ is a 4-cycle and $\ell_p$ is the 11d Planck length.

Let us stress that, in this work, we only consider GMSW
solutions of the form \eqref{GMSW_review_metric},
\eqref{GMSW_review_flux}
 in which none
of the Riemann surfaces $C_i$ is a torus,
so that $k_i \neq 0$.
According to the analysis of \cite{Gauntlett:2004zh},
in order to a have a regular solution 
at least one of the two Riemann surfaces must be a sphere.
We associate the label 1 to this sphere, 
while the label 2 is reserved to a Riemann surface of genus $g$,
with $g = 0$ or $g \ge 2$,
\beq
C_1 = S^2 \ , \qquad
C_2 = \Sigma_g   \ .
\eeq
We should emphasize that the 4-form $V_4$ in \eqref{GMSW_V4}
is understood to have integral fluxes
along 4-cycles. It follows from \eqref{G4_quantization_convention}
that the relation between $V_4$ and $G_4$ is
\beq
V_4 = \frac{G_4}{(2\pi \ell_p)^3} \ .
\eeq 
In the main text, we parametrized $V_4$ by writing
\beq
V_4 = 
\bigg[  d\gamma_\Sigma \, \frac{V_\Sigma}{2\pi}
+ d\gamma_{S^2} \, \frac{V_{S^2}}{2\pi} \bigg] \, \frac{D\psi}{2\pi}
- \Big[ 2\, \gamma_\Sigma + \chi  \, \gamma_{S^2}   \Big] \,
\frac{V_\Sigma}{2\pi}  \, \frac{V_{ S^2}}{2\pi}  \ .
\eeq
Comparison with \eqref{GMSW_review_flux} gives the identifications
\beq \label{fi_identifications}
\gamma_{S^2} = \frac{L^3}{ 2\pi \,  \ell_p^3} \, (\gamma_1 + s_1) \ , \qquad
\gamma_{\Sigma} = \frac{L^3}{ 2\pi \,  \ell_p^3} \, (\gamma_2 + s_2 ) \ ,
\qquad
2 \, s_2 + \chi \, s_1  = 2 \, \chi \, \gamma_0 \ ,
\eeq
where the constants $s_{1,2}$ 
can be chosen at will. In the text, 
this ambiguity in the precise definition of $\gamma_\Sigma$, $\gamma_{S^2}$
is resolved upon construction of $E_4$, when the condition
 \eqref{omegaI_condition} is enforced.

The holographic central charge for these solutions
was analyzed in \cite{Gauntlett:2006ai},
where the explicit value of $c$ is derived in the case $\gamma_0 = 0$,
$\Sigma_g = S^2$. 
In this situation, one verifies that 
$N^{\rm N} = - N^{\rm S}$,
or equivalently $M = 0$, see \eqref{north_and_south_fluxes}, \eqref{M_def}. 
In the notation of \cite{Gauntlett:2006ai} the central charge reads
\beq \label{their_c}
c = \frac{3^{3/2}}{2^6} \, \frac{9 \, (z+1)^3 - (3 \, z^2 + 4 \, z + 3) \, \sqrt X}{z^{3/2}} \, 
p^{3/2} \, q^{3/2} \, N^3 \ ,
\eeq
where the parameters $p$, $q$, $N$ are related to our quantities $N_{S^2}$,
$N_\Sigma$ by $N_{S^2} = N \, p$, $N_\Sigma = N \, q$.
The objects $X$, $z$ are defined as
\beq
X = 9 \, z^2 + 30 \, z + 9 \ , \qquad
z = \frac{2 \, q^2 - p \, q + 2 p^2 - 2 \, (p-q) \, \sqrt{p^2 + p \,q  + q^2}}{3 \, p \, q} \ .
\eeq
 In order to compare \eqref{their_c}  
to the central charge \eqref{central_charge} inferred from the inflow anomaly polynomial,
we need the following identity,
valid for positive numbers $p$, $q$,
\begin{align} \label{strange_identity}
\left(
2 \, p^2 - p \, q + 2 \, q^2 - 2 \, (p-q) \, \sqrt{p^2 + p \, q + q^2}
 \right)^{1/2}
 & = 
 q-p + \sqrt{p^2 + p  \, q + q^2}   \ .
\end{align}
With the help of \eqref{strange_identity},
the central charge \eqref{their_c} can be rewritten as
\beq
c = - \frac{9}{16} \, N^3 \, (p + q) \, (2 \, p^2 + p \, q + 2 \, q^2)
+ \frac 98 \, N^3 \, (p^2 + p \, q + q^2)^{3/2} \ ,
\eeq
which indeed matches \eqref{central_charge} for $\chi = 2$, $M = 0$.



\subsection{Details on the construction of $E_4$ for GMSW}  

Let us first discuss the construction of $V_4^{\rm eq}$. The isometries
of $M_6$ we gauge are the $U(1)$ symmetry associated to $\psi$
and the $SO(3)$ symmetry associated to $S^2$. The former has 
background connection $A^\psi$, while the latter is associated to $A^a$.

The claim that the 4-form $V_4^{\rm eq}$ is given by
\begin{align} \label{clever_V4eq}
V_4^{\rm eq} & = 
d\bigg[ 
\bigg(  \gamma_\Sigma \, \frac{V_\Sigma}{2\pi}
+ \gamma_{S^2} \, e_{2}^{S^2} \bigg) \, \frac{ ( D \psi)^{\rm g}}{2\pi}
\bigg]    \\
& = 
\bigg(  d\gamma_\Sigma \, \frac{V_\Sigma}{2\pi}
+ d\gamma_{S^2} \, e_2^{S^2}  \bigg) \, \frac{( D \psi)^{\rm g}}{2\pi}
+ \bigg( \gamma_\Sigma \, \frac{V_\Sigma}{2\pi}
+ \gamma_{S^2} \, e_2^{S^2}  \bigg)
\bigg(   - 2 \, e_2^{S^2} - \chi \, \frac{V_\Sigma}{2\pi} 
+ 2 \, \frac{F^\psi}{2\pi}  \bigg) 
\ . \nn
\end{align}
We have exploited the relation
\beq \label{gauged_psi_identity}
\frac{d   ( D \psi)^{\rm g}}{2\pi} = - 2 \, e_2^{S^2} - \chi \, \frac{V_\Sigma}{2\pi} 
+ 2 \, \frac{F^\psi}{2\pi}  \ ,
\eeq
which is derived below.
The 2-form $e_2^{S^2}$ is the global angular form for $SO(3)$, or equivalently
the closed and gauge-invariant completion of ${V_{S^2}}/(2\pi)$.
More explicitly,
\beq \label{e2_def}
e_2^{S^2} = \frac{1}{8\pi} \, ( \epsilon_{abc} \, Dz^a \, Dz^b \, z^c - 2 \, F_{a} \, z^a) 
= \frac{V_{S^2}}{2\pi} - \frac 12 \, \frac{d(z^a \, A_a)}{2\pi}
\ , \qquad
Dz^a = dz^a + \epsilon^{abc} \, A_b \, z_c \ .
\eeq 
A useful identity regarding $e_2^{S^2}$ is the Bott-Cattaneo formula \cite{bott1999integral},
\beq
\int_{S^2} (e_2^{S^2})^{2s+2} = 0 \ , \qquad
\int_{S^2} (e_2^{S^2})^{2s+1} = 2^{-2s} \, [p_1(SO(3))]^s \ , \qquad
s = 0,1,2,\dots
\eeq

The object $V_4^{\rm eq}$ is manifestly closed and gauge-invariant,
and reduces to $V_4$ if $F^\psi$ and $F^a$ are turned off.
Moreover, $V_4^{\rm eq}$ is globally well-defined.
Indeed, $S^2$ does not shrink anywhere on the $y$ interval,
and all terms with $ (D \psi)^{\rm g}$ are accompanied by a factor $dy$,
so that there is no singularity at the endpoints of the $y$ interval,
where $S^1_\psi$ shrinks.

Before analyzing $V_4^{\rm eq}$ further, let us derive the
identity \eqref{gauged_psi_identity}.
If all external connections are turned off,
\beq
\frac{D\psi}{2\pi} = \frac{d\psi}{2\pi} -2 \, \frac{A_{S^2}}{2\pi} - \chi \, \frac{A_\Sigma}{2\pi} \ , \qquad
dA_{S^2} = V_{S^2} \ , \qquad
dA_\Sigma= V_\Sigma \ .
\eeq
The 1-forms $A_{S^2}$, $A_\Sigma$ are antiderivatives of the volume
forms on $S^2$, $\Sigma$, and are only locally defined.
The gauging of the 1-form $d\psi$ is given by
\beq \label{dpsi_gauged}
(d\psi)^{\rm g} = d\psi + 2 \, A^\psi + A^a \, \Big[  z_a + 2\, \epsilon_{abc} \, z^b \, 
\nabla^\mu z^c \, (A_{S^2})_\mu
 \Big] \ .
\eeq 
The index $\mu$ is a curved 2d index on $S^2$ associated to local
coordinates $\zeta^\mu$, so that, for example,
$dz^a = \partial_\mu z^a \, d\zeta^\mu$,
$g^{S^2}_{\mu\nu} = \partial_\mu z^a \, \partial_\nu z_a$.
The symbol $\nabla$ is the Levi-Civita connection on $S^2$.
Notice the appearance in $(d\psi)^{\rm g}$ of terms proportional to $A^a$.
They are a consequence of 
the second term in the Killing 1-form
$k_a$ in \eqref{S2_base_isom}.

In order to compute $(D\psi)^{\rm g}$, we also need
\beq \label{AS2_gauged}
(A_{S^2})^{\rm g} = (A_{S^2})_\mu \, (d\zeta^\mu)^{\rm g}
= (A_{S^2})_\mu \, \Big[ d\zeta^\mu + \epsilon_{abc} \, z^b \, \nabla^\mu z^c \, A^a
\Big] \ .
\eeq
We are now in a position to write
\begin{align} \label{Dpsi_gauged}
\frac{(D\psi)^{\rm g}}{2\pi} &= \frac{(d\psi)^{\rm g}}{2\pi} -2 \, \frac{(A_{S^2})^{\rm g}}{2\pi} - \chi \, \frac{A_\Sigma}{2\pi}   = \frac{d\psi}{2\pi} + 2 \, \frac{A^\psi}{2\pi}
 - 2 \,\bigg[ \frac{A_{S^2}}{2\pi} - \frac 12 \, \frac{z^a \, A_a}{2\pi} \bigg]
- \chi \, \frac{A_\Sigma}{2\pi}  \ .
\end{align}
Notice the cancellation of the terms with $\epsilon_{abc}$ against
\eqref{dpsi_gauged} and \eqref{AS2_gauged}.
Making use of \eqref{e2_def}, it is now straightforward to check that
\eqref{Dpsi_gauged} implies \eqref{gauged_psi_identity}.

In order to make contact with the language of section \ref{sec_general_param},
we have to expand $V_4^{\rm eq}$ is powers of the external connections,
\begin{align}
V_4^{\rm eq} & = V_4^{\rm g} + F^\psi \, \omega_\psi^{\rm g}
+ F^a \, \omega_a
+ (F^\psi)^2 \, \sigma_{\psi \psi}
+ F^a \, F^b \, \sigma_{ab}
+ 2 \, F^\psi \, F^a \,\sigma_{\psi a} \ .
\end{align}
The $\omega$ and $\sigma$ quantities are extracted from comparison with
\eqref{clever_V4eq}. In what follows, we only need the expression of $\omega_\psi$
and $\omega_a$,
\begin{align}
\omega_\psi & = \frac{2}{2\pi} \, \bigg( \gamma_\Sigma \, \frac{V_\Sigma}{2\pi}
+ \gamma_{S^2} \, \frac{V_{S^2}}{2\pi} \bigg) \ , \nn \\
\omega_a & =  
\frac{z_a}{2\pi} \, \bigg[ 
- \frac 12 \, \bigg( 
d\gamma_{S^2} \, \frac{D\psi}{2\pi}
- (2 \, \gamma_\Sigma + \chi \, \gamma_{S^2}) \, \frac{V_\Sigma}{2\pi}
\bigg)
+ 2 \, \gamma_{S^2} \, \frac{V_{S^2}}{2\pi}
\bigg] \ .
\end{align}
We verify $\int_{M_6} V_4 \, \omega_a = 0$, while we compute
\beq
\int_{M_6} V_4 \, \omega_\psi = \frac{2}{2\pi} \Big[\gamma_{S^2} \, \gamma_{\Sigma} \Big]_{\rm S}^{\rm N} \ .
\eeq
This quantity must be set to zero. As a result, we can express
the four quantities $\gamma_{S^2,\Sigma}^{\rm N,S}$ in terms
of the three flux quanta $N_{S^2}$, $N_\Sigma$, $M$,
\beq
\gamma_{S^2}^{\rm N,S} = \frac{M \, N_{S^2}}{2 \, N_{\Sigma}  - \chi \, N_{S^2}} \pm \frac 12 \, N_{S^2} \ , \qquad
\gamma_{\Sigma}^{\rm N,S} = - \frac{M \, N_{\Sigma}}{2 \, N_{\Sigma}  - \chi \, N_{S^2}} \pm \frac 12 \, N_{\Sigma}  \ .
\eeq

Let us now discuss the equivariant completion of the
harmonic 2-forms $\omega_\alpha$ parametrized in \eqref{omega_alpha_param}.
It is given as
\begin{align}
\omega_\alpha^{\rm eq} & = d\bigg[  H_\alpha \, \frac{(D\psi)^{\rm g}}{2\pi}  \bigg]
+ t_{\alpha S^2} \, e_2^{S^2} + t_{\alpha \Sigma} \, \frac{V_\Sigma}{2\pi}
\nn \\
& = dH_\alpha \, \frac{(D\psi)^{\rm g}}{2\pi}
+ ( t_{\alpha S^2}  - 2 \, H_\alpha) \, e_2^{S^2}
+ ( t_{\alpha \Sigma}  - \chi \, H_\alpha) \,\frac{V_\Sigma}{2\pi}  
+ 2 \, H_\alpha \, \frac{F^\psi}{2\pi} \ .
\end{align}
This is manifestly closed and gauge-invariant and reduces to 
\eqref{omega_alpha_param} if all external connections are turned off.
Moreover, $\eqref{omega_alpha_param}$ is globally defined, since 
$(D\psi)^{\rm g}$ is accompanied by $dy$.


\bibliographystyle{./ytphys}
\bibliography{./refs}

\end{document}